\documentclass[a4paper,oneside]{article}
\usepackage[english]{babel}
\usepackage{graphicx, amsmath, amssymb, amsfonts, mathtools, graphics, url, tikz, hyperref, feynmp-auto, fullpage, cancel, bookmark, multirow, enumitem}
\usepackage[font=small,labelfont=bf]{caption}
\usepackage[numbers,elide]{natbib}
\title{Electric and heat transport in a charge two-channel Kondo device}
\author{G.A.R. van Dalum\textsuperscript{1}, A.K. Mitchell\textsuperscript{2} and L. Fritz\textsuperscript{1}\\{\small\textsuperscript{1}\emph{Institute for Theoretical Physics, Utrecht University, Princetonplein 5, 3584 CC Utrecht, The Netherlands}}\\{\small\textsuperscript{2}\emph{School of Physics, University College Dublin, Belfield, Dublin 4, Ireland}}}
\date{}

\newcounter{newsections}
\makeatletter
\@addtoreset{section}{newsections}
\makeatother

\begin{document}
\begin{fmffile}{diagrams}
\maketitle


\begin{abstract}
\noindent Motivated by the experimental realization of a multi-channel charge Kondo device [Iftikhar \emph{et al.}, Nature \textbf{526}, 233 (2015)], we study generic charge and heat transport properties of the charge two-channel Kondo model. We present a comprehensive discussion of the out-of-equilibrium and time-dependent charge transport, as well as thermal transport within linear response theory. The transport properties are calculated at, and also in the vicinity of, the exactly solvable Emery-Kivelson point, which has the form of a Majorana fermion resonant level model. We focus on regimes where our solution gives exact results for the physical quantum dot device, and highlight new predictions relevant to future experiments.
\end{abstract}


\section{Introduction}
The Kondo model is one of the paradigmatic models of strong correlation physics~\cite{hewson1997kondo}. In its original context, it was introduced to describe the physics of dilute magnetic impurities embedded in a metallic system~\cite{anderson1961localized,schrieffer1966relation}. The magnetic impurity is screened below an emergent low temperature scale $T_K$, the Kondo temperature, forming a non-trivial many-body singlet state which shows the behavior of a local Fermi liquid (FL)~\cite{nozieres1974fermi}. This scenario explained the unexpected increase in resistivity of such systems in the low temperature regime as a consequence of enhanced spin-flip scattering from the impurities~\cite{anderson1970poor}. More recently, it was realized that semiconductor quantum dot devices with strong local Coulomb interaction can also display Kondo physics~\cite{goldhaber1998kondo,*cronenwett1998tunable,kouwenhoven2001revival}. The Kondo model also played an important role on the theoretical side: it led to many new concepts and developments~\cite{wilson1975renormalization,wiegmann1981exact,andrei1980diagonalization} and still plays an important role as a testbed for techniques of strong correlations.

\begin{figure}[t]
  \centerline{\includegraphics[width=6.4cm]{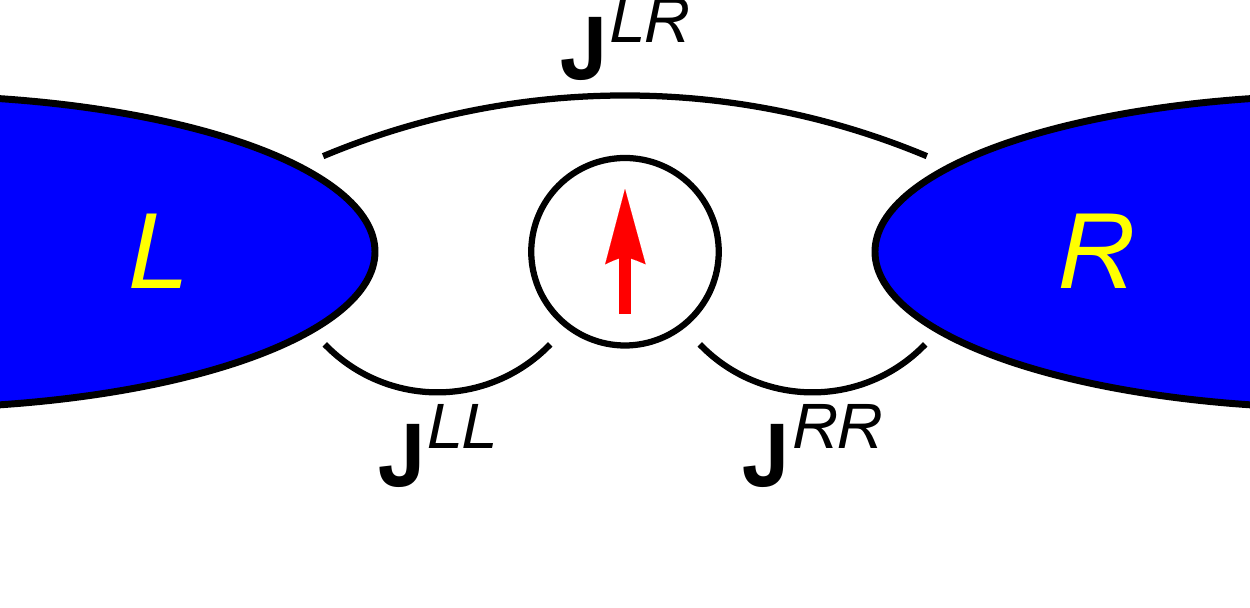}}
\caption{\label{fig:2CK_model}Schematic of the most general anisotropic 2CK model. The coupling constants are of the form $\mathbf{J}^{\alpha\beta}\equiv\left(J_x^{\alpha\beta}\;J_y^{\alpha\beta}\;J_z^{\alpha\beta}\right)^T$, also allowing for impurity-mediated exchange cotunneling between the leads via $\mathbf{J}^{LR}$.}
\end{figure}

The two-channel Kondo (2CK) model~\cite{nozieres} is a non-trivial extension of the Kondo model: two \emph{independent} metallic baths couple to a single impurity spin degree of freedom, and compete to screen it (see Fig.~\ref{fig:2CK_model} with $\mathbf{J}^{LR}=0$). In the case where one of the two baths couples more strongly, this bath eventually screens the impurity spin, while the less strongly coupled bath decouples asymptotically in the zero temperature limit, leading to an effective single channel Kondo effect with the ground state properties described as a FL. However, if both baths are coupled equally strongly (Fig.~\ref{fig:2CK_model} with $\mathbf{J}^{LR}=0$ and $\mathbf{J}^{LL}=\mathbf{J}^{RR}$) the Kondo screening is frustrated, and the ground state shows non-Fermi liquid (NFL) behavior with an impurity entropy indicative of a ground state degeneracy of $\sqrt{2}$, characteristic of a Majorana fermion. The effective Emery-Kivelson theory~\cite{emery} describing the critical point is indeed a local Majorana fermion, resonantly coupled to one-dimensional Majorana fermions.

Despite being a very interesting model showing non-Fermi liquid behavior, an experimental realization is notoriously difficult, but not impossible \cite{potok2007observation,keller2015universal,experiment,iftikhar2018tunable}. One of the main challenges is to ensure the strict independence of the two baths, meaning $\mathbf{J}^{LR}=0$ in Fig.~\ref{fig:2CK_model}.
For a single ultra-small quantum dot tunnel-coupled to two metallic leads, one has $\mathbf{J}^{LR}=\sqrt{\mathbf{J}^{LL}\mathbf{J}^{RR}}\ne 0$, and in this case a simple canonical transformation yields a pure one-channel model. Even for coupled quantum dot systems \cite{mitchell2011two} or single-molecule junctions \cite{mitchell2017kondo} in the spin-$\tfrac{1}{2}$ Kondo regime, $\mathbf{J}^{LR}$ is always finite and generates a crossover to a FL state on the lowest energy scales \cite{sela2011exact,*asymmag}.
However, replacing one lead with a quantum `box' (a large dot or grain) with finite capacitance suppresses inter-channel charge transfer \cite{zarand2006quantum}, such that $\mathbf{J}^{LR}=0$. This was demonstrated experimentally in Refs.~\cite{potok2007observation,keller2015universal} and the 2CK critical point was realized.

An alternative version of the 2CK model exploits a charge degeneracy in a large quantum dot instead of a spin degeneracy~\cite{matveev1, matveev2}: this setup is called the charge two-channel Kondo (C2CK) effect and is the focus of this work. In 2015, Iftikhar \emph{et al.}~\cite{experiment} realized the C2CK effect experimentally in a quantum dot device, enabling a spectacular experimental verification of theoretical predictions~\cite{matveev1,matveev2,andrew}. Indeed, this device was also able to probe the more exotic charge three-channel Kondo effect~\cite{iftikhar2018tunable}.

In this paper, we provide a comprehensive discussion of the transport properties of the C2CK system, and provide theoretical predictions for future transport measurements. The paper is organized as follows: in Sec.~\ref{sec:model} we start with a discussion of the two-channel Kondo model. We introduce it in its original spin version and also discuss aspects of the Emery-Kivelson mapping needed for the most technical parts of the paper. We then introduce the C2CK model and provide a simple dictionary to go back and forth between the spin and charge versions of the model. We end with a discussion of the limitations of the Emery-Kivelson solution and extensions. In Sec.~\ref{sec:preliminaries} we first discuss the technicalities of the non-equilibrium calculation for charge transport. We also introduce the general framework of linear response theory, which is required for the discussion of heat transport. In Sec.~\ref{sec:keldysh} we discuss exact charge transport properties of the Emery-Kivelson theory both for time-dependent and steady state situations, making explicit connection to the physical C2CK system at each stage. The dc solution discussed in this section is equivalent to the known solution of the spin 2CK model~\cite{sh2,eran1,eran2,andrew}, but adapted to the C2CK setup. On the other hand, in the ac case we use the framework from Ref.~\cite{sh1} (again adapted to the C2CK setup) to find an integral expression for the ac conductance, which we take a step further by evaluating it to find a closed-form expression. We go on to discuss the charge conductance within linear response using the Kubo formula. Although a voltage bias can be treated exactly within the Emery-Kivelson mapping, we explain why the full non-equilibrium calculation cannot be performed for heat transport. However, we can calculate heat transport properties within linear response. This is done in Sec.~\ref{sec:heat}, leading to a novel result for the heat conductance. We discuss the possibility to use heat transport to verify the Majorana character of the critical theory as well as the Wiedemann-Franz law at the NFL point in Sec.~\ref{sec:wf}, elaborating on results we published recently in Ref.~\cite{dalum2020wf}. In Sec.~\ref{sec:pert} we discuss the limits of validity of the Emery-Kivelson solution on a quantitative level and the possible corrections to this solution (the expressions for these corrections have previously been presented in Refs.~\cite{pert1,pert2,pert3}, but we provide a concise derivation for the reader's convenience). However, we emphasize already at this stage that our results for the NFL fixed point properties, and the subsequent crossovers to a FL state, are exact and not specific to the Emery-Kivelson approach used to obtain them. We conclude in Sec.~\ref{sec:conclusion}. Technical details are provided in extensive appendices.


\section{The anisotropic two-channel Kondo model}\label{sec:model}
We start by introducing the most general form of the anisotropic 2CK model, shown in Fig.~\ref{fig:2CK_model}. The model consists of two leads and a local part. For generality we also include a term describing an impurity magnetic field. The Hamiltonian then takes the form $\hat{H}=\hat{H}_\text{leads}+\hat{H}_\text{loc}+\hat{H}_{\text{mag}}$. We model the leads as effectively one-dimensional channels with Fermi velocity $v_F$ and a constant density of states. In the absence of any bias between the leads, we therefore have
\begin{equation}
\hat{H}_\text{leads}=\sum_{\alpha} \hat{H}_{\alpha}=i\hbar v_F\sum_\alpha\sum_\sigma\int\limits_{-\infty}^{\infty}\mathrm{d}x\,\psi^\dagger_{\alpha\sigma}(x)\partial_x\psi_{\alpha\sigma}(x) \;,
\end{equation}
where $\psi_{\alpha\sigma}(x)$ are the fermionic operators for lead $\alpha=L$ (left) or $R$ (right), with spin $\sigma=\uparrow$ or $\downarrow$. Meanwhile, the local part of the anisotropic 2CK model is described by the Hamiltonian~\cite{sh2}
\begin{equation}
\hat{H}_\text{loc}=\sum_{\alpha,\beta}\sum_\lambda J^{\alpha\beta}_\lambda s^\lambda_{\alpha\beta}\tau^\lambda \;,\label{eq:localH}
\end{equation}
with
\begin{equation}
\mathbf{s}_{\alpha\beta}\equiv\frac{1}{2}\sum_{\sigma,\sigma^\prime}\psi^\dagger_{\alpha\sigma}(0)\pmb{\sigma}_{\sigma\sigma^\prime}\psi_{\beta\sigma^\prime}(0) \;.
\end{equation}
Here $\alpha,\beta$ label the leads, $J^{\alpha\beta}_\lambda$ are the respective exchange coupling constants, $\mathbf{s}_{\alpha\beta}$ is the local electron spin density of the leads evaluated at the origin ($x=0)$, $\pmb{\sigma}$ is the vector of Pauli matrices, and $\lambda=x,y,z$. The operator for the impurity spin-$\tfrac{1}{2}$ degree of freedom, located at the origin, is denoted $\pmb{\tau}$. Finally, we take the constant magnetic field $B$ coupling to the impurity spin to be in the $z$-direction, giving $\hat{H}_{\text{mag}}=-B\tau^z$. The full Hamiltonian of the anisotropic 2CK model at zero bias is thus given by
\begin{equation}
\hat{H}=i\hbar v_F\sum_\alpha\sum_\sigma\int\limits_{-\infty}^{\infty}\mathrm{d}x\,\psi^\dagger_{\alpha\sigma}(x)\partial_x\psi_{\alpha\sigma}(x)+\sum_{\alpha,\beta}\sum_\lambda J^{\alpha\beta}_\lambda s^\lambda_{\alpha\beta}\tau^\lambda-B\tau^z \;.\label{eq:ham2}
\end{equation}


\subsection{Two-channel physics}\label{sec:2c_phys}
The addition of a second channel introduces behavior that is not present in the ordinary single channel Kondo model. Of particular interest is the situation with two independent baths ($\mathbf{J}^{LR}=0$), symmetric couplings ($\mathbf{J}^{LL}=\mathbf{J}^{RR}$) and no magnetic field ($B=0$). At this special point, both leads compete to form a Kondo singlet with the impurity at low temperatures; however the LR symmetry of the system frustrates complete screening. A signature of this is the finite residual impurity entropy $S_{\text{imp}}$ at temperature $T=0$ ({\it i.e.}, the entropy of the full system minus the entropy of the free leads). For this 2CK point, $S_\text{imp}=\frac{1}{2}\ln 2$ as the temperature goes to zero, characteristic of a Majorana degree of freedom~\cite{entropy}. At this special point in parameter space unconventional NFL behavior emerges, most notably in the temperature dependence of thermodynamic quantities. In particular, the heat capacity $\sim T\ln T$ and the magnetic susceptibility $\sim\ln T$ \cite{entropy,NFL1,universality}. Relaxing the above conditions and breaking these symmetries relieves the frustration and leads to a more conventional FL state, with vanishing residual entropy, linear temperature scaling of heat capacity, and constant low-temperature magnetic susceptibility. The symmetric model is therefore an NFL critical point separating different FL phases.

\begin{figure}[t]
  \centerline{\begin{tabular}{c|c|c}
    \includegraphics[width=5cm]{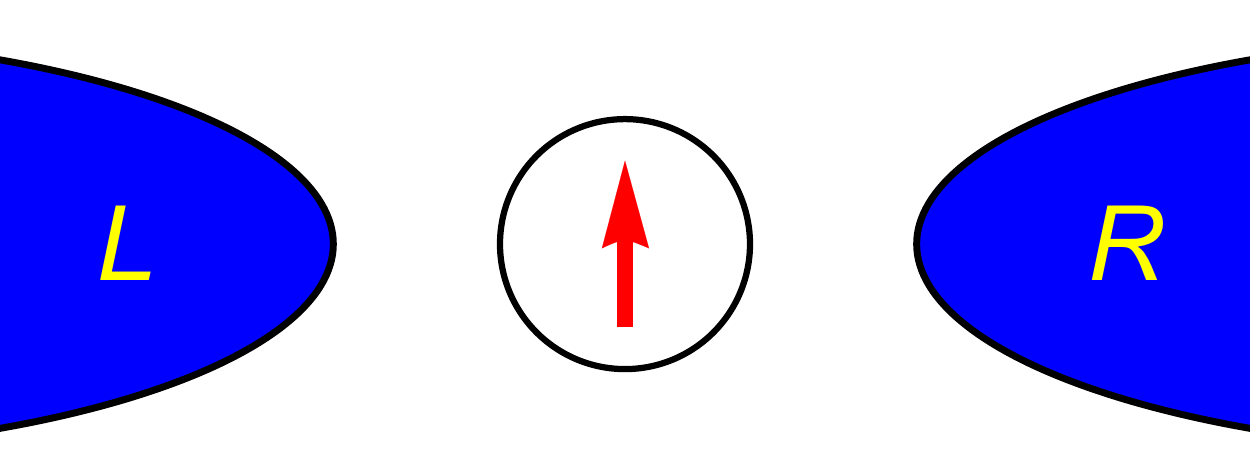} & \includegraphics[width=5cm]{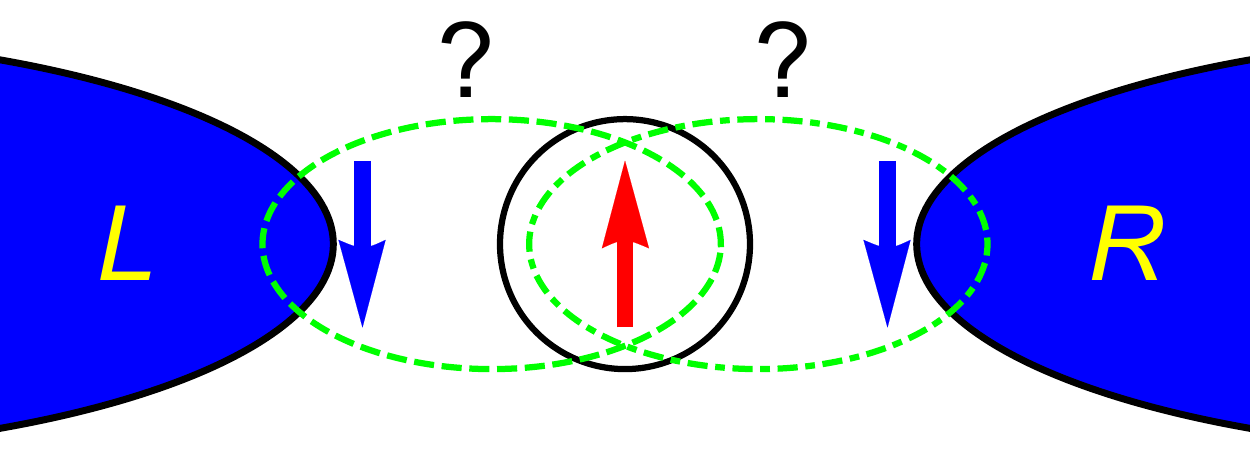} & \includegraphics[width=5cm]{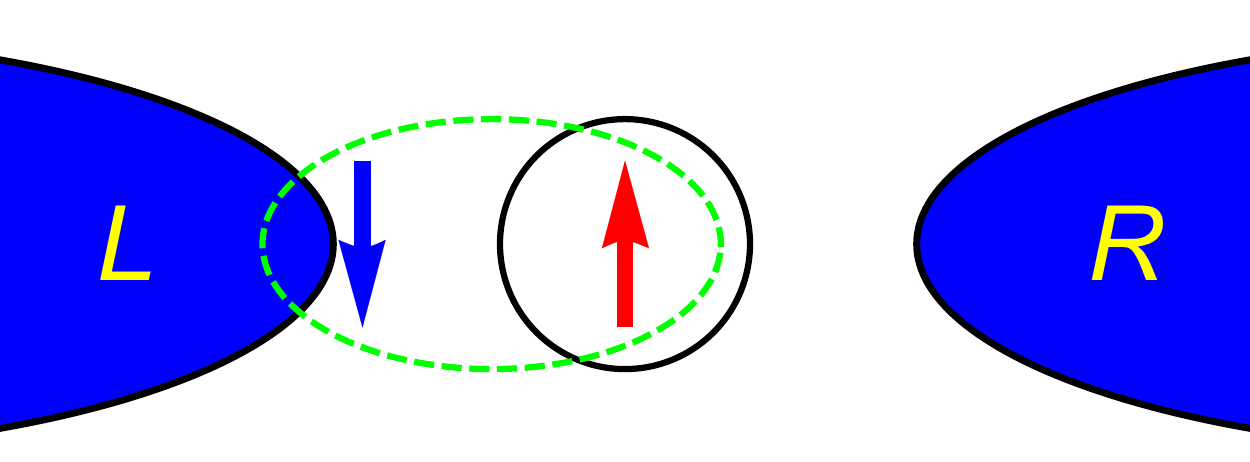}\\$T>T_K$ & $T^*<T<T_K$ & $T<T^*$\end{tabular}}
\caption{\label{fig:phases}Distinct temperature regimes of the 2CK model with $\mathbf{J}^{LR}=0$. Left: isolated local moment, with weak coupling and little heat transport. Middle: NFL critical point, characterized by frustration and unconventional heat transport signatures. Right: FL regime resulting in a transmission node. Illustrated is the particular case of small $LR$ coupling asymmetry such that one lead forms a Kondo singlet with the impurity, thereby decoupling the other lead and suppressing thermal transport (``Kondo blockade''). By contrast, note that electrical conductance is exactly zero at any temperature if $\mathbf{J}^{LR}=0$.}
\end{figure}

If the baths are independent ($\mathbf{J}^{LR}=0$), there is no charge transport between $L$ and $R$ leads, by construction (the total charge in $L$ and $R$ leads is separately conserved). However heat transport, due to a temperature difference between $L$ and $R$ leads, is in general finite due to spin-flip scattering (there is only global spin conservation, since the spin of $L$ and $R$ leads is not separately conserved). The model supports several regimes~\cite{FLNFL,andrew} illustrated in Fig.~\ref{fig:phases}, which have distinct thermal transport signatures.

At high temperatures, the effective (renormalized) coupling between the impurity and the leads is weak. As a result, the impurity forms a nearly free local moment, and heat transport between the leads through the impurity is perturbatively small. When the temperature is decreased below $T_K$ however, the Kondo effect sets in and the renormalized coupling between the impurity and the leads increases to a non-perturbative intermediate value. In this regime, the leads compete to screen the impurity spin. If the couplings are symmetric, {\it i.e.}, $\mathbf{J}^{LL}=\mathbf{J}^{RR}$, this results in frustration, as discussed above. The system then approaches the NFL critical point as $T\ll T_K$, and both leads remain coupled to the impurity. However, if there is a small detuning present ({\it e.g.}, a magnetic field $B\ne 0$, or an asymmetry in the couplings $\mathbf{J}^{LL}-\mathbf{J}^{RR}\ne 0$), an additional energy scale $k_BT^*$ emerges, below which the frustration is relieved. As $T\ll T^*$, the system instead flows towards the single channel FL ground state. This FL ground state does not support transport between the leads through the impurity. For example, a finite magnetic field locks the impurity into a single spin state blocking spin scattering, and asymmetry in the coupling between the impurity and the leads results in the decoupling of the less strongly coupled lead (the ``Kondo blockade'' scenario of Ref.~\cite{mitchell2017kondo}). Of particular interest is the crossover from the intermediate NFL region (where the temperature is still sufficiently large that the detuning perturbation can be neglected) to the FL regime (which pertains on the lowest temperature scales, where the renormalized detuning is large and dominates). Note  however, that this ``FL crossover'' only shows universal behavior when there is a clear separation of scales, $T^*\ll T_K$.

\begin{figure}[t]
  \centerline{\includegraphics[width=6cm]{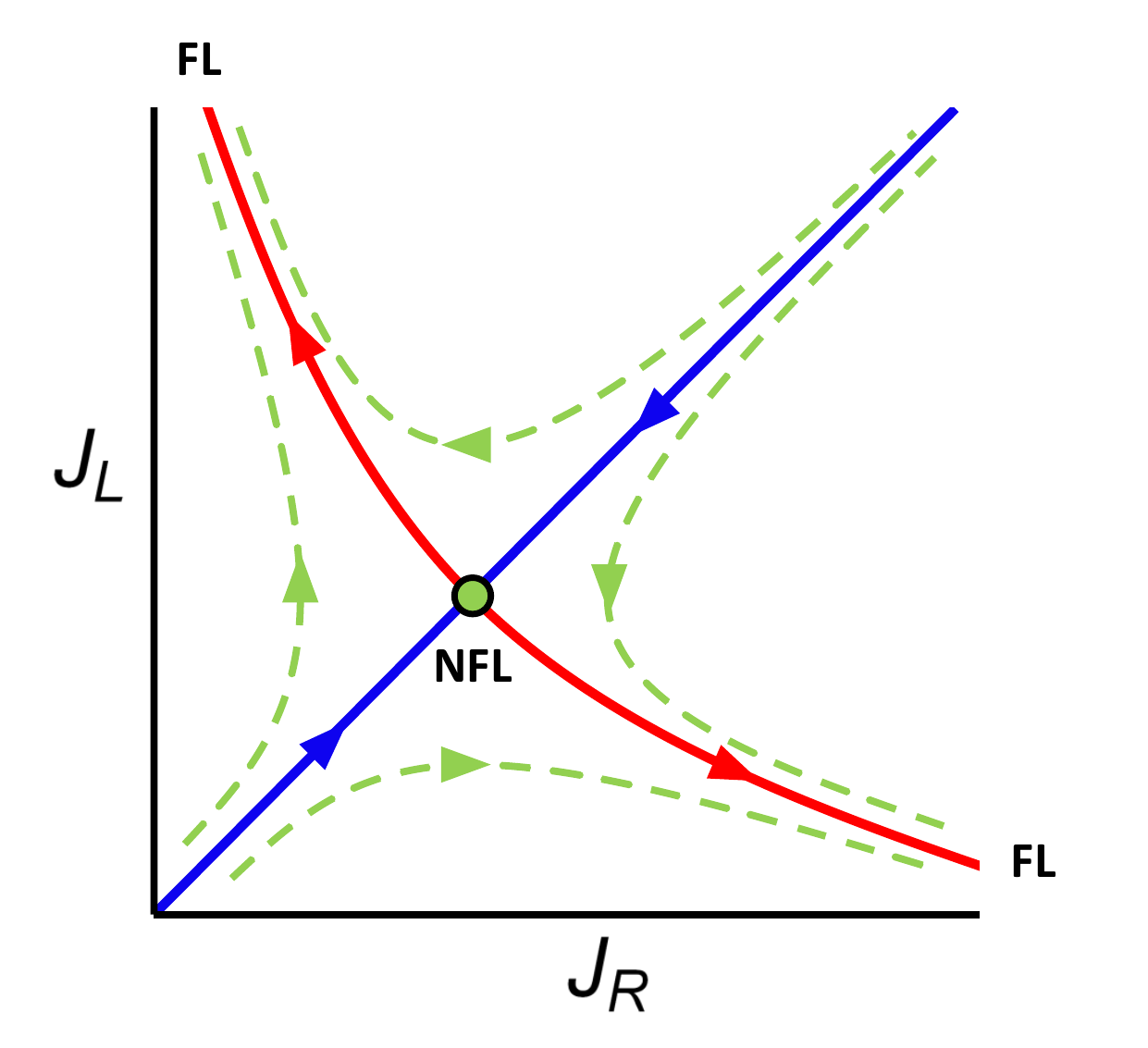}}
\caption{\label{fig:RG_flow}Schematic RG flow of the spin-isotropic 2CK model, in absence of exchange cotunneling between the leads. The solid inward-pointing arrows (blue) correspond to the $LR$ symmetric case, {\it i.e.}, the flow towards the NFL fixed point; the solid outward-pointing arrows (red) describe the flow away from this point, towards the FL regime, due to infinitesimal detuing perturbation $J_\Delta = J_L - J_R$ (this is a universal crossover between NFL and FL fixed points). The dashed lines (green) correspond to the behavior at finite $J_\Delta$.}
\end{figure}

In the spin-isotropic model (setting $J^{\alpha\alpha}_x=J^{\alpha\alpha}_y=J^{\alpha\alpha}_z\equiv J_\alpha$ and $\mathbf{J}^{LR}=0$), the above is summarized by the renormalization group (RG) flow illustrated in Fig.~\ref{fig:RG_flow}~\cite{nozieres,RGflow}. If the system has exact $LR$ symmetry, this is preserved under RG: the system flows along the blue solid lines (the diagonal) in Fig.~\ref{fig:RG_flow} towards the intermediate coupling NFL fixed point, starting either from weak or strong coupling. However, if there is a small $LR$ asymmetry, then the system first flows towards the NFL fixed point, but then flows away because the detuning grows under RG, and eventually a strong-coupling FL fixed point is reached (dashed lines). The coupling asymmetry $J_\Delta\equiv J_L-J_R$ is therefore an RG relevant perturbation. Similarly, magnetic field $B$ and exchange cotunneling $\mathbf{J}^{LR}$ are relevant. The smaller the detuning perturbation, the smaller $T^*$, so the closer the system flows along the solid lines. For $T^* \ll T_K$, the FL scale is quadratic in the perturbation strength, $T^* \sim J_\Delta^2$ (or $\sim B^2$) \cite{universality}. In the limit $T^*\ll T\ll T_K$, the system starts out very close to the NFL fixed point and follows the universal FL crossover line (solid red lines).

With the isotropic RG flow in mind, it should be noted that the flow diagram of the \emph{spin-anisotropic} 2CK model contains additional axes corresponding to the spin anisotropies, and the full SU(2) spin symmetry is broken.
Setting $J^{\alpha\beta}_x=J^{\alpha\beta}_y\equiv J^{\alpha\beta}_\bot$ but allowing $J^{\alpha\beta}_\bot \ne J^{\alpha\beta}_z$ reduces the spin symmetry to U(1), but a Kondo effect can still arise (no Kondo effect is possible if the symmetry is lowered further by allowing $J^{\alpha\beta}_x \ne J^{\alpha\beta}_y$). Therefore we consider the model with general $J^{\alpha\beta}_\bot$ and $J^{\alpha\beta}_z$. We also now set $J_z^{LR}=J_z^{RL}=0$. With this choice, Eq.~(\ref{eq:localH}) can be written as
\begin{equation}
\hat{H}_\text{loc}=\sum_{\alpha,\beta}\frac{J^{\alpha\beta}_\bot}{2}(s^+_{\alpha\beta}\tau^-+s^-_{\alpha\beta}\tau^+)+(J_z^{LL}s^z_{LL}+J_z^{RR}s^z_{RR})\tau^z\;,\label{eq:ham1}
\end{equation}
where $\tau^\pm=\tau^x\pm i\tau^y$ and $s^\pm_{\alpha\beta}=s^x_{\alpha\beta}\pm i s^y_{\alpha\beta}$ are the raising and lowering operators corresponding respectively to the impurity and lead spins. The first term of Eq.~(\ref{eq:ham1}) can thus be interpreted as spin-flip interactions, while the second term describes Ising type interactions. Returning to our discussion of the RG flow (again setting $\mathbf{J}^{LR}=0$), the flow diagram of this anisotropic model has two additional axes, representing the anisotropies $\Delta J_z^{\alpha\alpha}\equiv J_z^{\alpha\alpha}-J_\bot^{\alpha\alpha}$ for $\alpha=L$ or $R$. However, unlike the perturbations $J_{\Delta}$ or $B$, the anisotropies $\Delta J_z^{\alpha\alpha}$ are RG \emph{irrelevant} parameters~\cite{scaling}. As a result, the system will always end up flowing towards an isotropic fixed point upon scaling. This in turn means that Ising type interactions $J_z^{\alpha\alpha}$ are generated by the RG flow as the energy scale of the system $T/T_K$ goes to zero. In terms of Fig.~\ref{fig:RG_flow}, systems with different $J_z^{\alpha\alpha}$ start their flow out of the plane, but end up at a fixed point in the plane. Importantly, if $J_\Delta$ is very small, the system will flow first to the isotropic NFL fixed point, \textit{and then remain in the isotropic plane along the entire NFL to FL crossover}, independently of the spin anisotropy at the start of the flow. The FL crossover is therefore universal and pertains for any anisotropy in the bare model.

Indeed, there is a stronger sense in which the FL crossover is universal. The nature of the detuning leading to the FL crossover scale $T^*$ does not affect the FL crossover behavior itself. The same crossover, as a function of the rescaled $T/T^*$, is generated independently of the symmetry-breaking perturbation causing it \cite{sela2011exact,*asymmag} (only the crossover scale $T^*$ depends on the precise perturbations).

We exploit the emergent spin isotropy and the universality of the FL crossover in the following. Specifically, we utilize an exactly solvable point of the model,  corresponding to a specific value of the bare spin anisotropy, to access the NFL fixed point properties. Then we can study the universal FL crossover due to a small symmetry-breaking perturbation (we choose a finite impurity magnetic field here while maintaining $LR$ symmetry, as this case is the simplest to treat). Both the NFL fixed point properties and the FL crossover obtained in this way are valid for a bare model with different anisotropy and/or different perturbations (or even a combination of different perturbations). In particular, our results hold for the C2CK model, as shown below.


\subsection{Exactly solvable point of the model}\label{sec:toulouse}

We make use of the fact that Eq.~(\ref{eq:ham2}) describes an effective one-dimensional system to bosonize the model.
As per Eq.~(\ref{eq:ham1}), we take $J^{\alpha\beta}_x=J^{\alpha\beta}_y\equiv J^{\alpha\beta}_\bot$ and $J_z^{LR}=J_z^{RL}=0$. We now  additionally constrain $J_z^{LL}=J_z^{RR}\equiv J_z$. Importantly, it was shown by Emery and Kivelson~\cite{emery} that this 2CK model can be mapped onto a non-interacting resonant level model at a special point in parameter space -- namely, when $J_z = 2\pi h v_F$, where $v_F$ is the Fermi velocity of the leads. This procedure was generalized to a non-equilibrium situation, with a time-dependent bias voltage between the leads, by Schiller and Hershfield~\cite{sh2}. We make extensive use of these mappings in the following, and hence recapitulate the derivation below.

In short, the mapping presented in Refs.~\cite{emery,sh2} is done through a series of steps, starting with the bosonization of the fermionic fields, $\psi_{\alpha\sigma}(x)\propto e^{-i\Phi_{\alpha\sigma}(x)}$. Then, a change of basis (canonical transformation) is performed by taking new linear combinations of the old bosonic fields $\Phi_{L\uparrow}(x)$, $\Phi_{L\downarrow}(x)$, $\Phi_{R\uparrow}(x)$ and $\Phi_{R\downarrow}(x)$; the new fields are referred to as the charge, spin, flavor and spin-flavor modes, defined as
\begin{align}
\Phi_c(x)&\equiv\frac{1}{2}\big(\Phi_{L\uparrow}(x)+\Phi_{L\downarrow}(x)+\Phi_{R\uparrow}(x)+\Phi_{R\downarrow}(x)\big)\;,\label{eq:chargefield}\\
\Phi_s(x)&\equiv\frac{1}{2}\big(\Phi_{L\uparrow}(x)-\Phi_{L\downarrow}(x)+\Phi_{R\uparrow}(x)-\Phi_{R\downarrow}(x)\big)\;,\label{eq:spinfield}\\
\Phi_f(x)&\equiv\frac{1}{2}\big(\Phi_{L\uparrow}(x)+\Phi_{L\downarrow}(x)-\Phi_{R\uparrow}(x)-\Phi_{R\downarrow}(x)\big)\;,\label{eq:flavor}\\
\Phi_{sf}(x)&\equiv\frac{1}{2}\big(\Phi_{L\uparrow}(x)-\Phi_{L\downarrow}(x)-\Phi_{R\uparrow}(x)+\Phi_{R\downarrow}(x)\big)\;.\label{eq:spinflavor}
\end{align}
After rewriting the Hamiltonian in terms of these new bosonic fields and performing a unitary transformation, the model is refermionized to obtain
\begin{align}
&\hat{H}=i\hbar v_F\sum_\nu\int\limits_{-\infty}^{\infty}\mathrm{d}x\,\psi^\dagger_\nu(x)\partial_x\psi_\nu(x)+\frac{J^+}{2\sqrt{2\pi a_0}}\left(\psi_{sf}^\dagger(0)+\psi_{sf}(0)\right)\left(d^\dagger-d\right)\nonumber\\
&\qquad+\frac{J^{LR}_\bot}{2\sqrt{2\pi a_0}}\left(\psi_f^\dagger(0)-\psi_f(0)\right)\left(d^\dagger+d\right)+\frac{J^-}{2\sqrt{2\pi a_0}}\left(\psi_{sf}^\dagger(0)-\psi_{sf}(0)\right)\left(d^\dagger+d\right)\nonumber\\
&\qquad+\left(B-\left(J_z-2\pi\hbar v_F\right):\psi_s^\dagger(0)\psi_s(0):\right)\left(d^\dagger d-\frac{1}{2}\right) \;.\label{eq:ham3}
\end{align}
In the above expression, $\nu=c,s,f,sf$, the constant $a_0$ is an ultraviolet cut-off originating from the lattice spacing encountered in the bosonization procedure, $d=i\tau^+$ is a fermionic operator corresponding to the impurity spin, and the coupling constants $J^\pm$ are defined as
\begin{equation}
J^\pm\equiv\frac{1}{2}\left(J^{LL}_\bot\pm J^{RR}_\bot\right) \;.
\end{equation}
Eq.~(\ref{eq:ham3}) has two important features. Firstly, it immediately follows that the model is non-interacting at the point
\begin{equation}
J_z=2\pi\hbar v_F\;,\label{eq:toulouse}
\end{equation}
where the last term of Eq.~(\ref{eq:ham3}), the interaction term, vanishes. This exactly solvable point is a variation of the so-called Toulouse point of the one-channel Kondo model~\cite{toulouse}, and we will refer to this particular two-channel Toulouse point as the Emery-Kivelson (EK) point. At the EK point, the model is free and equivalent to a resonant level model. Secondly, we see that the leads are coupled to Majorana fermions on the impurity,
\begin{equation}
a\equiv\frac{1}{\sqrt{2}}\left(d^\dagger+d\right)\;,\qquad b\equiv\frac{1}{i\sqrt{2}}\left(d^\dagger-d\right)\;,
\end{equation}
and so the model is a \emph{Majorana} resonant level model at the EK point.


From Eq.~(\ref{eq:ham3}) at the EK point, we immediately see that for $B=0$ and $J^+=0$, the $b$ Majorana is strictly decoupled from the rest of the system. However, $J^+=0$ requires that $J_\bot^{LL}$ and $J_\bot^{RR}$ have different signs (\textit{i.e.}, one of the couplings is ferromagnetic). This is not the physical situation of interest, since Kondo couplings in real systems are generically antiferromagnetic. On the other hand $B=0$, $J^-=0$, and $J_\bot^{LR}=0$ results in the $a$ Majorana decoupling. Physically, this corresponds to the situation with $LR$-symmetric couplings (which are  antiferromagnetic) and no exchange cotunneling between the $L$ and $R$ leads, as desired. The implication of the free impurity Majorana is that we have a $T=0$ residual entropy of $S_{\text{imp}}=\tfrac{1}{2}\ln 2$. This is precisely the condition for the NFL critical point. Finite $B$, $J^-$, or $J_\bot^{LR}$ destabilizes the NFL fixed point by mixing in the other Majorana and ultimately quenching the residual entropy to give $S_{\text{imp}}=0$. These are therefore relevant perturbations. Note that finite $J_z^{LL}-J_z^{RR}$ is generated under RG by $J^-$, while finite $J_z^{LR}$ is generated under RG by $J_\bot^{LR}$, even though these are initially zero at the EK point.

We therefore now study Eq.~(\ref{eq:ham3}) at the point $J^-=0$ and $J_\bot^{LR}=0$, but retain the magnetic field term proportional to $B$ as a means of studying the FL crossover. The model then takes the simplified form
\begin{equation}
\hat{H}=\sum_\nu\sum_k\epsilon_k\psi^\dagger_{\nu,k}\psi_{\nu,k}+g_\bot\left(\psi^\dagger_{sf}(0)+\psi_{sf}(0)\right)\left(d^\dagger-d\right)+\frac{B}{2}\left(d^\dagger d-dd^\dagger\right) \;,\label{eq:Ham}
\end{equation}
where $\epsilon_k=\hbar v_Fk$ and $g_\bot\equiv J_\bot/2\sqrt{2\pi a_0}$. As we shall see in the next section the Hamiltonian, Eq.~(\ref{eq:Ham}), is relevant to describe the C2CK system.


\subsection{The charge two-channel Kondo model}
\begin{figure}[t]
  \centerline{\includegraphics[width=10cm]{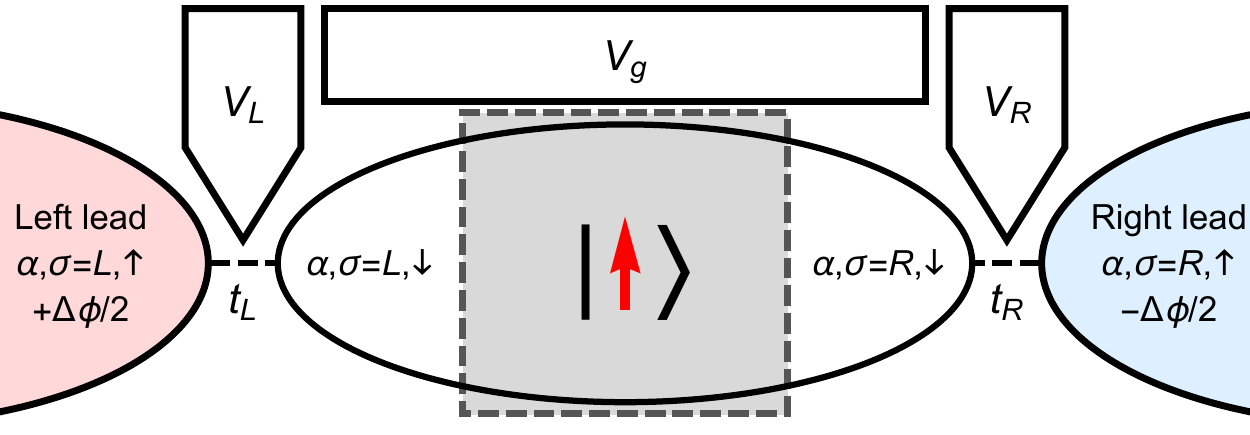}}
\caption{\label{fig:C2CK_setup}Schematic of the C2CK device, where $V_{L,R}$ govern the transmission coefficients $t_{L,R}$, and the gate voltage $V_g$ determines the charge on the dot. The dashed gray box denotes a metallic ``decoherer'' that ensures the left and right sides of the dot are essentially disconnected, {\it i.e.}, there is no coherent transport between them. Due to the large applied magnetic field, the electrons are spin-polarized and effectively spinless. Instead, we label itinerant electrons living on a lead as $\sigma=\uparrow$, and electrons located on the large dot as $\sigma=\downarrow$, such that the electron position on the lead or dot acts as a pseudospin. The two degenerate macroscopic charge states of the dot similarly act as a pseudospin, with $|N+1\rangle \equiv |\uparrow\rangle$ and $|N\rangle \equiv |\downarrow\rangle$. The left and right leads are maintained at a general ``potential'' $\pm\tfrac{1}{2}\Delta\phi$, where $\Delta\phi$ can be either a bias voltage or a temperature gradient.}
\end{figure}

Having covered the general anisotropic 2CK model and its exactly solvable point, we will now consider the C2CK device proposed in Refs.~\cite{matveev1,matveev2} and experimentally realized in Refs.~\cite{experiment,iftikhar2018tunable}. Before we discuss the corresponding effective model, we describe the components of the C2CK device as shown in Fig.~\ref{fig:C2CK_setup}. It consists of a large metallic island (acting as a quantum dot with a continuous spectrum) connected to two separate metallic leads through quantum point contacts with tunable transmission coefficients $t_{L}$ and $t_R$~\cite{matveev1, matveev2}. In a strong perpendicular magnetic field, two effects are utilized: (i) the leads and the dot are in the quantum Hall regime, providing unidirectional edge channels; (ii) spin degeneracy is broken both in the dot and the leads, producing spin-polarized fermions. Therefore we now omit the real spin index. The number of electrons on the quantum dot is controlled by a gate voltage $V_g$. This gate voltage imposes an electrostatic energy $\sim(Q+eN^\prime)^2$, where $N^\prime$ is a dimensionless parameter proportional to $V_g$, $e$ is the (positive) elementary charge, and $Q$ is the (negative) electric charge on the quantum dot. If $V_g$ is tuned such that $N^\prime$ is half-integer, we have a two-fold degeneracy with either $N=N^\prime-\tfrac{1}{2}$ or $N+1=N^\prime+\tfrac{1}{2}$ electrons on the dot. Given that the charging energy\footnote{The charging energy is the energy cost of having $N-1$ or $N+2$ rather than $N$ or $N+1$ electrons on the dot. This is equal to $E_C=\big((3e/2)^2-(e/2)^2\big)/2C=e^2/C$, where $C$ is the capacitance of the dot.} $E_C$ is sufficiently large ({\it i.e.}, $E_C\gg k_BT$) the dot states are effectively restricted to $|N\rangle$, $|N+1\rangle$. The last step towards achieving a two-channel situation is to ``disconnect'' the two sides of the dot and thereby the two leads. This is achieved by adding a large metallic ``decoherer'' on top of the dot, which serves to scatter  electrons, causing a long dwell time on the dot, and inhibiting coherent transport from the left to the right side of the dot. We therefore have essentially independent electronic systems, involving both dot and lead states, around the left and right quantum point contacts. However, the dynamics are correlated by the common dot charging energy.

In order to formulate an effective model for the C2CK device, we translate all components to the spin language that was also used for the general anisotropic 2CK model. First, we identify the macroscopic dot charge states $|N\rangle$, $|N+1\rangle$ as dot pseudospin states $|\downarrow\rangle$, $|\uparrow\rangle$. Additionally, we label the spinless itinerant electrons residing on the leads as ``spin up'' and those on the dot as ``spin down'' (see Fig.~\ref{fig:C2CK_setup}). We also distinguish between itinerant states on the left and on the right side of the dot, which is made possible by virtue of the decoherer. Charge transport between leads through the quantum dot proceeds by an electron tunneling from, say, the left lead onto the dot, and then another electron tunneling from the dot onto the right lead. In spin language, this process corresponds to a spin current: tunneling at the left quantum point contact corresponds to a pseudospin flip of the left conduction electrons and of the dot pseudospin, while subsequent tunneling at the right quantum point contact flips the dot pseudospin back at the same time as flipping the pseudospin of the right conduction electrons. Overall, the dot pseudospin is ``reset'', allowing the process to be repeated. These are the only allowed transport processes at low temperatures.  Charge transport through the quantum dot is therefore equivalent to a sequence of spin-flip processes, which are Kondo-enhanced.

The local part of the effective model describing the C2CK device is therefore given in pseudospin language by the first term of Eq.~(\ref{eq:ham1}), where the coupling constants $J^{\alpha\alpha}_\bot$ depend on the transmission coefficients $t_\alpha$, and $J^{LR}_\bot=0$. Terms proportional to $J_z$ are absent in the C2CK setup, and so the model has intrinsic spin anisotropy (although from the above discussion we know this to be irrelevant). The dot spin operators $\tau^\pm$ are included to enforce the constraints on the dot particle number, and as such can be thought of in terms of projectors, $\tau^+=|N+1\rangle\langle N|$ and $\tau^-=|N\rangle\langle N+1 |$.

The effect of a magnetic field $B$ on the dot pseudospin can be described by introducing a small detuning $\Delta V_g$ in the gate voltage (giving an energetic preference to one of the dot change states over the other), and is therefore proportional to $\tau^z$.

We conclude that the full Hamiltonian describing the C2CK system is given by the anisotopic 2CK model Eq.~(\ref{eq:ham2}), but with specific values of the parameters: $J^{LL}_\bot\sim t_L$, $J^{RR}_\bot \sim t_R$, $B\sim \Delta V_g$, $J^{LR}_\bot=J^{RL}_\bot=0$, and  $J^{\alpha\beta}_z=0$. The connection between the two models is summarized in Table~\ref{tab:dictionary}. This equivalence forms the basis for all calculations in this paper.

\begin{table}[t]
    \centering
    \begin{tabular}{|l||l|l|}
        \hline
         & Spin & Charge \\
        \hline
        Dot states & $|\uparrow\rangle$, $|\downarrow\rangle$ & $|N+1\rangle$, $|N\rangle$ \\
        Itinerant states & $\psi_{\alpha\sigma}$ & $\psi_{\alpha\uparrow}$ (leads), $\psi_{\alpha\downarrow}$ (dot) \\
        Spin-flip interactions & $J^{\alpha\alpha}_\bot$ & $t_\alpha$ \\
        Ising type interactions & $J_z$ & - \\
        Magnetic field & $B$ & $\Delta V_g$ \\
        \hline
    \end{tabular}
    \caption{\label{tab:dictionary}Summary of the relation between the anisotropic 2CK model and the C2CK device, further elaborated on in the main text. Note especially that the Ising type interactions $J_z$ are absent in the C2CK device.}
\end{table}

A few remarks are in order. (i) One should note that the EK point and the effective C2CK model are both special points of the anisotropic 2CK model, with the value for $J_z$ being the only difference between these two special points. Nevertheless, the distinction between the two is important: the anisotropic 2CK model is only non-interacting for a very specific finite value of $J_z$; the lack of Ising-type interactions in the C2CK model make it irreducibly strongly correlated. (ii) The redefinition of the spin label for the itinerant states requires careful consideration when applying a bias between the leads. In particular, for the spin 2CK model, both the spin up and the spin down states live in the leads and a bias affects both spin species; for the C2CK device, only the spin up states live in the leads, while the spin down states are located on the dot. In order to exploit the equivalence between the anisotropic 2CK model and the C2CK device, an applied bias in the C2CK device effectively involves only the (pseudo)spin-up electrons. Consequently, special care must be taken to translate the definition of the current operators into the pseudospin language.


\subsection{Low temperature limit of the C2CK model}\label{sec:lowtoulouse}
The C2CK model and the exactly-solvable EK point are very similar, both being special cases of the general anisotropic 2CK model, and in fact only differing in their values of $J_z$. Furthermore, as discussed in Sec.~\ref{sec:2c_phys}, both models flow to the same NFL fixed point since $J_z$ is RG irrelevant. Indeed the physics for all $T\ll T_K$ is the same in both models. In particular, for small detuning perturbations, the same FL crossover results.

To obtain universal results for the C2CK system in the regime $T\ll T_K$, we can therefore perform calculations for the anisotropic 2CK model at the EK point, and then send $T/T_K\rightarrow 0$. To access the universal FL crossover, we additionally require $T^*/T_K \rightarrow 0$. In practice, we achieve both by letting $T_K\rightarrow \infty$. Note however, that the crossover from the local-moment (free pseudospin) fixed point to the NFL fixed point of the C2CK system cannot be captured within the EK point calculation. Instead, correct results for the C2CK model for $T\simeq T_K$ can be accessed by expanding the EK point solution about $J_z=2\pi\hbar v_F$~\cite{sh3}. Here, such perturbations are a function of $T/T_K$, and only strictly vanish as $T/T_K\rightarrow 0$.

The above considerations allow us to use the Hamiltonian from Eq.~(\ref{eq:Ham}) as the starting point for all of the transport calculations that follow.


\section{Transport: preliminaries}\label{sec:preliminaries}
Here we summarize the preliminaries necessary to calculate transport properties in the C2CK model, using Eq.~(\ref{eq:Ham}). We first discuss general conserved charges that are coupled to a bias by a simple potential term, and introduce the necessary current operators. Then the Emery-Kivelson mapping is performed to obtain the effective current operators in the equivalent non-interacting theory. Finally, we discuss the special case of a temperature gradient, which cannot directly enter the effective Hamiltonian, and requires a different treatment within linear response theory.


\subsection{Potentials and current operators}\label{sec:currentop}
We consider quantum transport through the dot due to a potential gradient between the two leads. Therefore we shall examine how a general potential difference between the leads enters on the level of the Hamiltonian, and the form of the corresponding current operators. First we use the common example of a bias voltage and corresponding charge current.

We apply the bias voltage $V$ symmetrically such that the left lead feels a uniform potential of $V/2$, while the right lead feels $-V/2$. Given that $\sigma=\uparrow$ refers to the electrons in the leads, the additional term in the Hamiltonian due to a bias voltage is given by
\begin{equation}
\hat{H}_V=-\frac{eV}{2}\int\limits_{-\infty}^{\infty}\mathrm{d}x\left[\psi^\dagger_{L\uparrow}(x)\psi_{L\uparrow}(x)-\psi^\dagger_{R\uparrow}(x)\psi_{R\uparrow}(x)\right]\;,\label{eq:HV}
\end{equation}
which can simply be written as
\begin{equation}
\hat{H}_V=\hat{q}V\;,\qquad \hat{q}\equiv\frac{1}{2}\left(\hat{q}_L-\hat{q}_R\right)\;,\qquad \hat{q}_\alpha=-e\int\limits_{-\infty}^{\infty}\mathrm{d}x\,\psi^\dagger_{\alpha\uparrow}(x)\psi_{\alpha\uparrow}(x)\equiv -e\hat{N}_{\alpha}\;.
\label{eq:chargeex}
\end{equation}
Here, $\hat{q}_\alpha$ is an operator for the total charge on lead $\alpha$, while $\hat{N}_\alpha$ is the corresponding number operator. This can be generalized to a general, time-dependent ``charge'' operator $\hat{Q}_\alpha(t)$ for lead $\alpha$, coupled to a general time-dependent ``potential'' drop between the leads $\Delta\phi(t)$. The minimal coupling contribution to the Hamiltonian then reads
\begin{equation}
\hat{H}_\phi(t)=\hat{Q}(t)\Delta\phi(t)\;,\qquad \text{where}\qquad \hat{Q}(t)\equiv\frac{1}{2}\left(\hat{Q}_L(t)-\hat{Q}_R(t)\right)\;.\label{eq:genfield}
\end{equation}

Next, we define the general current operator $\hat{I}_Q(t)$, corresponding to the general charge $\hat{Q}(t)$. Applying the continuity equation and imposing total charge conservation, the current leaving lead $\alpha$ is given by $\hat{I}_{Q_\alpha}=-\frac{\mathrm{d}\hat{Q}_\alpha}{\mathrm{d}t}$. A natural way to define the current flowing through the dot region is as the average of the current \emph{leaving} the left lead and the current \emph{entering} the right lead. This gives
\begin{equation}
\hat{I}_Q\equiv\frac{1}{2}\left(-\frac{\mathrm{d}\hat{Q}_L}{\mathrm{d}t}+\frac{\mathrm{d}\hat{Q}_R}{\mathrm{d}t}\right)=-\frac{\mathrm{d}\hat{Q}}{\mathrm{d}t}=\frac{i}{\hbar}\big[\hat{Q},\hat{H}\big]\;,\label{eq:currentdef}
\end{equation}
with $\hat{H}$ being the full Hamiltonian (Eq.~(\ref{eq:Ham}) with the addition of $\hat{H}_\phi$). For charge transport, we thus have $\hat{I}_c=\frac{e}{2}\frac{\mathrm{d}}{\mathrm{d}t}\big(\hat{N}_L-\hat{N}_R\big)$, while for energy transport the current is given by $\hat{I}_E=-\frac{1}{2}\frac{\mathrm{d}}{\mathrm{d}t}\big(\hat{H}_L-\hat{H}_R\big)$, where $\hat{H}_\alpha$ is the part of the Hamiltonian corresponding to lead $\alpha$. Now, from the first law of thermodynamics at constant volume, $\mathrm{d}E=\mathrm{d}Q+\mu\,\mathrm{d}N$ ($\mathrm{d}Q$ referring to heat), it follows that the heat current operator is given by $\hat{I}_h=\hat{I}_E-\mu\hat{I}_N$, where $\mu$ is the chemical potential in the leads.


\subsection{Emery-Kivelson mapping of the current operators}\label{sec:emery}
As discussed in Sec.~\ref{sec:lowtoulouse}, the strategy employed in this paper is to utilize the exactly solvable EK point to calculate observables. It is therefore necessary to apply the Emery-Kivelson mapping~\cite{emery} (as briefly outlined in Sec.~\ref{sec:toulouse}) to the current operators. First we perform the mapping on the generalized ``charge'' operators $\hat{Q}_c=-\frac{e}{2}\big(\hat{N}_L-\hat{N}_R\big)$ and $\hat{Q}_E=\frac{1}{2}\big(\hat{H}_L-\hat{H}_R\big)$. The current operators then follow from the commutators of these operators with the full (mapped) Hamiltonian using Eq.~(\ref{eq:Ham}).\footnote{It is also possible to calculate the commutators first and only then going through the mapping procedure, but that turns out to be much more cumbersome.} More details on the bosonization and refermionization~\cite{giamarchi,senechal2004introduction,gogolin} used in the mapping procedure can be found in Appendix~\ref{ap:bosonization}.

The first part of the mapping procedure is the introduction of a bosonic field $\Phi_{\alpha\sigma}(x)$ for each of the fermionic fields $\psi_{\alpha\sigma}(x)$,
\begin{equation}
\psi_{\alpha\sigma}(x)=\frac{1}{\sqrt{2\pi a_0}}e^{i\phi_{\alpha\sigma}}e^{-i\Phi_{\alpha\sigma}(x)} \;,
\end{equation}
where $e^{i\phi_{\alpha\sigma}}$ are Klein factors to ensure the correct anticommutation relations between the fermionic fields. Following the usual bosonization prescription, the various components of the charge operators transform according to
\begin{align}
\int\limits_{-\infty}^{\infty}\mathrm{d}x\,\psi^\dagger_{\alpha\sigma}(x)\psi_{\alpha\sigma}(x)&=\frac{1}{2\pi}\int\limits_{-\infty}^{\infty}\mathrm{d}x\,\partial_x\Phi_{\alpha\sigma}(x) \;,\label{eq:bosonization1}\\
\int\limits_{-\infty}^{\infty}\mathrm{d}x\,\psi^\dagger_{\alpha\sigma}(x)\partial_x\psi_{\alpha\sigma}(x)&=-\frac{i}{4\pi}\int\limits_{-\infty}^{\infty}\mathrm{d}x\left(\partial_x\Phi_{\alpha\sigma}(x)\right)^2 \;,\label{eq:bosonization2}
\end{align}
where normal ordering of the fermionic fields is implied. Substituting these expression into the definitions of the charge operators, and writing in terms of the $\nu=c,s,f,sf$ fields from Eqs.~(\ref{eq:chargefield})-(\ref{eq:spinflavor}), we find
\begin{align}
\hat{Q}_c&=-\frac{e}{4\pi}\int\limits_{-\infty}^{\infty}\mathrm{d}x\big(\partial_x\Phi_f(x)+\partial_x\Phi_{sf}(x)\big)\;,\\
\hat{Q}_E&=\frac{\hbar v_F}{8\pi}\int\limits_{-\infty}^{\infty}\mathrm{d}x\big(\partial_x\Phi_c(x)+\partial_x\Phi_s(x)\big)\big(\partial_x\Phi_f(x)+\partial_x\Phi_{sf}(x)\big)\;.
\end{align}
The next step of the Emery-Kivelson mapping procedure is the unitary transformation $\hat{\mathcal{O}}\rightarrow\hat{U}\hat{\mathcal{O}}\hat{U}^\dagger$, with $\hat{U}=e^{i\chi_s\tau^z}$ and $\chi_s\equiv\Phi_s(0)-\phi_s$. Using the commutation relation
\begin{equation}
\left[\Phi_\mu(x),\partial_x\Phi_\nu(x^\prime)\right]=2\pi i\,\delta_{\mu,\nu}\,\delta(x-x^\prime) \;,\label{eq:commutation}
\end{equation}
together with $d=i\tau^+$ (such that $\tau^z=-(d^\dagger d-1/2)$), it is straightforward to show that
\begin{align}
\hat{Q}_E&\rightarrow\frac{\hbar v_F}{8\pi}\int\limits_{-\infty}^{\infty}\mathrm{d}x\big(\partial_x\Phi_c(x)+\partial_x\Phi_s(x)\big)\big(\partial_x\Phi_f(x)+\partial_x\Phi_{sf}(x)\big)\nonumber\\
&\quad\,+\frac{\hbar v_F}{4}\left(d^\dagger d-\frac{1}{2}\right)\big(\partial_x\Phi_f(x)+\partial_x\Phi_{sf}(x)\big)\Big|_{x=0} \;,
\end{align}
under this unitary transformation, while $\hat{Q}_c$ remains unchanged. The final step of the mapping procedure consists of refermionization. Using relations similar to those involved in the initial bosonization step and noting that
\begin{equation}
\int\limits_{-\infty}^{\infty}\mathrm{d}x:\psi^\dagger_\mu(x)\psi_\mu(x)\psi^\dagger_\nu(x)\psi_\nu(x):\,=\frac{1}{4\pi^2}\int\limits_{-\infty}^{\infty}\mathrm{d}x\big(\partial_x\Phi_\mu(x)\big)\big(\partial_x\Phi_\nu(x)\big)\label{eq:bosonization3}
\end{equation}
for $\mu\neq\nu$ (as shown in Appendix~\ref{ap:bosonization}), the charge operators can be written as
\begin{align}
\hat{Q}_c&=-\frac{e}{2}\int\limits_{-\infty}^{\infty}\mathrm{d}x\Big(\psi^\dagger_f(x)\psi_f(x)+\psi^\dagger_{sf}(x)\psi_{sf}(x)\Big)\;,\\
\hat{Q}_E&=\frac{\pi\hbar v_F}{2}\int\limits_{-\infty}^{\infty}\mathrm{d}x:\Big(\psi_c^\dagger(x)\psi_c(x)+\psi_s^\dagger(x)\psi_s(x)\Big)\Big(\psi_f^\dagger(x)\psi_f(x)+\psi_{sf}^\dagger(x)\psi_{sf}(x)\Big):\nonumber\\
&\quad\,+\frac{\pi\hbar v_F}{2}\Big(:\psi_f^\dagger(0)\psi_f(0):+:\psi_{sf}^\dagger(0)\psi_{sf}(0):\Big)\left(d^\dagger d-\frac{1}{2}\right) \;.
\end{align}

We now determine the current operators by Fourier transforming the charge operators to momentum space and evaluating the commutators with the Hamiltonian from Eq.~(\ref{eq:Ham}). Starting with the current operator corresponding to electric charge:
\begin{align}
\hat{I}_c&=-\frac{ie}{2\hbar}\sum_k\left[\psi^\dagger_{f,k}\psi_{f,k}+\psi^\dagger_{sf,k}\psi_{sf,k},\hat{H}\right]\nonumber\\
&=-\frac{ieg_\bot}{2\hbar\sqrt{L}}\sum_k\left(\psi^\dagger_{sf,k}-\psi_{sf,k}\right)\left(d^\dagger-d\right)\;,\label{eq:Ic}
\end{align}
where $L$ is a length scale originating from Fourier transforming the $\psi_\nu$ fields ({\it i.e.}, the lattice constant times the total number of lattice sites on a given lead). Although more cumbersome, the energy current can be obtained in the same way:
\begin{align}
\hat{I}_E&=\frac{i\pi v_Fg_\bot}{2L^{3/2}}\sum\limits_{\mathclap{k,k^\prime,k^{\prime\prime}}}\left(\psi_{c,k^\prime}^\dagger\psi_{c,k^{\prime\prime}}+\psi_{s,k^\prime}^\dagger\psi_{s,k^{\prime\prime}}\right)\left(\psi_{sf,k}^\dagger-\psi_{sf,k}\right)\left(d^\dagger-d\right)\nonumber\\
&\quad\,+\frac{i\pi v_Fg_\bot}{4L^{3/2}}\sum\limits_{\mathclap{k,k^\prime,k^{\prime\prime}}}\left(2\,\psi_{f,k^\prime}^\dagger\psi_{f,k^{\prime\prime}}+\delta_{k^\prime,k^{\prime\prime}}\right)\left(\psi_{sf,k}^\dagger+\psi_{sf,k}\right)\left(d^\dagger+d\right)\nonumber\\
&\quad\,+\frac{i\pi v_F}{4L}\sum\limits_{\mathclap{k,k^\prime}}\left(\epsilon_{k^\prime}-\epsilon_k\right)\left(\psi_{f,k}^\dagger\psi_{f,k^\prime}+\psi_{sf,k}^\dagger\psi_{sf,k^\prime}\right)\left(d^\dagger d-dd^\dagger\right) \;.\label{eq:IE}
\end{align}
Strikingly, the energy current operator (and therefore the heat current operator) is much more complicated than the charge current operator. This originates from the fact that heat transport itself is a more complicated concept: while electric transport only involves charge-carrying excitations, heat transport involves \emph{all} modes supported by the system. As a result, the mapping of a strongly interacting system to an effective non-interacting model comes at the price of a significantly more complicated heat current operator. In terms of the Emery-Kivelson mapping procedure, this fundamental difference between charge and heat transport emerges during the unitary transformation. In particular, the operator corresponding to electric charge does not pick up additional terms due to the fact that the spin modes $\Phi_s$ do not carry charge and therefore commute with $\hat{Q}_c$. On the other hand, the spin modes \emph{do} carry energy, resulting in several additional terms entering into $\hat{Q}_E$ upon performing the unitary transformation. The second and third lines of Eq.~(\ref{eq:IE}) originate from this step.

With the general charge $Q$ coupled to a bias according to Eq.~(\ref{eq:genfield}), the observable time- and temperature-dependent current can now be calculated by taking the expectation value of the corresponding current operator $\hat{I}_{Q}$ with respect to the full Hamiltonian. In the case of charge transport, the full Hamiltonian (including the minimal coupling term) is quadratic and can be treated exactly using the Keldysh formalism. This is done in Sec.~\ref{sec:keldysh}.


\subsection{Linear response theory}\label{sec:linres}
With the full non-equilibrium current $\langle \hat{I}_Q \rangle$ at hand, one can take the zero-bias limit $\Delta \phi \rightarrow 0$ to find the conductance $d\langle \hat{I}_Q \rangle/d\Delta \phi$ in linear response. However, the strategy outlined in the previous section requires that the bias term enters directly in the Hamiltonian, and can be transformed in the effective model through the Emery-Kivelson mapping. The expectation value of the transformed current operator can then be evaluated directly in the transformed model. This all works perfectly in the case of a voltage bias, Eq.~(\ref{eq:HV}) \cite{sh1}.

However, a temperature gradient cannot be dealt with in this way, and heat transport is much more subtle.
 One cannot directly calculate the expectation value of the physical heat current operator in the Emery-Kivelson model for two reasons. First, the temperature gradient cannot enter the Hamiltonian in the same way as the bias voltage, since temperature is a boundary condition. The usual solution for this problem is to instead give the leads a different temperature in their Fermi-Dirac distributions. This brings us to the second problem: as will become clear in Sec.~\ref{sec:keldysh}, direct calculation of the current depends on the flavor and spin-flavor modes being in thermal equilibrium. This means that they both must obey the Fermi-Dirac distribution with a well-defined temperature. However, the flavor and spin-flavor modes are composite modes, with contributions living on both leads (see Eqs.~(\ref{eq:flavor}) and (\ref{eq:spinflavor})). Therefore there is no well-defined thermal equilibrium for these modes if left and right leads are themselves at different temperatures. We conclude that the full non-equilibrium calculation of thermal transport is impossible within the Emery-Kivelson framework at the exactly-solvable EK point. To calculate thermal transport, we need to circumvent these problems and use a different approach.

 In linear response, an alternative approach is to calculate the linear susceptibilities directly from perturbation theory in the bias. For charge transport this method reproduces the zero-bias limit results of the full non-equilibrium calculation. However, as we will see below, it also allows us to overcome the problems associated with calculating thermal transport. In particular, when working within linear response theory, the linear susceptibility is obtained from the equilibrium solution in absence of the bias ~\cite{moca}.  In this case, a well-defined temperature can be assigned to the composite flavor and spin-flavor modes, which are in thermal equilibrium.

As a starting point, we again consider a general ``charge" $\hat{Q}$ coupled to a general ``potential" $\Delta\phi$, previously considered in Eq.~(\ref{eq:genfield}). For real time $t$, the  expectation value of the current corresponding to $\hat{Q}$ is given by
\begin{equation}
\big\langle\hat{I}_Q\big\rangle(t)=\int\limits_{-\infty}^{\infty}\mathrm{d}t^\prime\chi(t,t^\prime)\Delta\phi(t^\prime)+\mathcal{O}\left(\Delta\phi^2\right) \;.
\end{equation}
If the system is time-independent (in steady state, such that the susceptibility obeys $\chi(t,t^\prime)=\chi(t-t^\prime)$), the Fourier transform of this equation follows simply from the convolution theorem as
\begin{equation}
\big\langle\hat{I}_Q\big\rangle(\omega)=\chi(\omega)\Delta\phi(\omega)+\mathcal{O}\left(\Delta\phi^2\right) \;.
\end{equation}
Furthermore defining $\langle\ldots\rangle_0$ to be the expectation value in absence of a potential gradient ({\it i.e.}, the static equilibrium case), the susceptibility can be obtained from
\begin{equation}
\chi(\omega)=\frac{i}{\hbar\omega}\left(C^\text{R}(\omega)-C^\text{R}(0)\right) \;,\label{eq:kubo}
\end{equation}
with $C^\text{R}(\omega)$ being the Fourier transform of the retarded current autocorrelator,
\begin{equation}\label{eq:CRdef}
C^\text{R}(\omega)=\int\limits_{-\infty}^{\infty}\mathrm{d}\Delta t\,C^\text{R}(\Delta t)e^{i\omega\Delta t}=\int\limits_{-\infty}^{\infty}\mathrm{d}\Delta t\left(-i\theta(\Delta t)\big\langle\big[\hat{I}_Q(\Delta t),\hat{I}_Q(0)\big]\big\rangle_0\right)e^{i\omega\Delta t} \;.
\end{equation}
The above is known as the Kubo formula~\cite{green,kubo}; a short derivation of this formula can be found in Appendix~\ref{ap:kubo}. It provides a way to calculate the linear susceptibility of some current $I_Q$ to a potential drop $\Delta\phi$ between the leads, purely in terms of ``bare'' equilibrium quantities. In order to evaluate the right-hand side of Eq.~(\ref{eq:kubo}), we will first calculate the imaginary time correlation function, defined as
\begin{equation}
C^\tau(\tau_1-\tau_2)\equiv-\big\langle T_\tau\hat{I}_Q(\tau_1)\hat{I}_Q(\tau_2)\big\rangle_0 \;,\label{eq:corfun}
\end{equation}
where $T_\tau$ is the time ordering operator and $\tau=it$. From here, it is most convenient to switch to bosonic Matsubara frequencies $\Omega_n\equiv\frac{2\pi n}{\hbar\beta}$ since the current operators only contain even powers of fermionic operators,
\begin{equation}
C^\tau(i\Omega_n)=\int\limits_0^{\hbar\beta}\mathrm{d}\tau\,C^\tau(\tau)e^{i\Omega_n\tau} \;.
\end{equation}
The susceptibility in terms of real frequency $\omega$ is now found by performing analytic continuation on the correlation function, writing $C^\tau(i\Omega_{n>0})\rightarrow C(\omega+i0^+)\equiv C^\text{R}(\omega)$~\cite{altland}, where we note that the positive Matsubara frequencies are sufficient.\footnote{The poles and branch cuts of the the analytically continued function $C(z\in\mathbb{C})$ are all located on the real axis, such that $C(z)$ can be a different analytic function for $\text{Im}[z]>0$ and $\text{Im}[z]<0$. Since we are interested in points with $\text{Im}[z=\omega+i0^+]>0$, we only have to consider the points on the positive imaginary axis, {\it i.e.}, $i\Omega_{n>0}$.} Finally, note that the dc limit is obtained by taking $\omega\rightarrow 0$, that is $\chi_\text{dc}=\lim\limits_{\omega\rightarrow 0}\chi(\omega)$.

However, to calculate thermal transport, we still have the problem of how to incorporate the temperature gradient as a source term in the Hamiltonian. The solution to this problem was first proposed by Luttinger in 1964~\cite{luttinger}. The idea is that temperature is not the only field that couples to the energy density: a gravitational field couples to the energy density as well. The advantage of a gravitational field is that it \emph{can} enter the Hamiltonian in the general way outlined in Eq.~(\ref{eq:genfield}). In the absence of a chemical potential $\mu$, the heat current is phenomenologically given by
\begin{equation}
I_h=\chi_T\frac{\Delta T}{T}+\chi_\psi\Delta\psi\;,
\end{equation}
where $\Delta T$ and $\Delta\psi$ denote the drop in temperature and gravitational field between the leads, respectively. Luttinger showed that the corresponding linear susceptibilities must be equal to each other, {\it i.e.}, $\chi_T=\chi_\psi$. Therefore, one can calculate the susceptibility due to a gravitational field $\chi_\psi$ in absence of a temperature gradient, and then use this result to find the current due to a temperature gradient in absence of a gravitational field. To summarize, we can find the heat current due to a temperature gradient by first calculating $\chi_\psi$ (which is in turn done by considering a contribution to the Hamiltonian of the form of Eq.~(\ref{eq:genfield})), then setting $\chi_T=\chi_\psi$ and calculating $I_h=\chi_T\Delta T/T$. While the full heat current is no longer exact (neglecting the $\mathcal{O}\left((\Delta T/T)^2\right)$ terms), the linear susceptibility $\chi_T$ can be obtained exactly.

Finally, we consider the linear response currents in the presence of both a bias voltage and a temperature gradient. The equations for the charge and heat currents can be written as
\begin{equation}
\begin{pmatrix}I_c\\I_h\end{pmatrix}=\begin{pmatrix}\chi_{11} & \chi_{12}\\ \chi_{21} & \chi_{22}\end{pmatrix}\begin{pmatrix}V\\ \Delta T/T\end{pmatrix} \;,\label{eq:offdiag}
\end{equation}
where $\chi_{11}\sim\langle\hat{I}_c\hat{I}_c\rangle_0$ and $\chi_{22}\sim\langle\hat{I}_E\hat{I}_E\rangle_0$ are respectively the isolated charge and heat susceptibilities, while $\chi_{12}\sim\langle\hat{I}_c\hat{I}_E\rangle_0$ and $\chi_{21}\sim\langle\hat{I}_E\hat{I}_c\rangle_0$ represent thermopower \cite{luttinger}. In this more general situation, the heat current is consequently given by $I_h=\chi_{21}V+\chi_{22}\Delta T/T$. Defining the heat conductance $\kappa$ through $I_h\equiv\kappa\Delta T$, the heat conductance can assume two different forms: (i) in absence of a bias voltage, the heat conductance satisfies $T\kappa\big|_{V=0}=\chi_{22}$; (ii) in absence of an electric current, the heat conductance is given by $T\kappa\big|_{I_c=0}=\chi_{22}-\chi_{12}\chi_{21}/\chi_{11}$. In the latter case, a bias voltage of $V=-(\chi_{12}/\chi_{11})\Delta T/T$ has been applied to cancel the thermopower that emerges as a result of the non-zero temperature gradient. In general, it is therefore necessary to specify which quantity ({\it i.e.}, either $V$ or $I_c$) is set to zero when evaluating the heat conductance.


\subsection{Propagators}\label{sec:propagators}
As we have seen in the previous sections, finding the actual observable currents requires calculating expectation values of either the current operators themselves, or current-current correlation functions. This in turn requires finding the propagators of the model. For notational convenience, from now on we will use the similarities with a regular resonant level model to identify ``spin-flavor'' as ``left'', and ``flavor'' as ``right'' (within this convention, the left and right propagators below are labeled as $L$ and $R$). This distinction is not necessary for the case of channel symmetry (as the flavor modes are then decoupled from the rest of the system), but we retain it here for completeness. We emphasize that the left/right labels used here are unrelated to the original left and right leads entering in the definition of the original model. Following the usual functional integral formalism to construct the action of the model, we then obtain the following expression for the full Green function of the system,
\begin{equation}
\mathbf{G}\equiv\begin{pmatrix}\mathbf{L} & \mathbf{G}_{ld} & \mathbf{G}_{lr}\\ \mathbf{G}_{dl} & \mathbf{D} & \mathbf{G}_{dr}\\ \mathbf{G}_{rl} & \mathbf{G}_{rd} & \mathbf{R}\end{pmatrix}=\begin{pmatrix}\mathbf{L}_0^{-1} & -\mathbf{g}_\bot/\hbar & 0\\ -\mathbf{g}^\dagger_\bot/\hbar & \mathbf{D}_0^{-1} & 0\\ 0 & 0 & \mathbf{R}_0^{-1}\end{pmatrix}^{-1}\;,\label{eq:greensinv}
\end{equation}
independent of the basis of the components. Here, $\mathbf{L}$, $\mathbf{R}$, and $\mathbf{D}$ are the full Green functions corresponding to the spin-flavor modes, the flavor modes, and the dot, respectively, while $\mathbf{L}_0$, $\mathbf{R}_0$ and $\mathbf{D}_0$ are the corresponding ``bare" Green functions in the absence of the dot-lead hybridization.\footnote{The word ``bare" can either mean a system in absence of a bias, $\Delta\phi=0$, or alternatively a system without dot-lead hybridization, $g_\bot=0$. We make clear the precise meaning when it is not clear from context.} Here $\mathbf{g}_\bot$ governs the coupling between the spin-flavor modes and the dot. Block inversion of the right-hand side of Eq.~(\ref{eq:greensinv}) gives
\begin{align}
\mathbf{D}&=\left(\mathbf{D}_0^{-1}-\pmb{\Sigma}_d\right)^{-1}\;,\qquad \pmb{\Sigma}_d\equiv\frac{1}{\hbar^2}\,\mathbf{g}_\bot^\dagger\cdot\mathbf{L}_0\cdot\mathbf{g}_\bot\;,\label{eq:dfull}\\
\mathbf{G}_{ld}&=\frac{1}{\hbar}\,\mathbf{L}_0\cdot\mathbf{g}_\bot\cdot\mathbf{D}\;,\label{eq:Gld}\\
\mathbf{L}&=\mathbf{L}_0+\frac{1}{\hbar^2}\,\mathbf{L}_0\cdot\mathbf{g}_\bot\cdot\mathbf{D}\cdot\mathbf{g}_\bot^\dagger\cdot\mathbf{L}_0\;,\label{eq:Gll}
\end{align}
where $\pmb{\Sigma}_d$ can be identified as the self-energy of the dot. All full propagators can thus be calculated from the full Green function on the dot, together with bare quantities. This essentially reduces the problem of finding the currents to obtaining a single Green function.

In order to determine the necessary Green functions, it is important to incorporate the fact that all tunneling happens via the Majorana modes $a\equiv(d^\dagger+d)/\sqrt{2}$ and $b\equiv(d^\dagger-d)/i\sqrt{2}$. This Majorana character can be properly incorporated by switching to the Nambu spinor basis, for example working with $\mathbf{d}^\dagger\equiv(d^\dagger\;d)$. Doing so, we find the following action,
\begin{equation}
S=\frac{\hbar}{2}\int\limits_{-\infty}^{\infty}\mathrm{d}t\,\bar{\psi}\cdot\mathbf{G}^{-1}\cdot\psi\;,\label{eq:Snambu}
\end{equation}
where $\psi,\bar{\psi}$ are vectors containing all of the Grassmann fields (the factor $1/2$ accounts for the doubling on going to the Nambu basis). In momentum space, all $L$ components of the hybridization matrix (labeled by index $k$) can be deduced from Eq.~(\ref{eq:Ham}), and are given by
\begin{equation}
\mathbf{g}_{\bot,k}=\frac{g_{\bot}}{\sqrt{L}}\begin{pmatrix}-1 & 1 \\ -1 & 1\end{pmatrix}\equiv\frac{g_\bot}{\sqrt{L}}\,\mathbf{g} \;,\label{eq:gbotm}
\end{equation}
independent of $k$. Similarly, the momentum space components of all Green functions are also $2\times 2$ matrices. Using the observation that the bare Hamiltonian ({\it i.e.}, in absence of dot-lead tunneling) is symmetric in $\nu=f,sf$, together with the fact that it does not contain superconducting pairing terms such as $dd$ or $d^\dagger d^\dagger$, the components of the bare propagators are found to be of the form
\begin{equation}
\mathbf{L}_{0,kk^\prime}=\mathbf{R}_{0,kk^\prime}=\delta_{k,k^\prime}\mathbf{L}_{0,k}=\delta_{k,k^\prime}\begin{pmatrix}L_{0,k,1} & 0 \\ 0 & L_{0,k,2}\end{pmatrix} \;,\qquad\mathbf{D}_0=\begin{pmatrix}D_{0,1} & 0 \\ 0 & D_{0,2}\end{pmatrix} \;.\label{eq:greensform}
\end{equation}
In momentum space, the Green functions given by Eqs.~(\ref{eq:Gld}) and (\ref{eq:Gll}) become
\begin{align}
\mathbf{G}_{ld,k}&=\frac{g_\bot}{\hbar\sqrt{L}}\,\mathbf{L}_{0,k}\cdot\mathbf{g}\cdot\mathbf{D}\;,\label{eq:Gldk}\\
\mathbf{L}_{kk^\prime}&=\delta_{k,k^\prime}\mathbf{L}_{0,k}+\frac{g_\bot^2}{\hbar^2L}\,\mathbf{L}_{0,k}\cdot\mathbf{g}\cdot\mathbf{D}\cdot\mathbf{g}^\dagger\cdot\mathbf{L}_{0,k^\prime}\;,
\end{align}
with the dot self-energy
\begin{equation}
\pmb{\Sigma}_d=\frac{g_\bot^2}{\hbar^2}\,\mathbf{g}^\dagger\cdot\Big(\frac{1}{L}\sum_k\mathbf{L}_{0,k}\Big)\cdot\mathbf{g}\equiv\frac{g_\bot^2}{\hbar^2}\,\mathbf{g}^\dagger\cdot\mathbf{L}_0^\prime\cdot\mathbf{g} \;.\label{eq:sigmad}
\end{equation}
It should be noted that all of the above fields and Green functions have an implied time-dependence.

In the case of linear response theory, the required expectation values involve only equilibrium propagators, and we may use Matsubara techniques. In the absence of a bias and in terms of fermionic Matsubara frequencies $\omega_n$, the necessary Green functions are given by
\begin{align}
\mathbf{L}_{0,k}(i\omega_n)&=\hbar\begin{pmatrix}(i\hbar\omega_n-\epsilon_k)^{-1} & 0 \\ 0 & (i\hbar\omega_n+\epsilon_k)^{-1}\end{pmatrix}\;,\label{eq:L0k}\\
\mathbf{D}(i\omega_n)\equiv\mathbf{G}_{dd}(i\omega_n)&=\int\limits_{-\infty}^\infty\mathrm{d}\epsilon\,\frac{\pmb{\rho}(\epsilon)}{i\hbar\omega_n-\epsilon} \;,\qquad\pmb{\rho}(\epsilon)\equiv-\frac{1}{\pi}\text{Im}\!\left[\mathbf{D}^\text{R}(\epsilon)\right] \;,\label{eq:Dw}
\end{align}
where $\pmb{\rho}$ can be interpreted as a density of states~\cite{altland}, and the retarded dot Green function is given by
\begin{equation}
\mathbf{D}^\text{R}(\epsilon)=\frac{\hbar}{\epsilon(\epsilon+i\Gamma)-B^2}\begin{pmatrix}\epsilon+B+\frac{i}{2}\Gamma & \frac{i}{2}\Gamma \\ \frac{i}{2}\Gamma & \epsilon-B+\frac{i}{2}\Gamma\end{pmatrix} \;.\label{eq:DR}
\end{equation}
Here, the parameter $\Gamma$ has been introduced for notational convenience and for later reference; it is defined according to
\begin{equation}
\Gamma\equiv 2g_\bot^2\frac{\mathrm{d}k}{\mathrm{d}\epsilon_k}=\frac{2g_\bot^2}{\hbar v_F}=\frac{J_\bot^2}{4\pi a_0\hbar v_F}\;.
\end{equation}
A full derivation of the dot propagator from Eq.~(\ref{eq:DR}) can be found in Appendix~\ref{ap:dotgreen}.


\section{Exact results for charge transport}\label{sec:keldysh}
We now discuss the exact solution of the 2CK model at the EK point, in the presence of a generalized time-dependent bias voltage that drives the system out of equilibrium. The methods discussed here are an application (and in some cases a generalization) of the methods introduced by Jauho \emph{et al.} in Ref.~\cite{jauho}, and by Schiller and Hershfield in Refs.~\cite{sh2,sh1}.

Applying the mapping procedure from Sec.~\ref{sec:emery} to the voltage bias term from Eq.~(\ref{eq:HV}), and adding the result to Eq.~(\ref{eq:Ham}), we obtain the full model in Emery-Kivelson form at the EK point,
\begin{equation}
\hat{H}=\sum_{\nu=f,sf}\sum_k\left(\epsilon_k-\frac{eV(t)}{2}\right)\psi^\dagger_{\nu,k}\psi_{\nu,k}+\frac{g_\bot}{\sqrt{L}}\sum_k\left(\psi^\dagger_{sf,k}+\psi_{sf,k}\right)\left(d^\dagger-d\right)+\frac{B}{2}\left(d^\dagger d-dd^\dagger\right)\;.\label{eq:HamKel}
\end{equation}
Here, the $\nu=c,s$ modes have been omitted (integrated out) because they do not couple to the potential or the impurity, and therefore do not affect transport properties. To solve this model, we now take the wide-band limit, $\epsilon_k=\hbar v_Fk$ for all momenta $k$ ranging from $-\infty$ to $\infty$. The continuum limit then corresponds to
\begin{equation}
\frac{1}{L}\sum_k\rightarrow \int\limits_{-\infty}^\infty\frac{\mathrm{d}k}{2\pi}=\frac{1}{v_F}\int\limits_{-\infty}^\infty\frac{\mathrm{d}\epsilon_k}{2\pi\hbar}\; \;.
\end{equation}
As the Hamiltonian contains an explicit time-dependence, standard equilibrium techniques cannot be used, and we  instead use Keldysh techniques~\cite{kamenev} to calculate the necessary correlators. More information about the Keldysh structure employed in this section can be found in Appendix~\ref{ap:kel}.

According to the Keldysh prescription, each Green function gains an additional matrix structure,
\begin{equation}
\mathbf{G}=\begin{pmatrix}\mathbf{G}^\text{R} & \mathbf{G}^\text{K}\\ 0 & \mathbf{G}^\text{A}\end{pmatrix}\;,\label{eq:kelstruc}
\end{equation}
where $\mathbf{G}^\text{R/A}$ are the retarded and advanced Green functions, while $\mathbf{G}^\text{K}$ are the so-called Keldysh components of the Green functions. The desired two-point functions are proportional to the Keldysh Green functions and can in general be obtained from
\begin{equation}
\big\langle \psi_\mu \psi^\dagger_\nu\big\rangle=\frac{i}{2}G^\text{K}_{\mu\nu}\;.\label{eq:greendef}
\end{equation}
Returning to the current operator from Eq.~(\ref{eq:Ic}), we can now write the expectation value of the charge current as
\begin{equation}
I_c(t)\equiv\big\langle \hat{I}_c\big\rangle(t)=-\frac{eg_\bot}{4\hbar\sqrt{L}}\sum_k\left(G_{ld,k,11}^\text{K}+G_{ld,k,22}^\text{K}-G_{ld,k,12}^\text{K}-G_{ld,k,21}^\text{K}\right)(t,t)\;.\label{eq:chargecurrent2}
\end{equation}
Here, the first two indices of the Green functions, $ld$,  denote the block of the full Green function $\mathbf{G}$ being considered. The final two indices refer to the Nambu spinor component. Together with Eq.~(\ref{eq:Gldk}), we find the relevant Green function to be\footnote{Here, matrix multiplication of the form $\left(\mathbf{A}\cdot \mathbf{B}\right)(t,t^\prime)$ is shorthand notation for $\int\limits_{-\infty}^{\infty}\mathrm{d}t^{\prime\prime}\mathbf{A}(t,t^{\prime\prime})\cdot \mathbf{B}(t^{\prime\prime},t^\prime)$.}
\begin{equation}
\frac{1}{\sqrt{L}}\sum_k\mathbf{G}_{ld,k}(t,t)=\frac{g_\bot}{\hbar}\left(\mathbf{L}_0^\prime\cdot\mathbf{g}\cdot\mathbf{D}\right)(t,t)\;,\label{eq:Gldsum}
\end{equation}
with the dot self-energy being given by Eq.~(\ref{eq:sigmad}). It should be noted that in the steady state dc limit ($V(t)=\text{const.}$), the system is completely time-independent, such that Green functions assume the form $G(t,t^\prime)=G(t-t^\prime)$. As a result, the current is also time-independent.

The difficulty in finding propagators in any non-equilibrium problem is related to finding the corresponding non-equilibrium density matrix. In thermal equilibrium, the density matrix is given by $\hat{\rho}_0=\text{exp}[-\beta(\hat{H}-\mu\hat{N})]$, while out of equilibrium one has to solve the quantum Boltzmann equation. The latter is usually not possible in an exact manner. We will circumvent this problem by assuming that the bare flavor and spin-flavor modes are in thermal equilibrium, with the bias voltage only acting on the tunnel junctions between the leads and the dot. As we have seen in the previous section, the only \emph{full} Green function that we need for the calculation of the currents is the one on the dot. While this interacting dot is still very much out of equilibrium, we can now make use of Eqs.~(\ref{eq:dfull}) and (\ref{eq:sigmad}) to see that the non-equilibrium behavior can be expressed in terms of \emph{bare} Green functions, thereby avoiding any direct calculation of the non-equilibrium density matrix.

The required Keldysh Green functions in Eq.~(\ref{eq:chargecurrent2}) are components of the Green function matrix in Eq.~(\ref{eq:Gldsum}). To extract them, we utilize an identity following from Eq.~(\ref{eq:kelstruc}),
\begin{equation}
\left(\mathbf{A}\cdot \mathbf{B}\right)^\text{K}=\mathbf{A}^\text{R}\cdot \mathbf{B}^\text{K}+\mathbf{A}^\text{K}\cdot \mathbf{B}^\text{A}\;.\label{eq:RKKA}
\end{equation}
In order to evaluate such expressions, we employ standard methods for the retarded and advanced Green functions, while the Keldysh components are obtained using the general relation
\begin{equation}
\mathbf{G}^\text{K}=\mathbf{G}^\text{R}\cdot\mathbf{F}-\mathbf{F}\cdot\mathbf{G}^\text{A} \;,\label{eq:keldyshcomp}
\end{equation}
where the Hermitian matrix $\mathbf{F}$ can in principle be found by solving the quantum Boltzmann equation. In thermal equilibrium, the \emph{fluctuation-dissipation theorem} (FDT) holds \cite{kamenev},
\begin{equation}
\mathbf{F}(\epsilon)=\big(1-2n_F(\epsilon)\big)\mathbb{I}\equiv f(\epsilon)\mathbb{I}\;,\label{eq:FDT}
\end{equation}
where $n_F(\epsilon)$ is the Fermi-Dirac distribution, and $\mathbb{I}$ is the identity matrix. We emphasize that the above expression for the matrix $\mathbf{F}$ is \emph{only} valid in thermal equilibrium and cannot be used in general non-equilibrium conditions. However, as discussed above, the flavor and spin-flavor modes both act as baths in the thermodynamic limit, such that the bare Green functions corresponding to these modes can be assumed to satisfy the FDT. For these modes themselves, the time-dependent bias voltage can simply be interpreted as a time-dependent shift in the chemical potential~\cite{jauho}.

To proceed, we must now calculate the retarded, advanced and Keldysh Green functions of both the flavor modes and the spin-flavor modes, as well as the retarded and advanced components on the dot. For all of the bare retarded and advanced Green functions, we use the following relation,
\begin{equation}
\left(\delta(t-t^\prime)(i\partial_{t^\prime}-\epsilon_k(t^\prime)/\hbar\pm i0^+)\right)^{-1}=\mp i\theta\left(\pm(t-t^\prime)\right)e^{-\frac{i}{\hbar}\int\limits^t_{t^\prime}\mathrm{d}t^{\prime\prime}\,\epsilon_k(t^{\prime\prime})}\;.
\end{equation}
We consider first the Green function $\big(\mathbf{L}_0^\prime\big)^\text{R/A}$,
\begin{align}
\big(\mathbf{L}_0^\prime\big)^\text{R/A}(t,t^\prime)\Big|_{V=0}&=\frac{1}{v_F}\int\limits_{-\infty}^{\infty}\frac{\mathrm{d}\epsilon_k}{2\pi\hbar}\begin{pmatrix}\big(\delta(t-t^\prime)(i\partial_{t^\prime}-\epsilon_k/\hbar\pm i0^+)\big)^{-1} & 0 \\ 0 & \big(\delta(t-t^\prime)(i\partial_{t^\prime}+\epsilon_k/\hbar\pm i0^+)\big)^{-1}\end{pmatrix}\nonumber\\
&=\mp\frac{i}{v_F}\theta\left(\pm(t-t^\prime)\right)\int\limits_{-\infty}^{\infty}\frac{\mathrm{d}\epsilon_k}{2\pi\hbar}\begin{pmatrix}e^{-\frac{i\epsilon_k}{\hbar}(t-t^\prime)} & 0 \\ 0 & e^{\frac{i\epsilon_k}{\hbar}(t-t^\prime)}\end{pmatrix}\nonumber\\
&=\mp\frac{i}{2v_F}\delta(t-t^\prime)\mathbb{I}_2 \;,
\end{align}
where $\mathbb{I}_2$ is the $2\times 2$ identity matrix. Turning on the bias voltage does not change this result,  since
\begin{equation}
\big(\mathbf{L}_0^\prime\big)^\text{R/A}(t,t^\prime)=\mp\frac{i}{2v_F}\delta(t-t^\prime)\begin{pmatrix}e^{\frac{ie}{2\hbar}\int\limits^t_{t^\prime}\mathrm{d}t^{\prime\prime}\,V(t^{\prime\prime})} & 0 \\ 0 & e^{-\frac{ie}{2\hbar}\int\limits^t_{t^\prime}\mathrm{d}t^{\prime\prime}\,V(t^{\prime\prime})}\end{pmatrix}=\mp\frac{i}{2v_F}\delta(t-t^\prime)\mathbb{I}_2 \;.\label{eq:L0pRA}
\end{equation}
For the calculation of the Keldysh components, we turn to Eqs.~(\ref{eq:keldyshcomp}) and (\ref{eq:FDT}). Dropping the subscript $k$ from the integration variable, we find
\begin{align}
\big(\mathbf{L}_0^\prime\big)^\text{K}(t,t^\prime)&=-\frac{i}{v_F}\int\limits_{-\infty}^{\infty}\frac{\mathrm{d}\epsilon}{2\pi\hbar}\begin{pmatrix}f(\epsilon)\,e^{-\frac{i\epsilon}{\hbar}(t-t^\prime)+\frac{ie}{2\hbar}\int\limits^t_{t^\prime}\mathrm{d}t^{\prime\prime}\,V(t^{\prime\prime})} & 0 \\ 0 & f(-\epsilon)\,e^{\frac{i\epsilon}{\hbar}(t-t^\prime)-\frac{ie}{2\hbar}\int\limits^t_{t^\prime}\mathrm{d}t^{\prime\prime}\,V(t^{\prime\prime})}\end{pmatrix}\nonumber\\
&=-\frac{i}{v_F}\int\limits_{-\infty}^{\infty}\frac{\mathrm{d}\epsilon}{2\pi\hbar}f(\epsilon)e^{-\frac{i\epsilon}{\hbar}(t-t^\prime)}\begin{pmatrix}e^{\frac{ie}{2\hbar}\int\limits^t_{t^\prime}\mathrm{d}t^{\prime\prime}\,V(t^{\prime\prime})} & 0 \\ 0 & e^{-\frac{ie}{2\hbar}\int\limits^t_{t^\prime}\mathrm{d}t^{\prime\prime}\,V(t^{\prime\prime})}\end{pmatrix} \;.\label{eq:L0K}
\end{align}

We can now use the above results and properties to evaluate the charge current. To do this, we introduce Majorana Green functions on the dot, corresponding to the Majorana fermions $a$ and $b$. These are given by
\begin{align}
D_{aa}&=\frac{1}{2}\left(D_{11}+D_{12}+D_{21}+D_{22}\right)\;,\label{eq:majoranagreens1}\\
D_{bb}&=\frac{1}{2}\left(D_{11}-D_{12}-D_{21}+D_{22}\right)\;,\\
D_{ab}&=\frac{1}{2i}\left(D_{11}-D_{12}+D_{21}-D_{22}\right)\;,\\
D_{ba}&=\frac{1}{2i}\left(-D_{11}-D_{12}+D_{21}+D_{22}\right)\;,\label{eq:majoranagreens2}
\end{align}
where $D_{ij}$ are the original components of the $2\times 2$ matrix $\mathbf{D}$. In terms of these Majorana propagators, Eq.~(\ref{eq:Gldsum}) becomes
\begin{align}
\frac{1}{\sqrt{L}}\sum_k\mathbf{G}_{ld,k}&=\frac{g_\bot}{\hbar}\begin{pmatrix}L_{0,1}^\prime\left(D_{21}-D_{11}\right) & L_{0,1}^\prime\left(D_{22}-D_{12}\right)\\ L_{0,2}^\prime\left(D_{21}-D_{11}\right) & L_{0,2}^\prime\left(D_{22}-D_{12}\right)\end{pmatrix}\nonumber\\
&=\frac{g_\bot}{\hbar}\begin{pmatrix}L_{0,1}^\prime\left(-D_{bb}+iD_{ba}\right) & L_{0,1}^\prime\left(D_{bb}+iD_{ba}\right)\\ L_{0,2}^\prime\left(-D_{bb}+iD_{ba}\right) & L_{0,2}^\prime\left(D_{bb}+iD_{ba}\right)\end{pmatrix}\;.
\end{align}
An expression for the charge current now follows by inserting these results into Eq.~(\ref{eq:chargecurrent2}),
\begin{align}
I_c(t)&=\frac{eg_\bot^2}{2\hbar^2}\left(\left(L_{0,1}^\prime-L_{0,2}^\prime\right)D_{bb}\right)^\text{K}(t,t)\nonumber\\
&=\frac{eg_\bot^2}{2\hbar^2}\int\limits_{-\infty}^{\infty}\mathrm{d}t^\prime\left(\big(L_{0,1}^\prime\big)^\text{K}(t,t^\prime)-\big(L_{0,2}^\prime\big)^\text{K}(t,t^\prime)\right)D^\text{A}_{bb}(t^\prime,t)\nonumber\\
&=\frac{ieg_\bot^2}{2\hbar^2v_F}\int\limits_{-\infty}^{\infty}\frac{\mathrm{d}\epsilon}{2\pi\hbar}f(\epsilon)\int\limits_{-\infty}^{\infty}\mathrm{d}t^\prime\,e^{-\frac{i\epsilon}{\hbar}(t-t^\prime)}\bigg(e^{-\frac{ie}{2\hbar}\int\limits^t_{t^\prime}\mathrm{d}t^{\prime\prime}\,V(t^{\prime\prime})}-e^{\frac{ie}{2\hbar}\int\limits^t_{t^\prime}\mathrm{d}t^{\prime\prime}\,V(t^{\prime\prime})}\bigg)D^\text{A}_{bb}(t^\prime,t)  \;,\label{eq:chargecurrent4}
\end{align}
where we used that $\big(\mathbf{L}_0^\prime\big)^\text{R}\propto\mathbb{I}_2$ to find that the term proportional to $D_{bb}^\text{K}$ vanishes. Motivated by the work on a regular resonant level model from Ref.~\cite{jauho}, a different (and in many cases more convenient) way of writing the charge current is obtained by noting that $\hat{I}_c$ is a Hermitian operator, together with the observation that $\big(L_{0,2}^\prime\big)^\text{K}(t,t^\prime)=\big(\big(L_{0,1}^\prime\big)^\text{K}(t,t^\prime)\big)^*$. The latter is a consequence of the fact that $f(\epsilon)$ is an odd function in $\epsilon$. As a result, the second line of Eq.~(\ref{eq:chargecurrent4}) reveals that the Majorana dot Green function $D^\text{A}_{bb}(t,t^\prime)$ must be completely imaginary. This implies
\begin{align}
I_c(t)&=\frac{eg_\bot^2}{\hbar^2}\int\limits_{-\infty}^\infty\mathrm{d}t^\prime\,\text{Im}\!\left[-\frac{i}{v_F}\int\limits_{-\infty}^{\infty}\frac{\mathrm{d}\epsilon}{2\pi\hbar}f(\epsilon)e^{-\frac{i\epsilon}{\hbar}(t-t^\prime)+\frac{ie}{2\hbar}\int\limits^t_{t^\prime}\mathrm{d}t^{\prime\prime}\,V(t^{\prime\prime})}iD^\text{A}_{bb}(t^\prime,t)\right]\nonumber\\
&=\frac{e\Gamma}{2\hbar}\text{Im}\!\left[\int\limits_{-\infty}^{\infty}\frac{\mathrm{d}\epsilon}{2\pi\hbar}f(\epsilon)A(\epsilon,t)\right] \;,\label{eq:chargecurrent3}
\end{align}
with
\begin{equation}
A(\epsilon,t)\equiv\int\limits_{-\infty}^\infty\mathrm{d}t^\prime\,e^{-\frac{i\epsilon}{\hbar}(t-t^\prime)+\frac{ie}{2\hbar}\int\limits^t_{t^\prime}\mathrm{d}t^{\prime\prime}\,V(t^{\prime\prime})}D^\text{A}_{bb}(t^\prime,t)\;.\label{eq:A}
\end{equation}
This equation is the most general expression for the charge current, which depends only on the time-dependent form of the bias voltage $V(t)$, and the Majorana Green function on the dot, $D^\text{A}_{bb}(t^\prime,t)$. As such, the problem of finding the charge current for any time-dependent bias voltage reduces to the problem to finding the function $A(\epsilon,t)$.

Since the bare dot Green function $\mathbf{D}_0$ has not yet been specified, the results are still valid even for more general on-site dot behavior. However, we will restrict ourselves to the model at hand, where the bare on-site dot behavior is fully determined by the magnetic field $B$. As is shown in Appendix~\ref{ap:dotgreen}, the full Majorana dot Green function $\mathbf{D}$ is given by
\begin{align}
D^\text{R/A}_{bb}(t,t^\prime)&=\mp i\theta\left(\pm(t-t^\prime)\right)e^{\mp\frac{\Gamma}{2\hbar}(t-t^\prime)}\left[\cosh\left(\frac{1}{2\hbar}\sqrt{\Gamma^2-4B^2}(t-t^\prime)\right)\right.\nonumber\\
&\quad\,\mp\left.\frac{\Gamma}{\sqrt{\Gamma^2-4B^2}}\sinh\left(\frac{1}{2\hbar}\sqrt{\Gamma^2-4B^2}(t-t^\prime)\right)\right] \;,
\end{align}
while its Fourier transform is simply
\begin{equation}
D^\text{R/A}_{bb}(\epsilon)=\frac{\hbar\epsilon}{\epsilon(\epsilon\pm i\Gamma)-B^2} \;.\label{eq:Dbbe}
\end{equation}
Having derived a general framework to solve this out-of-equilibrium problem, we will now apply the framework to several example bias voltages that are relevant to experiments.


\subsection{The dc solution}\label{sec:dc}
Let us first consider the dc solution with $V(t)=V$. In this case the function $A(\epsilon,t)$ reduces to $D^\text{A}_{bb}(\epsilon-eV/2)$, as shown in  Appendix~\ref{ap:dotgreen}. Using Eq.~(\ref{eq:chargecurrent4}) we find
\begin{equation}
I_c(t)=\frac{ie\Gamma}{4\hbar}\int\limits_{-\infty}^{\infty}\frac{\mathrm{d}\epsilon}{2\pi\hbar}\left[f\left(\epsilon-\frac{eV}{2}\right)-f\left(\epsilon+\frac{eV}{2}\right)\right]D^\text{A}_{bb}(\epsilon)\;.
\end{equation}
The combination $f(\epsilon-eV/2)-f(\epsilon+eV/2)$ is even in $\epsilon$, so the odd part of $D^\text{A}_{bb}(\epsilon)$ does not contribute to the overall integral. Furthermore, the explicit expression in Eq.~(\ref{eq:Dbbe}) implies that the even part of $D^\text{A}_{bb}(\epsilon)$ is simply the imaginary part (this is physically sensible since the expectation of the current should in the end be pure real). Therefore we find
\begin{equation}
I_c(t)=\frac{e\Gamma}{2}\int\limits_{-\infty}^{\infty}\frac{\mathrm{d}\epsilon}{2\pi\hbar}\left[n_F\left(\epsilon-\frac{eV}{2}\right)-n_F\left(\epsilon+\frac{eV}{2}\right)\right]\left(-\text{Im}\!\left[\frac{\epsilon}{\epsilon(\epsilon+i\Gamma)-B^2}\right]\right) \;,\label{eq:Icdc}
\end{equation}
where we have used $f(\epsilon-eV/2)-f(\epsilon+eV/2)=2\left(n_F(\epsilon+eV/2)-n_F(\epsilon-eV/2)\right)$, together with $\text{Im}\!\left[D^\text{A}_{bb}(\epsilon)\right]=-\text{Im}\!\left[D^\text{R}_{bb}(\epsilon)\right]$. Note that the latter object is simply $\pi$ times the spectral function corresponding to the $b$ Majorana fermion and that the expression is indeed independent of $t$. Moreover, Eq.~(\ref{eq:Icdc}) is consistent with the known results for the anisotropic spin 2CK model\footnote{The expressions for the dc charge current
in the spin and charge 2CK models are however not identical. This is because, in the case of the spin 2CK model, the bias voltage couples to both spin up and spin down electrons in the leads, whereas in the charge 2CK model, the voltage only couples to the effective spin up lead electrons. Importantly, the spin 2CK model does \emph{not} support any charge transport for $J^-=0$ and $J^{LR}_\bot=0$, while the charge 2CK has non-zero and in fact strongly Kondo-enhanced conductance at this point.   All subsequent references to ``known" results refer to the spin 2CK model, and it should be understood that differences arise on going to the charge 2CK case.\label{fn:knownresults}} from Ref.~\cite{sh2}.

We now go further and evaluate the integral in Eq.~(\ref{eq:Icdc}) to find a closed-form expression for the full non-equilibrium charge current for this system in the dc limit. We do this by making use of the Matsubara representation of the Fermi-Dirac distribution,
\begin{equation}
n_F(\epsilon)=\frac{1}{\beta}\sum\limits_{\mathclap{n=-\infty}}^\infty\frac{e^{i\omega_n0^+}}{i\hbar\omega_n-\epsilon}\;,\qquad\omega_n\equiv\frac{\pi\left(2n+1\right)}{\hbar\beta} \;,\label{eq:matsrep}
\end{equation}
where $\omega_n$ are the fermionic Matsubara frequencies. The chemical potential $\mu$ is absent from this expression, due to our implicit choice to measure all energies with respect to it. Plugging this back into Eq.~(\ref{eq:Icdc}) and splitting the sum into two parts, we find
\begin{equation}
I_c=\frac{e\Gamma}{2\pi\hbar\beta}\sum\limits_{n=0}^\infty\text{Re}\!\left[\int\limits_{-\infty}^\infty\mathrm{d}\epsilon\frac{\Gamma\epsilon^2}{\epsilon^4+\left(\Gamma^2-2B^2\right)\epsilon^2+B^4}\left(\frac{e^{i\omega_n0^+}}{i\hbar\omega_n-\left(\epsilon-eV/2\right)}-\frac{e^{i\omega_n0^+}}{i\hbar\omega_n-\left(\epsilon+eV/2\right)}\right)\right]\;,
\end{equation}
where we used the observation that the sum over $n$ from $-\infty$ to $-1$ is simply the complex conjugate of the sum from $0$ to $\infty$. We evaluate the remaining integral using contour integration. Closing the contour in the negative imaginary plane and assuming $4B^2<\Gamma^2$, the only enclosed poles are located at
\begin{equation}
\label{eq:epm}
-\!i\epsilon_\pm\equiv-i\Gamma\sqrt{\frac{1}{2}-\left(\frac{B}{\Gamma}\right)^2\pm\sqrt{\frac{1}{4}-\left(\frac{B}{\Gamma}\right)^2}}\;.
\end{equation}
The corresponding residue is given by
\begin{equation}
\text{Res}\!\left(\frac{\Gamma\epsilon^2}{\epsilon^4+\left(\Gamma^2-2B^2\right)\epsilon^2+B^4},-i\epsilon_\pm\right)=\pm\frac{i\epsilon_\pm}{2\Gamma\sqrt{1-4\left(\frac{B}{\Gamma}\right)^2}}\;.
\end{equation}
Using the residue theorem, we now find
\begin{equation}
I_c=\frac{e}{2\hbar\beta}\frac{1}{\sqrt{1-4\left(\frac{B}{\Gamma}\right)^2}}\sum\limits_{\mathclap{\alpha=\pm 1}}\alpha\epsilon_\alpha\sum\limits_{n=0}^\infty\text{Re}\!\left[\frac{1}{i\hbar\omega_n+\left(i\epsilon_\alpha+eV/2\right)}-\frac{1}{i\hbar\omega_n+\left(i\epsilon_\alpha-eV/2\right)}\right]\;,
\end{equation}
where we discarded the factor $e^{i\omega_n0^+}$. This is allowed because this factor only becomes important in the large $n$ limit, while the remainder of the summand scales with $n^{-2}$. Now to finish the derivation, we make use of the \emph{digamma function}, defined in terms of the gamma function as $\Psi(z) =\mathrm{d}\ln\Gamma(z)/\mathrm{d}z = [\Psi(z^*)]^*$, and note the following identity
\begin{equation}
\Psi(a)-\Psi(b)=(a-b)\sum\limits_{n=0}^\infty\frac{1}{(n+a)(n+b)} \;,\label{eq:digammadiff}
\end{equation}
which is a very useful property for all calculations at non-zero temperatures that are to follow. More information about this digamma function, including a derivation of the latter identity, can be found in Appendix~\ref{ap:digamma}. It then follows that
\begin{align}
\sum\limits_{n=0}^\infty\text{Re}\!\left[\frac{1}{i\hbar\omega_n+\left(i\epsilon_\alpha\pm eV/2\right)}\right]&=\left(\frac{\beta}{2\pi}\right)^2\sum\limits_{n=0}^\infty\frac{\pm eV/2}{\left(n+\frac{1}{2}+\frac{\epsilon_\alpha\mp ieV/2}{2\pi k_BT}\right)\left(n+\frac{1}{2}+\frac{\epsilon_\alpha\pm ieV/2}{2\pi k_BT}\right)}\nonumber\\
&=\frac{i\beta}{4\pi}\left[\Psi\left(\frac{1}{2}+\frac{\epsilon_\alpha\mp ieV/2}{2\pi k_BT}\right)-\Psi\left(\frac{1}{2}+\frac{\epsilon_\alpha\pm ieV/2}{2\pi k_BT}\right)\right]\nonumber\\
&=\frac{\beta}{2\pi}\text{Im}\!\left[\Psi\left(\frac{1}{2}+\frac{\epsilon_\alpha\pm ieV/2}{2\pi k_BT}\right)\right]\;.
\end{align}
This gives the final expression for the finite-temperature and non-equilibrium dc current at the EK point of the 2CK model, which is exact:
\begin{equation}
I_c=\frac{e}{2\pi\hbar}\frac{1}{\sqrt{1-4\left(\frac{B}{\Gamma}\right)^2}}\left(\epsilon_+\text{Im}\!\left[\Psi\left(\frac{1}{2}+\frac{\epsilon_++ieV/2}{2\pi k_BT}\right)\right]-\epsilon_-\text{Im}\!\left[\Psi\left(\frac{1}{2}+\frac{\epsilon_-+ieV/2}{2\pi k_BT}\right)\right]\right)\;.\label{eq:Idc}
\end{equation}

\begin{figure}[t]
  \centerline{\begin{tabular}{cc}
    \includegraphics[height=4.8 cm]{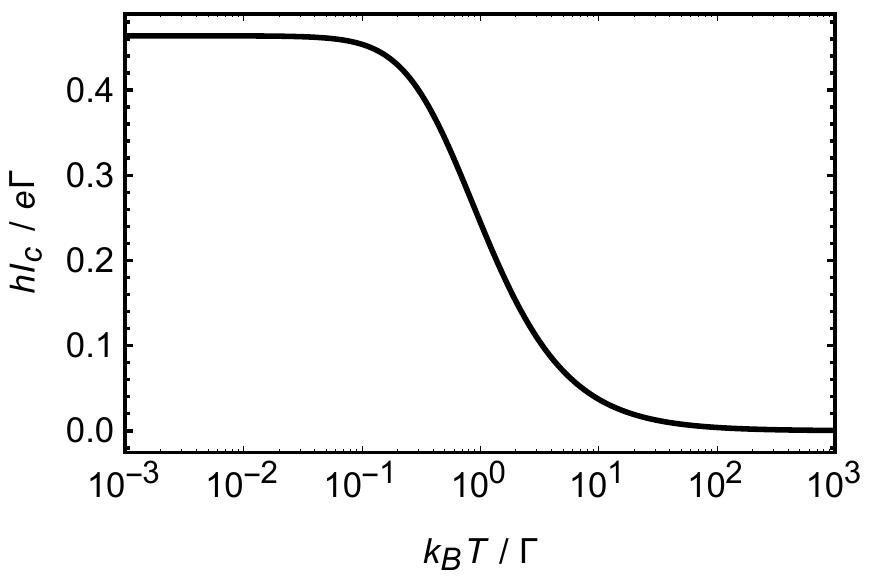} & \includegraphics[height=4.8 cm]{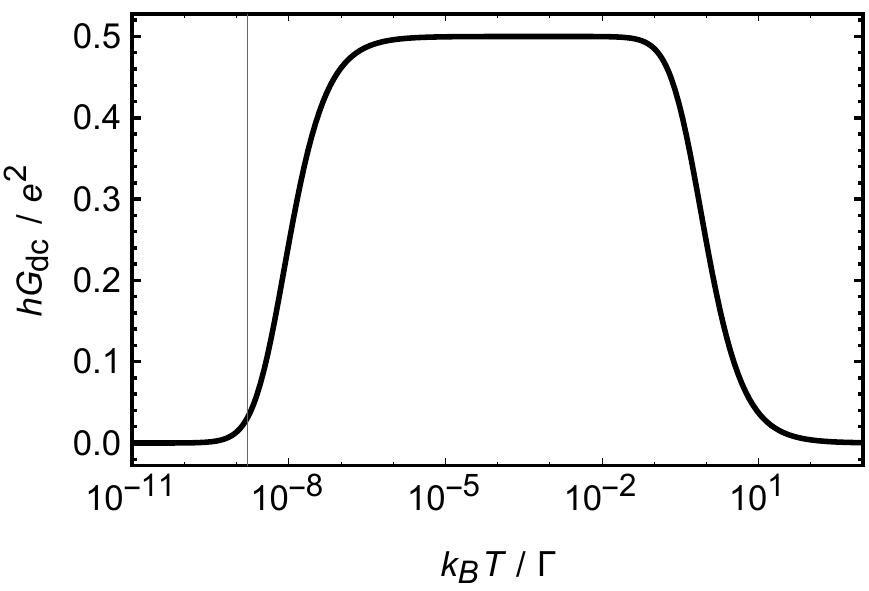}\end{tabular}}
\caption{\label{fig:dctemp}Dc electric transport of the C2CK model at the EK point. Left: dc current as a function of temperature for $eV/\Gamma=1$ and $B^2/\Gamma^2=0$. Right: dc differential conductance in the limit $V\rightarrow 0$, for $B^2/\Gamma^2=10^{-8}$. The vertical line corresponds to the FL crossover temperature, $T^*$.}
\end{figure}

The \emph{differential conductance}, $G$, can then be obtained. For our purposes, it is defined as
\begin{equation}
G\equiv\left\langle\frac{\mathrm{d}I_c(t)}{\mathrm{d}V(t)}\right\rangle_t \;,\label{eq:diffcond}
\end{equation}
where $\langle\ldots\rangle_t$ denotes the time average. In the dc case, this gives
\begin{align}
G_\text{dc}&=\frac{e^2}{4\pi\hbar}\frac{1}{\sqrt{1-4\left(\frac{B}{\Gamma}\right)^2}}\left(\frac{\epsilon_+}{2\pi k_BT}\text{Re}\!\left[\psi^{(1)}\left(\frac{1}{2}+\frac{\epsilon_++ieV/2}{2\pi k_BT}\right)\right]-\frac{\epsilon_-}{2\pi k_BT}\text{Re}\!\left[\psi^{(1)}\left(\frac{1}{2}+\frac{\epsilon_-+ieV/2}{2\pi k_BT}\right)\right]\right)\nonumber\\
&\!\!\!\stackrel{V\rightarrow 0}{=}\frac{e^2}{4\pi\hbar}\frac{1}{\sqrt{1-4\left(\frac{B}{\Gamma}\right)^2}}\left(\frac{\epsilon_+}{2\pi k_BT}\psi^{(1)}\left(\frac{1}{2}+\frac{\epsilon_+}{2\pi k_BT}\right)-\frac{\epsilon_-}{2\pi k_BT}\psi^{(1)}\left(\frac{1}{2}+\frac{\epsilon_-}{2\pi k_BT}\right)\right)\;,\label{eq:GTs}
\end{align}
with $\psi^{(1)}(z)$ being the trigamma function, {\it i.e.}, the derivative of $\Psi(z)$. Fig.~\ref{fig:dctemp} shows examples of the dc current and conductance as functions of temperature. At zero field (left panel), the Kondo effect leads to enhanced current flow through the dot at finite bias, on the temperature scale of $T_K$. This corresponds to the non-equilibrium crossover from the local moment fixed point to the NFL fixed point, and is seen from  Fig.~\ref{fig:dctemp} to arise for $k_B T_K \sim \Gamma$. For finite magnetic field $B\ne 0$ (right panel), the NFL fixed point is destabilized and a FL crossover is generated. This crossover shows up in the zero-bias conductance on the temperature scale of $T^*$ (gray vertical line), which can be read off as $k_B T^* \sim B^2/\Gamma$.

With $B=0$, we have $\epsilon_+=\Gamma$ and  $\epsilon_-=0$ from Eq.~(\ref{eq:epm}). If additionally $V=0$, then Eq.~(\ref{eq:GTs}) shows that there is a single characteristic scale in the problem, $\epsilon_+$. We identify this with the Kondo scale, defining $k_B T_K \equiv \Gamma/2\pi$.
Within the effective Majorana resonant level description, the Kondo scale is therefore simply proportional to the effective dot-lead hybridization.\footnote{Note that this expression for $T_K$ is a peculiarity of the non-interacting EK point: the Kondo scale is exponentially small in the dot-lead exchange coupling in the isotropic 2CK model and indeed the true C2CK system.}
The zero bias conductance in this limit is a universal function of the single rescaled parameter $x=T/T_K$, which follows from Eq.~(\ref{eq:GTs}) as $G_{\text{dc}}(x)=(e^2/2h)\tfrac{1}{x}\psi^{(1)}\left(\tfrac{1}{2}+\tfrac{1}{x}\right )$.
Note that this expression for the conductance has a well-defined limit as $x\rightarrow 0$, corresponding to low temperatures compared with $T_K$, and gives $G_{\text{dc}}^\text{NFL}=e^2/2h$ at the NFL fixed point.

As explained in Sec.~\ref{sec:lowtoulouse}, the above results only capture the physics of the real C2CK quantum dot device in the limit $T\ll T_K$ (or equivalently $x\to 0$), since then both the anisotropic 2CK model at the EK point and the C2CK model both have flowed under RG to the same isotropic 2CK fixed point. Thus, we conclude that $G_{\text{dc}}^\text{NFL}=e^2/2h$ applies for $T\ll T_K$ at the critical point of the real C2CK system.

Turning now to finite $B$ and the resulting FL crossover, Eq.~(\ref{eq:epm}) gives $\epsilon_+=\Gamma $ and $\epsilon_-=B^2/\Gamma $ in the limit $B^2\ll\Gamma^2$. From Eq.~(\ref{eq:GTs}) we may still identify the Kondo scale as $k_B T_K = \Gamma/2\pi$, but now we have a second scale in the problem, $k_B T^* = B^2/(2\pi\Gamma)$, such that $T^*\ll T_K$. Taking the limit $T/T_K \to 0$ while keeping $T/T^*$ finite yields an expression for the crossover on the temperature scale of $T^*$. This is the FL crossover, and is a universal function of the single parameter $y=T/T^*$, provided there is good scale separation $T^*\ll T_K$. Importantly, since $T\ll T_K$ along this entire FL crossover, it again describes the physical C2CK system of interest (Sec.~\ref{sec:lowtoulouse}).
These considerations lead us to the main result of this section: the \emph{exact} dc charge conductance of the C2CK model along the FL crossover \cite{sh2,eran1,eran2,andrew},
\begin{equation}
G_\text{dc}=\frac{e^2}{2h}\left(1-\frac{T^*}{ T}\text{Re}\!\left[\psi^{(1)}\left(\frac{1}{2}+\frac{T^*}{T}+\frac{ieV}{4\pi k_BT}\right)\right]\right) \;.\label{eq:Gdcinf}
\end{equation}
This result holds in the full non-equilibrium situation at finite voltage bias, provided $eV \ll k_B T_K$ as well as $T\ll T_K$. Indeed, this result has been confirmed  directly in the C2CK experiment of Ref.~\cite{iftikhar2018tunable} in the linear response regime by scanning the dot gate voltage across the Coulomb peak. The gate detuning away from the dot charge degeneracy point in this system corresponds in pseudospin language to the magnetic field $B$, and is responsible for generating the FL crossover.

Finally we comment on the FL crossover generated by other symmetry-breaking perturbations, such as channel asymmetry, rather than by magnetic field as considered explicitly above. In fact, Eq.~(\ref{eq:Gdcinf}) is universal in the stronger sense that the same conductance behavior is obtained along the FL crossover as a function of $T/T^*$, independent of the perturbations generating the scale $T^*$. Although we do not repeat the calculation here, we have explicitly confirmed Eq.~(\ref{eq:Gdcinf}) in the case of channel asymmetry, where we find $ k_BT^*=(J_\bot^{LL}-J_\bot^{RR})^2/32\pi^2 a_0\hbar v_F$. In practice in the experimental context, the precise strength of perturbations (or indeed the combination of perturbations) will not be known; instead, $T^*$ can simply be related to the conductance half-width-at-half-maximum.


\subsection{The ac solution}\label{sec:acsol}
We now proceed to the time-dependent case of an ac bias voltage, $V(t)=V_0+\Delta V\cos(\omega_0 t)$. We employ a method similar to that of Floquet theory. In this case, Eq.~(\ref{eq:A}) yields
\begin{equation}
A(\epsilon,t)=\int\limits_{-\infty}^\infty\mathrm{d}t^\prime\,e^{-\frac{i(\epsilon-eV_0/2)}{\hbar}(t-t^\prime)+\frac{ie\Delta V}{2\hbar}\int\limits_{t^\prime}^t\mathrm{d}t^{\prime\prime}\cos(\omega_0t^{\prime\prime})}D^\text{A}_{bb}(t^\prime,t) \;. \label{eq:Aepst}
\end{equation}
The trick to evaluate this expression is to note that all terms of $D^\text{A}_{bb}(t^\prime,t)$ are of the form $c\theta(t-t^\prime)e^{-z(t-t^\prime)}$, with $\text{Re}[z]>0$ (see Appendix~\ref{ap:dotgreen}). To find an analytic expression for the function $A(\epsilon,t)$, we use the following identity~\cite{jauho} involving Bessel functions of the first kind $J_n(\alpha)$ (see also Appendix~\ref{ap:bessel}),
\begin{equation}
\sum\limits_{\mathclap{n=-\infty}}^\infty e^{-in\omega t}J_n(\alpha)=e^{-i\alpha\sin(\omega t)} \;.\label{eq:besselsum}
\end{equation}
This allows us to write all terms in $A(\epsilon,t)$ in terms of Bessel functions, using
\begin{align}
&e^{-\left(z+i(\epsilon-eV_0/2)/\hbar\right)t+\frac{ie\Delta V}{2\hbar\omega_0}\sin(\omega_0t)}\int\limits_{-\infty}^t\mathrm{d}t^\prime\,e^{\left(z+i(\epsilon-eV_0/2)/\hbar\right)t^\prime-\frac{ie\Delta V}{2\hbar\omega_0}\sin(\omega_0t^\prime)}\nonumber\\
=&\,e^{-\left(z+i(\epsilon-eV_0/2)/\hbar\right)t+\frac{ie\Delta V}{2\hbar\omega_0}\sin(\omega_0t)}\sum_{\mathclap{n=-\infty}}^\infty J_n\left(\frac{e\Delta V}{2\hbar\omega_0}\right)\int\limits_{-\infty}^t\mathrm{d}t^\prime\,e^{\left(z+i(\epsilon-eV_0/2-n\hbar\omega_0)/\hbar\right)t^\prime}\nonumber\\
=&\,\sum_{\mathclap{n=-\infty}}^\infty J_n\left(\frac{e\Delta V}{2\hbar\omega_0}\right)\frac{e^{i\left(\frac{e\Delta V}{2\hbar\omega_0}\sin(\omega_0t)-n\omega_0t\right)}}{z+i(\epsilon-eV_0/2-n\hbar\omega_0)/\hbar} \;. \label{eq:deltaVcase}
\end{align}
In the dc limit $\Delta V=0$, Eq.~(\ref{eq:deltaVcase}) reduces to
\begin{equation}
\frac{1}{z+i(\epsilon-eV_0/2)/\hbar}\equiv \sum_{\mathclap{n=-\infty}}^\infty J_n\left(\frac{e\Delta V}{2\hbar\omega_0}\right)\frac{e^{i\left(\frac{e\Delta V}{2\hbar\omega_0}\sin(\omega_0t)-n\omega_0t\right)}}{z+i(\epsilon-eV_0/2)/\hbar} \;. \label{eq:deltaV0case}
\end{equation}
Comparing Eqs.~(\ref{eq:deltaVcase}) and (\ref{eq:deltaV0case}), and noting that
$A_\text{dc}(\epsilon)=D^\text{A}_{bb}(\epsilon-eV_0/2)$ from Eq.~(\ref{eq:Aepst}), we find
\begin{equation}
A(\epsilon,t)=\sum_{\mathclap{n=-\infty}}^\infty J_n\left(\frac{e\Delta V}{2\hbar\omega_0}\right)e^{i\left(\frac{e\Delta V}{2\hbar\omega_0}\sin(\omega_0t)-n\omega_0t\right)}D^\text{A}_{bb}\left(\epsilon-\frac{eV_0}{2}-n\hbar\omega_0\right) \;.
\end{equation}
Returning to Eq.~(\ref{eq:chargecurrent3}) and applying this result, we obtain
\begin{equation}
I_c(t)=\frac{e\Gamma}{2\hbar}\text{Im}\!\left[e^{i\frac{e\Delta V}{2\hbar\omega_0}\sin(\omega_0t)}\sum_{\mathclap{n=-\infty}}^\infty e^{-in\omega_0t}J_n\left(\frac{e\Delta V}{2\hbar\omega_0}\right)\int\limits_{-\infty}^{\infty}\frac{\mathrm{d}\epsilon}{2\pi\hbar}f\left(\epsilon+\frac{eV_0}{2}+n\hbar\omega_0\right)D^\text{A}_{bb}(\epsilon)\right] \;.
\end{equation}
Another trick we can use is noting that we can replace $f(\epsilon+eV_0/2+n\hbar\omega_0)$ by $f(\epsilon+eV_0/2+n\hbar\omega_0)-f(\epsilon)$. The reason we can do this is because this additional term is odd in $\epsilon$, while not containing any $n$-dependence. Lacking any $n$-dependence, all the prefactors in front of the integral of this additional term are equal to $1$, so this term is proportional to the integral over $f(\epsilon)$ (an odd function) times the imaginary part of $D^\text{A}_{bb}(\epsilon)$. As we discussed before, the latter is even, so the integral vanishes and this additional term is therefore equal to zero. Applying this trick, we obtain
\begin{align}
I_c(t)&=e\Gamma\,\text{Im}\Bigg[e^{i\frac{e\Delta V}{2\hbar\omega_0}\sin(\omega_0t)}\sum_{\mathclap{n=-\infty}}^\infty e^{-in\omega_0t}J_n\left(\frac{e\Delta V}{2\hbar\omega_0}\right)\nonumber\\
&\quad\,\times\int\limits_{-\infty}^{\infty}\frac{\mathrm{d}\epsilon}{2\pi\hbar}\left[n_F(\epsilon)-n_F\left(\epsilon-\frac{eV_0}{2}-n\hbar\omega_0\right)\right]\frac{\epsilon}{\epsilon(\epsilon+i\Gamma)-B^2}\Bigg] \;,\label{eq:Itac}
\end{align}
agreeing with the ac results for the anisotropic spin 2CK from Ref.~\cite{sh1}.

\begin{figure}[t]
  \centerline{\begin{tabular}{cc}
    \includegraphics[height=4.8cm]{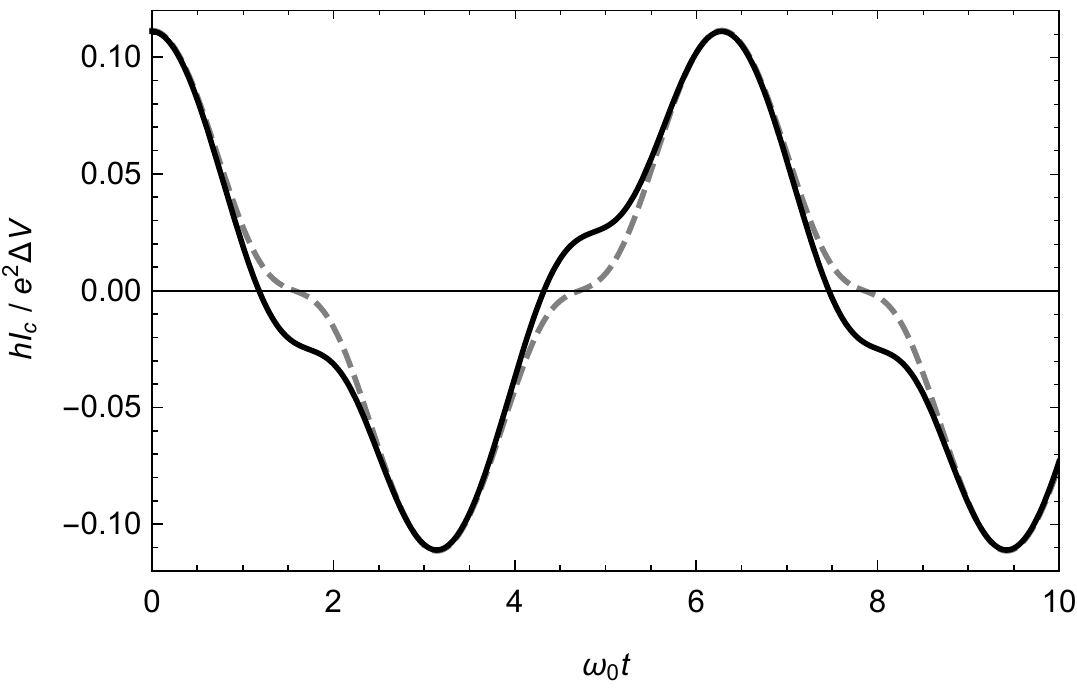} &
    \includegraphics[height=5.2 cm]{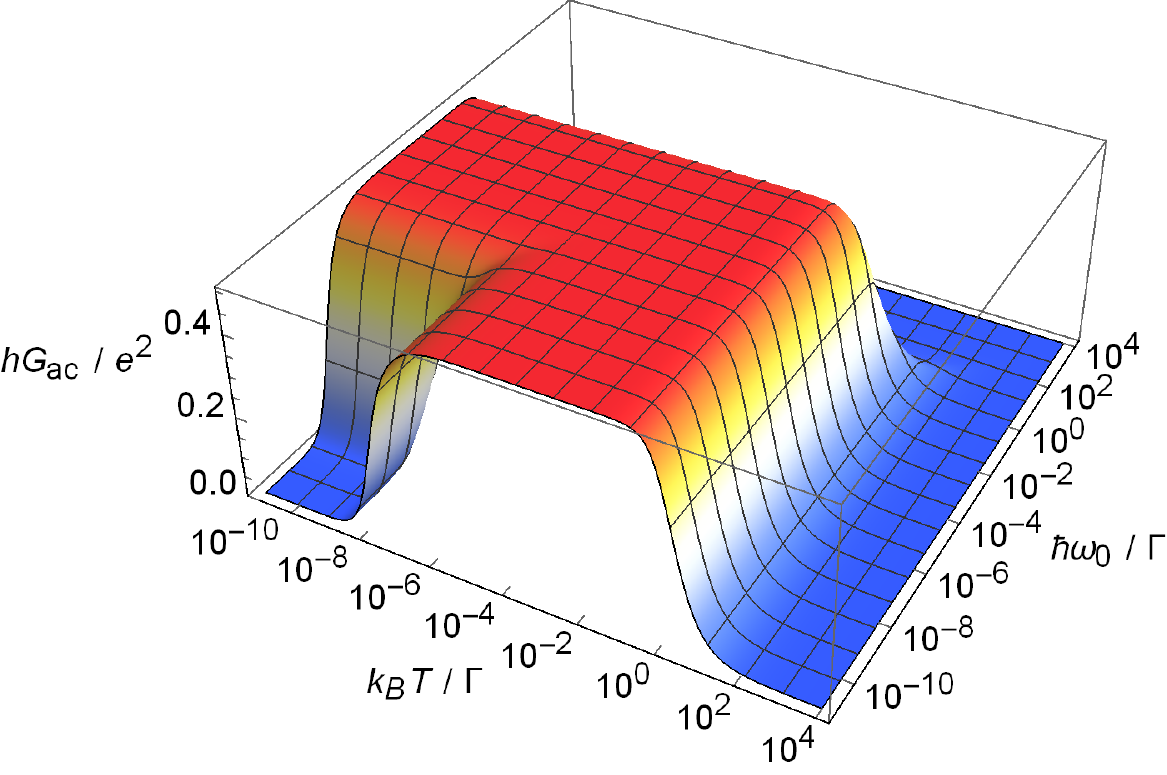}
  \end{tabular}}
\caption{\label{fig:accond}Ac electric transport of the 2CK model at the EK point. Left: example of the ac current (solid) versus its adiabatic limit $I_{c,\text{dc}}(V(t))$ (dashed) as a function of time for $T_K\rightarrow\infty$, $V_0=0$, $T/(2\pi T^*)=\hbar\omega_0/(2\pi k_BT^*)=0.1$ and $e\Delta V/(2\pi k_BT^*)=2$. Right: ac differential conductance in the limit $V\rightarrow 0$, with $T^*/T_K=10^{-8}$.}
\end{figure}

The sum over the Bessel functions in the expression prevents further simplification and a direct closed-form solution. However we may extract further analytic insight from Eq.~(\ref{eq:Itac}) by writing $I_c(t)$ in terms of its Fourier components. We use the convention
\begin{equation}
I_c(t)=\sum\limits_{\mathclap{n=-\infty}}^\infty I_n e^{in\omega_0t}\qquad\Longleftrightarrow\qquad I_n=\frac{1}{T}\int\limits_{\mathclap{-T/2}}^{\mathclap{T/2}}\mathrm{d}t\,I_c(t)e^{-in\omega_0t}  \;,
\end{equation}
where $T$ is the period of the oscillating current, and $\omega_0$ is the corresponding frequency (which is the same as our previous $\omega_0$ due to the periodicity of the original Hamiltonian). Now we Fourier transform the time-dependences of the current, using Eq.~(\ref{eq:besselsum}):
\begin{align}
\frac{1}{T}\int\limits_{\mathclap{-T/2}}^{\mathclap{T/2}}\mathrm{d}t\,e^{-i(n\pm n^\prime)\omega_0t\pm\frac{ie\Delta V}{2\hbar\omega_0}\sin(\omega_0t)}&=\frac{1}{T}\sum\limits_{\mathclap{m=-\infty}}^\infty\quad\int\limits_{\mathclap{-T/2}}^{\mathclap{T/2}}\mathrm{d}t\,e^{i(\pm m-n\mp n^\prime)\omega_0t}J_m\left(\frac{e\Delta V}{2\hbar\omega_0}\right)\nonumber\\
&=\sum\limits_{\mathclap{m=-\infty}}^\infty\delta_{\pm m,n\pm n^\prime}J_m\left(\frac{e\Delta V}{2\hbar\omega_0}\right)=J_{n^\prime\pm n}\left(\frac{e\Delta V}{2\hbar\omega_0}\right) \;.
\end{align}
Additionally, we use $\text{Im}[f(t)z]=\left(f(t)z-(f(t)z)^*\right)/2i$ to find that the Fourier transform of a term of this form is simply $\left(f_nz-(f^*)_nz^*\right)/2i$, where $(f^*)_n$ is the Fourier transform of the complex conjugate of $f(t)$. The Fourier components of the current are therefore given by
\begin{align}
I_n&=-\frac{ie\Gamma}{2}\sum_{\mathclap{n^\prime=-\infty}}^\infty J_{n^\prime}\left(\frac{e\Delta V}{2\hbar\omega_0}\right)\int\limits_{-\infty}^{\infty}\frac{\mathrm{d}\epsilon}{2\pi\hbar}\left[n_F(\epsilon)-n_F\left(\epsilon-\frac{eV_0}{2}-n^\prime\hbar\omega_0\right)\right]\nonumber\\
&\quad\,\times\left(J_{n^\prime+n}\left(\frac{e\Delta V}{2\hbar\omega_0}\right)\frac{\epsilon}{\epsilon(\epsilon+i\Gamma)-B^2}-J_{n^\prime-n}\left(\frac{e\Delta V}{2\hbar\omega_0}\right)\frac{\epsilon}{\epsilon(\epsilon-i\Gamma)-B^2}\right)\nonumber\\
&=\frac{ie\Gamma}{2}\sum_{\mathclap{n^\prime=-\infty}}^\infty J_{n^\prime}\left(\frac{e\Delta V}{2\hbar\omega_0}\right)J_{n^\prime+n}\left(\frac{e\Delta V}{2\hbar\omega_0}\right)\int\limits_{-\infty}^{\infty}\frac{\mathrm{d}\epsilon}{2\pi\hbar}\bigg(\frac{2i\Gamma\epsilon^2n_F(\epsilon)}{\epsilon^4+\left(\Gamma^2-2B^2\right)\epsilon^2+B^4}\nonumber\\
&\quad\,+\frac{\epsilon\,n_F(\epsilon-eV_0/2-n^\prime\hbar\omega_0)}{\epsilon(\epsilon+i\Gamma)-B^2}-\frac{\epsilon\,n_F(\epsilon-eV_0/2-(n^\prime+n)\hbar\omega_0)}{\epsilon(\epsilon-i\Gamma)-B^2}\bigg) \;.\label{eq:Inint}
\end{align}
It is now possible to fully evaluate the remaining integrals. However, the resulting expressions are rather cumbersome and do not contain great physical significance (see also Sec.~\ref{sec:lowtoulouse}). We therefore omit that calculation here, but direct the interested reader to Appendix~\ref{ap:accurrent} for the full evaluation of the current at $B=0$. As an illustration of the current dynamics in this system, a plot of the current at a point along the FL crossover line is in shown in the left panel of Fig.~\ref{fig:accond}.

We now focus on the differential conductance, defined in  Eq.~(\ref{eq:diffcond}). Although Eq.~(\ref{eq:Inint}) is general and exact, for simplicity and concreteness we now consider the small $\Delta V$ behavior around $V_0=0$ (\textit{i.e.}, the linear response regime due to an ac bias voltage, in the zero-bias limit). First, we expand the current in powers of $\Delta V$. To do this, we employ the following expansions obtained in Appendix~\ref{ap:bessel},
\begin{equation}
J_0(\alpha)=1+\mathcal{O}(\alpha^2)\;,\qquad J_{n\neq 0}(\alpha)=\frac{\left(\text{sgn}(n)\right)^{|n|}}{|n|!}\left(\frac{\alpha}{2}\right)^{|n|}+\mathcal{O}(\alpha^{|n|+2}) \;.\label{eq:besselexp}
\end{equation}
Inserting these into Eq.~(\ref{eq:Inint}), it immediately follows that
\begin{equation}
I_n\propto\left(\frac{e\Delta V}{2\hbar\omega_0}\right)^{|n|}+\mathcal{O}\bigg(\left(\frac{e\Delta V}{2\hbar\omega_0}\right)^{|n|+2}\bigg) \;.
\end{equation}
Returning to the full time-dependent current, we thus find
\begin{align}
I_c(t)&=I_0+I_1e^{i\omega_0t}+I_1^*e^{-i\omega_0t}+\mathcal{O}\bigg(\left(\frac{e\Delta V}{2\hbar\omega_0}\right)^2\bigg)\nonumber\\
&=I_0+2\,\text{Re}[I_1]\cos(\omega_0t)-2\,\text{Im}[I_1]\sin(\omega_0t)+\mathcal{O}\bigg(\left(\frac{e\Delta V}{2\hbar\omega_0}\right)^2\bigg) \;,
\end{align}
where we used the reality condition for the current $I_c(t)$ to write $I_{-1}=I_1^*$. From here, we can calculate the linear response differential conductance:
\begin{align}
G_\text{ac}&=\left\langle\frac{\partial I_c(t)}{\partial t}\Big/\frac{\partial V(t)}{\partial t}\right\rangle_t\nonumber\\
&=\frac{2\,\text{Re}[I_1]}{\Delta V}+\frac{2\,\text{Im}[I_1]}{\Delta V}\cancelto{0}{\left\langle\frac{\cos(\omega_0t)}{\sin(\omega_0t)}\right\rangle_t} \;.\label{eq:acconddef}
\end{align}
Next, we expand $I_1$, using
\begin{align}
J_n(\alpha)J_{n+1}(\alpha)&=\left(\delta_{n,0}+\frac{\alpha}{2}(\delta_{n,1}-\delta_{n,-1})+\mathcal{O}(\alpha^2)\right)\left(\delta_{n,-1}+\frac{\alpha}{2}(\delta_{n,0}-\delta_{n,-2})+\mathcal{O}(\alpha^2)\right)\nonumber\\
&=\frac{\alpha}{2}(\delta_{n,0}-\delta_{n,-1})+\mathcal{O}(\alpha^2) \;.\label{eq:bessel1approx}
\end{align}
Combining all of the above and simplifying the result, we obtain the following linear response differential conductance due to a pure ac bias voltage:
\begin{align}
G_\text{ac}&=\frac{e^2\Gamma}{4\hbar\omega_0}\text{Im}\Bigg[\int\limits_{-\infty}^{\infty}\frac{\mathrm{d}\epsilon}{2\pi\hbar}\bigg(\frac{\epsilon\,n_F(\epsilon-\hbar\omega_0)}{\epsilon(\epsilon-i\Gamma)-B^2}+\frac{\epsilon\,n_F(\epsilon+\hbar\omega_0)}{\epsilon(\epsilon+i\Gamma)-B^2}-2\text{Re}\!\left[\frac{\epsilon\,n_F(\epsilon)}{\epsilon(\epsilon+i\Gamma)-B^2}\right]\bigg)\Bigg]\nonumber\\
&=\frac{e^2\Gamma}{4\hbar\omega_0}\int\limits_{-\infty}^{\infty}\frac{\mathrm{d}\epsilon}{2\pi\hbar}\big(n_F(\epsilon-\hbar\omega_0)-n_F(\epsilon+\hbar\omega_0)\big)\frac{\Gamma\epsilon^2}{\epsilon^4+\left(\Gamma^2-2B^2\right)\epsilon^2+B^4}\;.
\end{align}
The latter integral is of a very similar form to Eq.~(\ref{eq:Icdc}) for the dc current. Applying the same techniques as before, we straightforwardly obtain an exact expression for the ac differential conductance,
\begin{equation}
G_\text{ac}=\frac{e^2}{4\pi\hbar^2\omega_0}\frac{1}{\sqrt{1-4\left(\frac{B}{\Gamma}\right)^2}}\left(\epsilon_+\text{Im}\!\left[\Psi\left(\frac{1}{2}+\frac{\epsilon_++i\hbar\omega_0}{2\pi k_BT}\right)\right]-\epsilon_-\text{Im}\!\left[\Psi\left(\frac{1}{2}+\frac{\epsilon_-+i\hbar\omega_0}{2\pi k_BT}\right)\right]\right) \;.
\end{equation}
Note that the dc limit $\omega_0\rightarrow 0$ of this expression indeed reproduces Eq.~(\ref{eq:GTs}). Moreover, as is shown in the right panel of Fig.~\ref{fig:accond}, the conductance depends on the frequency in a similar way to how it depends on temperature. A driving voltage therefore has a similar effect as thermal fluctuations in terms of setting the scale for the onset of Kondo correlations.

Finally, we again take the limit $T_K\rightarrow\infty$, and use the same identification of temperature scales as in the previous section. Using the series expansion  $\Psi(z)=\ln(z)+\mathcal{O}(1/z)$, we obtain the following exact expression for the linear response ac conductance along the FL crossover:
\begin{equation}
G_\text{ac}=\frac{e^2}{2h}\left(1-\frac{2\pi k_BT^*}{\hbar\omega_0}\text{Im}\!\left[\Psi\left(\frac{1}{2}+\frac{T^*}{ T}+\frac{i\hbar\omega_0}{2\pi k_BT}\right)\right]\right)\;.\label{eq:Gacinf}
\end{equation}
As before, provided $k_BT$, $eV$ and $\hbar \omega_0$ are all $\ll k_BT_K$, the above expressions give the exact FL crossover behavior in the physical C2CK model; they are essentially an exact closed-form evaluation of the known integral expressions from Ref.~\cite{sh1} (see again footnote~\ref{fn:knownresults}).

The framework developed in this section can be straightforwardly generalized to provide exact solutions for charge transport in other time-dependent situations. Although we focused here on the steady state, the same methodology can be applied to calculate transient behavior, including the relaxation to a new equilibrium steady state after, \textit{e.g.}, a quantum quench. For a sudden quench in the dc voltage, relaxation is found to occur on a timescale $\sim \hbar/\Gamma$.


\subsection{Linear response charge transport from the Kubo formula}\label{sec:charge}
In the previous subsections, we saw how electric transport in the C2CK model can be treated exactly using Keldysh techniques. Here we employ instead the Kubo formula for the linear susceptibility, Eq.~(\ref{eq:kubo}) of  Sec.~\ref{sec:linres}. Note that in this case, all necessary expectation values are taken in the absence of a potential gradient and will be evaluated at the non-interacting EK point. As a result, Wick's theorem can be applied to write all $2n$-point functions in terms of propagators. In the following, we exploit this to calculate the linear response susceptibility corresponding to the charge current along the FL crossover.

For four-point functions of Grassmann variables, Wick's theorem reads
\begin{equation}
\langle abcd\rangle_0=\langle ab\rangle_0\langle cd\rangle_0+\langle ad\rangle_0\langle bc\rangle_0-\langle ac\rangle_0\langle bd\rangle_0 \;,
\end{equation}
where $\langle\ldots\rangle_0$ again refers to the expectation value in absence of a bias. Referring back to Eqs.~(\ref{eq:Ic}) and (\ref{eq:corfun}), one can now immediately express the required four-point function as
\begin{align}
C_c^\tau(\tau>0)&=\frac{e^2g_\bot^2}{4\hbar^2L}\sum\limits_{\mathclap{k,k^\prime}}\bigg(\left\langle\big(\psi^\dagger_{sf,k}(\tau)-\psi_{sf,k}(\tau)\big)\big(d^\dagger(\tau)-d(\tau)\big)\right\rangle_0\left\langle\big(\psi^\dagger_{sf,k^\prime}(0)-\psi_{sf,k^\prime}(0)\big)\big(d^\dagger(0)-d(0)\big)\right\rangle_0\nonumber\\
&\quad\,+\left\langle\big(\psi^\dagger_{sf,k}(\tau)-\psi_{sf,k}(\tau)\big)\big(d^\dagger(0)-d(0)\big)\right\rangle_0\left\langle\big(d^\dagger(\tau)-d(\tau)\big)\big(\psi^\dagger_{sf,k^\prime}(0)-\psi_{sf,k^\prime}(0)\big)\right\rangle_0\nonumber\\
&\quad\,-\left\langle\big(\psi^\dagger_{sf,k}(\tau)-\psi_{sf,k}(\tau)\big)\big(\psi^\dagger_{sf,k^\prime}(0)-\psi_{sf,k^\prime}(0)\big)\right\rangle_0\Big\langle\big(d^\dagger(\tau)-d(\tau)\big)\big(d^\dagger(0)-d(0)\big)\Big\rangle_0\bigg) \;. \label{eq:Cc}
\end{align}
Here, the first term is equal to $-\big\langle\hat{I}_c(\tau)\big\rangle_0\big\langle\hat{I}_c(0)\big\rangle_0=0$, which vanishes because no current flows in the absence of a potential. It corresponds to bubble diagrams that cancel by expanding the partition function, see Appendix~\ref{ap:kubo}. All remaining propagators from the above expression are of the form
\begin{align}
\big\langle\!\left(\alpha^\dagger(\tau)-\alpha(\tau)\right)\left(\beta^\dagger(\tau^\prime)-\beta(\tau^\prime)\right)\!\big\rangle_0&=\big\langle \alpha^\dagger(\tau)\beta^\dagger(\tau^\prime)\big\rangle_0+\big\langle \alpha(\tau)\beta(\tau^\prime)\big\rangle_0-\big\langle \alpha(\tau)\beta^\dagger(\tau^\prime)\big\rangle_0-\big\langle \alpha^\dagger(\tau)\beta(\tau^\prime)\big\rangle_0\nonumber\\
&=-G_{\alpha\beta,21}(\tau-\tau^\prime)-G_{\alpha\beta,12}(\tau-\tau^\prime)+G_{\alpha\beta,11}(\tau-\tau^\prime)+G_{\alpha\beta,22}(\tau-\tau^\prime)\nonumber\\
&\equiv\sum_{\mu\nu}^\prime G_{\alpha\beta,\mu\nu}(\tau-\tau^\prime) \;,
\end{align}
where $\mu,\nu$ denote the components in the Nambu basis, while the prime signifies a signed sum over the components. Inserting this into Eq.~(\ref{eq:Cc}), noting that the bubble diagrams vanish, and writing the Green functions in terms of Matsubara frequencies, we obtain
\begin{align}
C_c^\tau(\tau>0)&=-\frac{e^2g_\bot^2}{4\hbar^2L}\sum\limits_{\mathclap{k,k^\prime}}\sum_{\mu\nu}^\prime\sum_{\rho\sigma}^\prime\big(G_{ld,k,\mu\nu}(\tau)G_{ld,k^\prime,\rho\sigma}(-\tau)+G_{ll,kk^\prime,\mu\nu}(\tau)G_{dd,\rho\sigma}(\tau)\big)\nonumber\\
&=-\frac{e^2g_\bot^2}{4\hbar^2L}\sum\limits_{\mathclap{k,k^\prime}}\sum_{\mu\nu}^\prime\sum_{\rho\sigma}^\prime\frac{1}{(\hbar\beta)^2}\sum\limits_{\mathclap{n,n^\prime=-\infty}}^\infty\Big(G_{ld,k,\mu\nu}(i\omega_n)G_{ld,k^\prime,\rho\sigma}(i\omega_{n^\prime})e^{-i(\omega_n-\omega_{n^\prime})\tau}\nonumber\\
&\quad\,+G_{ll,kk^\prime,\mu\nu}(i\omega_n)G_{dd,\rho\sigma}(i\omega_{n^\prime})e^{-i(\omega_n+\omega_{n^\prime})\tau}\Big) \;.\label{eq:Ctmat}
\end{align}
Transforming the entire expression to Matsubara frequencies, this becomes
\begin{align}
C_c^\tau(i\Omega_n)&=-\frac{e^2g_\bot^2}{4\hbar^2L}\frac{1}{\hbar\beta}\sum\limits_{\mathclap{k,k^\prime}}\sum_{\mu\nu}^\prime\sum_{\rho\sigma}^\prime\sum\limits_{n^\prime=-\infty}^\infty
\bigg(G_{ld,k,\mu\nu}(i\omega_{n^\prime})G_{ld,k^\prime,\rho\sigma}(i\omega_{n^\prime-n})+G_{ll,kk^\prime,\mu\nu}(i\omega_{n^\prime})G_{dd,\rho\sigma}(-i\omega_{n^\prime-n})\bigg)\;,
\end{align}
where it should be noted that $\omega_{n^\prime-n}=\omega_{n^\prime}-\Omega_n$. For analytic continuation to real frequencies,  $C_c^\tau(i\Omega_{n>0})\rightarrow C_c^\text{R}(\omega)$, the positive Matsubara frequencies are sufficient, and so we will restrict ourselves to $n>0$ from now on. Next, we use the expressions for the Green functions derived in Sec.~\ref{sec:propagators}. Performing the matrix multiplications, the required Green functions are given by
\begin{align}
\sum_{\mu\nu}^\prime G_{dd,\mu\nu}(i\omega_n)&=2D_{bb}(i\omega_n)\;,\qquad
\sum_{\mu\nu}^\prime G_{ld,k,\mu\nu}(i\omega_n)=\frac{4g_\bot}{\sqrt{L}}\frac{\epsilon_k}{(\hbar\omega_n)^2+\epsilon_k^2}D_{bb}(i\omega_n) \;,\nonumber\\
\sum_{\mu\nu}^\prime G_{ll,kk^\prime,\mu\nu}(i\omega_n)&=-2i\hbar\,\delta_{k,k^\prime}\frac{\hbar\omega_n}{(\hbar\omega_n)^2+\epsilon_k^2}+\frac{8g_\bot^2}{L}\frac{\epsilon_k}{(\hbar\omega_n)^2+\epsilon_k^2}\frac{\epsilon_{k^\prime}}{(\hbar\omega_n)^2+\epsilon_{k^\prime}^2}D_{bb}(i\omega_n) \;.\label{eq:signedsums}
\end{align}
Taking the continuum limit for the sums over $k,k^\prime$ and using the fact that all terms that are odd in either $\epsilon_k$ or $\epsilon_{k^\prime}$ vanish upon integration, the four-point function simplifies to
\begin{equation}
C_c^\tau(i\Omega_{n>0})=\frac{ie^2\Gamma}{2}\frac{1}{\hbar\beta}\sum\limits_{n^\prime=-\infty}^\infty\int\limits_{-\infty}^\infty\frac{\mathrm{d}\epsilon_k}{2\pi\hbar}\frac{\hbar\omega_{n^\prime}}{(\hbar\omega_{n^\prime})^2+\epsilon_k^2}D_{bb}(-i\omega_{n^\prime-n}) \;.\label{eq:Cchargeloop}
\end{equation}
We note that the above autocorrelator can be interpreted as a one-loop bubble diagram, with one half of the loop corresponding to a Majorana component of $\mathbf{L}_{0,k}(i\omega_{n^\prime})$, and the other half to $D_{bb}(-i\omega_{n^\prime-n})$. 

Let us now consider the remaining sum and integral. Evaluating the integral\footnote{This result for the integral assumes that $\omega_{n^\prime}$ remains finite, which is no longer true when considering the full sum. The actual expression involves $\arctan(\Lambda/\hbar\omega_{n^\prime})$, where $\Lambda$ is the energy bandwidth (which is usually taken to infinity whenever possible), effectively introducing a cut-off $N$ in the sum over $n^\prime$. Although the naive introduction of a hard cut-off $N$ does lead to errors in the expression for the current autocorrelator $C_c^\tau(i\Omega_{n>0})$, the desired dc limit of the linear susceptibility is still exact due to the fact that the erroneous region $\hbar\omega_{n^\prime}\sim \Lambda$ does not contribute to the linear order term in $n$. The latter follows from the fact that the autocorrelator only contains the combination $D_{bb}(-i\omega_{n^\prime-n})-D_{bb}(i\omega_{n^\prime+n})$: for terms in the region $\hbar\omega_{n^\prime}\sim \Lambda\rightarrow\infty$ ({\it i.e.}, $n^\prime\gg n$), this combination is both analytic and even in $n$, see Eq.~(\ref{eq:eint}). The errors introduced by writing $\arctan(\Lambda/\hbar\omega_{n^\prime})\rightarrow \text{sgn}(\omega_{n^\prime})\,\pi/2$ therefore only depend on even powers of $n$.\label{fn:limits}}
over $\epsilon_k$,
\begin{align}
C_c^\tau(i\Omega_{n>0})&=\frac{ie^2\Gamma}{4\hbar^2\beta}\sum\limits_{\mathclap{n^\prime=-\infty}}^\infty\text{sgn}(\omega_{n^\prime})D_{bb}(-i\omega_{n^\prime-n})\nonumber\\
&=\frac{ie^2\Gamma}{4\hbar^2\beta}\sum\limits_{\mathclap{n^\prime=0}}^\infty\big(D_{bb}(-i\omega_{n^\prime-n})-D_{bb}(i\omega_{n^\prime+n})\big)\;, \label{eq:Ccmat}
\end{align}
where we used the definition of the fermionic Matsubara frequencies to rewrite the sum over the negative frequencies as a sum over positive ones. To make further progress we require an explicit expression for the $bb$ component of the dot Green function. According to Eqs.~(\ref{eq:Dw}) and (\ref{eq:Dbbe}), this component is given by
\begin{equation}
D_{bb}(i\omega_n)=\frac{\hbar\Gamma}{\pi}\int\limits_{-\infty}^\infty\mathrm{d}\epsilon\,\frac{\epsilon^2}{\epsilon^4+(\Gamma^2-2B^2)\epsilon^2+B^4}\frac{1}{i\hbar\omega_n-\epsilon} \;.
\end{equation}
We evaluate this integral using contour integration. If $\omega_n>0$, we choose a semicircle in the negative imaginary plane to close the contour. Doing so, and assuming that $4B^2<\Gamma^2$, we see that the contour integral is essentially the same as in Sec.~\ref{sec:dc}, with the poles being located at $\epsilon=-i\epsilon_\pm$. Meanwhile, if $\omega_n<0$, we choose to close the contour in the positive imaginary plane, and the poles are located at $\epsilon=i\epsilon_\pm$. The corresponding residue also picks up an additional minus sign, that is again cancelled by taking into account the change in integration direction. Using our results from Sec.~\ref{sec:dc}, we find,
\begin{equation}
D_{bb}(i\omega_n)=-\frac{i\hbar}{\Gamma}\frac{1}{\sqrt{1-4\left(\frac{B}{\Gamma}\right)^2}}\left(\frac{\epsilon_+}{\hbar\omega_n+\text{sgn}(\omega_n)\epsilon_+}-\frac{\epsilon_-}{\hbar\omega_n+\text{sgn}(\omega_n)\epsilon_-}\right) \;.\label{eq:eint}
\end{equation}
Inserting this result into Eq.~(\ref{eq:Ccmat}) we obtain,
\begin{align}
C_c^\tau(i\Omega_{n>0})&=-\frac{e^2}{4\hbar\beta}\frac{1}{\sqrt{1-4\left(\frac{B}{\Gamma}\right)^2}}\sum\limits_{\mathclap{\alpha=\pm 1}}\alpha\epsilon_\alpha\sum\limits_{n^\prime=0}^\infty\left(\frac{1}{\hbar\omega_{n^\prime-n}+\text{sgn}(\omega_{n^\prime-n})\epsilon_\alpha}+\frac{1}{\hbar\omega_{n^\prime+n}+\epsilon_\alpha}\right)\nonumber\\
&=-\frac{e^2}{4\hbar\beta}\frac{1}{\sqrt{1-4\left(\frac{B}{\Gamma}\right)^2}}\sum\limits_{\mathclap{\alpha=\pm 1}}\alpha\frac{\beta\epsilon_\alpha}{2\pi}\sum\limits_{n^\prime=0}^\infty\left(\frac{1}{n^\prime-n+\frac{1}{2}+\text{sgn}\!\left(n^\prime-n+\frac{1}{2}\right)\frac{\beta\epsilon_\alpha}{2\pi}}+\frac{1}{n^\prime+n+\frac{1}{2}+\frac{\beta\epsilon_\alpha}{2\pi}}\right)\nonumber\\
&=-\frac{e^2}{2\hbar\beta}\frac{1}{\sqrt{1-4\left(\frac{B}{\Gamma}\right)^2}}\sum\limits_{\mathclap{\alpha=\pm 1}}\alpha\frac{\beta\epsilon_\alpha}{2\pi}\left(\sum\limits_{n^\prime=0}^\infty\frac{1}{n^\prime+\frac{1}{2}+\frac{\beta\epsilon_\alpha}{2\pi}}-\sum\limits_{n^\prime=0}^{n-1}\frac{1}{n^\prime+\frac{1}{2}+\frac{\beta\epsilon_\alpha}{2\pi}}\right)\;.\label{eq:Ccsplit}
\end{align}
The first sum of the final equality diverges, being proportional to $\ln(\Lambda)$, where $\Lambda$ is the energy bandwidth. This term is a constant independent of the external Matsubara frequency $\Omega_n$, such that it does not contribute to the linear susceptibility after performing analytic continuation. Working out the second sum (see Appendix~\ref{ap:digamma} for more details):
\begin{equation}
C_c^\tau(i\Omega_{n>0})=\text{const.}+\frac{e^2}{2\hbar\beta}\frac{1}{\sqrt{1-4\left(\frac{B}{\Gamma}\right)^2}}\sum\limits_{\mathclap{\alpha=\pm 1}}\alpha\frac{\beta\epsilon_\alpha}{2\pi}\Psi\left(\frac{1}{2}+\frac{\beta\epsilon_\alpha}{2\pi}+n\right)\;.
\end{equation}
Finally, we perform the analytic continuation to real frequencies to find
\begin{equation}
C_c^\text{R}(\omega)=\text{const.}+\frac{e^2}{2\hbar\beta}\frac{1}{\sqrt{1-4\left(\frac{B}{\Gamma}\right)^2}}\sum\limits_{\mathclap{\alpha=\pm 1}}\alpha\frac{\beta\epsilon_\alpha}{2\pi}\Psi\left(\frac{1}{2}+\frac{\beta\epsilon_\alpha}{2\pi}-i\frac{\beta\hbar\omega}{2\pi}\right)\;.
\end{equation}

Returning to Eq.~(\ref{eq:kubo}) and taking the limit $\omega\rightarrow 0$, we find the dc susceptibility of the charge current:
\begin{equation}
\chi_{c,\text{dc}}=\frac{e^2}{2h}\frac{1}{\sqrt{1-4\left(\frac{B}{\Gamma}\right)^2}}\left(\frac{\beta\epsilon_+}{2\pi}\psi^{(1)}\left(\frac{1}{2}+\frac{\beta\epsilon_+}{2\pi}\right)-\frac{\beta\epsilon_-}{2\pi}\psi^{(1)}\left(\frac{1}{2}+\frac{\beta\epsilon_-}{2\pi}\right)\right)\;.\label{eq:chictou}
\end{equation}
This result is identical to Eq.~(\ref{eq:GTs}), confirming that the Kubo formula indeed gives the same results as Keldysh formalism in the zero-bias limit.

Now taking the limits $B^2\ll\Gamma^2$ (such that we can identify $\epsilon_+\simeq \Gamma$ as the Kondo temperature $2\pi k_BT_K$ and $\epsilon_-\simeq B^2/\Gamma$ as the FL crossover temperature $2\pi k_BT^*$) and $T\ll T_K$, we again recover the known charge conductance $G$ of the C2CK model, but now evaluated directly in linear response $V\to 0$,
\begin{equation}
G=\chi_{c,\text{dc}}=\frac{e^2}{2h}\left(1-\frac{T^*}{ T}\psi^{(1)}\left(\frac{1}{2}+\frac{T^*}{ T}\right)\right)\;.\label{eq:chargec}
\end{equation}
The first equality follows from the definition of $G$ from Eq.~(\ref{eq:diffcond}) combined with the linear response current $I_c=\chi_{c,\text{dc}}V$. The above is the expected behavior of the linear dc charge conductance along the FL crossover in the C2CK system.


\section{Exact results for heat transport}\label{sec:heat}
We now turn to heat transport. As explained in Sec.~\ref{sec:linres}, the methods employed in the previous section for the full non-equilibrium charge transport calculations at the EK point of the 2CK model cannot be used to find the heat conductance due to a temperature gradient between the leads. Therefore in this section, we restrict our attention to linear response theory. The method of calculation here proceeds in a similar fashion to that described in Sec.~\ref{sec:charge} for the charge transport using the Kubo formula.

Setting $\mu=0$ now for simplicity ({\it i.e.}, measuring all energies with respect to the chemical potential of the leads), the heat current operator is equal to the energy current operator from Eq.~(\ref{eq:IE}). The heat current operator is considerably more complicated than the charge current operator. We begin by decomposing it into five terms which we will treat separately. Specifically,  $\hat{I}_h=\sum_{i=1}^5\hat{I}_i$, with
\begin{align}
\hat{I}_1&=-\frac{\pi v_Fg_\bot}{\sqrt{2}L^{3/2}}\sum\limits_{\mathclap{k,k^\prime,k^{\prime\prime}}}\left(\psi_{c,k^\prime}^\dagger\psi_{c,k^{\prime\prime}}+\psi_{s,k^\prime}^\dagger\psi_{s,k^{\prime\prime}}\right)\left(\psi_{sf,k}^\dagger-\psi_{sf,k}\right)b\;,\label{eq:I1}\\
\hat{I}_2&=\frac{i\pi v_Fg_\bot}{\sqrt{2}L^{3/2}}\sum\limits_{\mathclap{k,k^\prime,k^{\prime\prime}}}\psi_{f,k^\prime}^\dagger\psi_{f,k^{\prime\prime}}\left(\psi_{sf,k}^\dagger+\psi_{sf,k}\right)a\;,\label{eq:I2}\\
\hat{I}_3&=\frac{i\pi v_Fg_\bot}{\left(2L\right)^{3/2}}\sum\limits_{\mathclap{k,k^\prime,k^{\prime\prime}}}\delta_{k^\prime,k^{\prime\prime}}\left(\psi_{sf,k}^\dagger+\psi_{sf,k}\right)a=\frac{i\Lambda g_\bot}{2^{3/2}\hbar\sqrt{L}}\sum\limits_{\mathclap{k}}\left(\psi_{sf,k}^\dagger+\psi_{sf,k}\right)a\;,\label{eq:I3}\\
\hat{I}_4&=\frac{\pi v_F}{2L}\sum\limits_{\mathclap{k,k^\prime}}\left(\epsilon_{k^\prime}-\epsilon_k\right)\psi_{f,k}^\dagger\psi_{f,k^\prime}ab\;,\label{eq:I4}\\
\hat{I}_5&=\frac{\pi v_F}{2L}\sum\limits_{\mathclap{k,k^\prime}}\left(\epsilon_{k^\prime}-\epsilon_k\right)\psi_{sf,k}^\dagger\psi_{sf,k^\prime}ab\;.\label{eq:I5}
\end{align}
Here, $a$ and $b$ again refer to the dot Majorana operators, and $\Lambda$ is the energy cut-off that is introduced when writing $\int_{-\infty}^\infty\mathrm{d}\epsilon_k\rightarrow\int_{-\Lambda}^\Lambda\mathrm{d}\epsilon_k$. Additionally, it is useful to decompose the current autocorrelator in a similar way:
\begin{equation}
C^\tau_h(\tau>0)=-\sum\limits_{\mathclap{i,j=1}}^5\big\langle\hat{I}_i(\tau)\hat{I}_j(0)\big\rangle_0\equiv\sum\limits_{\mathclap{i,j=1}}^5C_{ij}(\tau)\;.
\end{equation}
The main task of this section is thus the identification and subsequent evaluation of all non-zero components of $C_{ij}(\tau)$, most of which are complicated eight-point functions. The complexity of this task makes it more difficult to calculate the heat conductance along the FL crossover exactly. Instead, we will restrict ourselves to the NFL fixed point properties for all calculations involving heat transport. In the following, we therefore consider explicitly the channel-symmetric case with $B=0$, such that the FL scale $T^*=0$.

We start by identifying the vanishing components of $C_{ij}(\tau)$. The first useful observation is that the $\nu=c,s,f$ modes are all decoupled from the rest of the system in the absence of a potential gradient. As shown in Appendix~\ref{ap:bubble}, the bubble diagrams of the form $\sum_{k,k^\prime}\big\langle\psi^\dagger_{\nu,k}(\tau)\psi_{\nu,k^\prime}(\tau)\big\rangle_0$ ({\it i.e.}, the excitation densities) with $\nu\neq sf$ are therefore all equal to zero. Using Wick's theorem, this already eliminates $12$ of the $25$ components, namely $C_{1i}$ with $i\neq 1$, $C_{23}$, $C_{25}$, and their conjugates. Moreover, the flavor modes only contribute to the kinetic energy, such that fields with different momenta are uncorrelated. Therefore the correlator $\big\langle\psi^\dagger_{f,k}(\tau)\psi_{f,k^\prime}(\tau)\big\rangle_0$ is proportional to $\delta_{k,k^\prime}$, and the product of this correlator with $(\epsilon_{k^\prime}-\epsilon_k)$ also vanishes. This eliminates the components $C_{34}$ and $C_{45}$ as well as their conjugates. Finally, in absence of a magnetic field the combination $(C_{35}+C_{53})$ vanishes as a consequence of the fact that they contain bubble diagrams. This is somewhat subtle, as explained in Appendix~\ref{ap:bubble}.

This leaves the diagonal components $C_{ii}$ and the combination $(C_{24}+C_{42})$. In fact, the only term that contributes to the heat conductance is $C_{11}$. The other terms are finite but at least \textit{quadratic} in frequency, and therefore do not survive the $\omega\to 0$ dc limit of Eq.~(\ref{eq:kubo}). Extensive details of these explicit calculations are given in Appendix~\ref{ap:heatcor}.
Here we focus on the single surviving component that gives a finite contribution to the linear response heat transport.

    Exploiting the fact that the charge and spin modes are decoupled from the spin-flavor modes and the dot, the $ C_{11}(\tau)$ component can be written as
    \begin{align}
    C_{11}(\tau)&=\frac{(\pi v_Fg_\bot)^2}{4L^3}\sum\limits_{\mathclap{\substack{k,k^\prime,k^{\prime\prime}\\q,q^\prime,q^{\prime\prime}}}}\left\langle\big(\psi_{c,k^\prime}^\dagger(\tau)\psi_{c,k^{\prime\prime}}(\tau)+\psi_{s,k^\prime}^\dagger(\tau)\psi_{s,k^{\prime\prime}}(\tau)\big)\big(\psi_{c,q^\prime}^\dagger(0)\psi_{c,q^{\prime\prime}}(0)+\psi_{s,q^\prime}^\dagger(0)\psi_{s,q^{\prime\prime}}(0)\big)\right\rangle_0\nonumber\\
    &\quad\,\times\left\langle\big(\psi^\dagger_{sf,k}(\tau)-\psi_{sf,k}(\tau)\big)\big(d^\dagger(\tau)-d(\tau)\big)\big(\psi^\dagger_{sf,q}(0)-\psi_{sf,q}(0)\big)\big(d^\dagger(0)-d(0)\big)\right\rangle_0 \;.
    \end{align}
    To simplify the first line, we refer to the previous observation that the excitation densities corresponding to both the charge modes and the spin modes are equal to zero. As a result, the cross terms do not contribute. Meanwhile, the second line is identical to the charge autocorrelator (up to a constant prefactor) that was evaluated in Sec.~\ref{sec:charge}. Simplifying the first line and applying the result from Eq.~(\ref{eq:Ctmat}) to the second line, we find
    \begin{align}
    C_{11}(\tau)&=-\frac{(\pi v_Fg_\bot)^2}{4L^3}\sum\limits_{\mathclap{\substack{k,k^\prime,k^{\prime\prime}\\q,q^\prime,q^{\prime\prime}}}}\big(G_{cc,k^\prime q^{\prime\prime},22}(\tau)G_{cc,k^{\prime\prime}q^\prime,11}(\tau)+G_{ss,k^\prime q^{\prime\prime},22}(\tau)G_{ss,k^{\prime\prime}q^\prime,11}(\tau)\big)\nonumber\\
    &\quad\,\times\sum_{\mu\nu}^\prime\sum_{\rho\sigma}^\prime\big(G_{ld,k,\mu\nu}(\tau)G_{ld,q,\rho\sigma}(-\tau)+G_{ll,kq,\mu\nu}(\tau)G_{dd,\rho\sigma}(\tau)\big) \;.
    \end{align}
    From Eq.~(\ref{eq:signedsums}), it follows that the first term of the second line is odd in both $k$ and $q$, and therefore vanishes upon summation over these momenta. Transformed to Matsubara frequencies, the above thus becomes
    \begin{align}
    C_{11}(i\Omega_n)&=-\frac{(\pi v_Fg_\bot)^2}{4L^3}\frac{1}{(\hbar\beta)^3}\sum\limits_{\substack{k,k^\prime,k^{\prime\prime}\\q,q^\prime,q^{\prime\prime}}}\sum_{\mu\nu}^\prime\sum_{\rho\sigma}^\prime\sum\limits_{n^\prime,n^{\prime\prime},n^{\prime\prime\prime}}G_{ll,kq,\mu\nu}\big(-i(\omega_{n^\prime}+\omega_{n^{\prime\prime}}+\omega_{n^{\prime\prime\prime}}-\Omega_n)\big)G_{dd,\rho\sigma}(i\omega_{n^{\prime\prime\prime}})\nonumber\\
    &\quad\,\times\bigg(G_{cc,k^\prime q^{\prime\prime},22}(i\omega_{n^\prime})G_{cc,k^{\prime\prime}q^\prime,11}(i\omega_{n^{\prime\prime}})+G_{ss,k^\prime q^{\prime\prime},22}(i\omega_{n^\prime})G_{ss,k^{\prime\prime}q^\prime,11}(i\omega_{n^{\prime\prime}})\bigg)\;,
    \end{align}
    where the sums over $n^\prime,n^{\prime\prime},n^{\prime\prime\prime}$ all run over $\mathbb{Z}$. Since the charge and spin modes are completely decoupled, the corresponding Green functions satisfy $\mathbf{G}_{cc,kk^\prime}(i\omega_n)=\mathbf{G}_{ss,kk^\prime}(i\omega_n)=\delta_{k,k^\prime}\mathbf{L}_{0,k}(i\omega_n)$, see Eq.~(\ref{eq:L0k}). Plugging in the expressions from Eq.~(\ref{eq:signedsums}), omitting the terms that are odd in any of the momenta and relabelling the remaining momenta:
    \begin{align}
    C_{11}(i\Omega_n)&=-\frac{2(\pi v_Fg_\bot)^2}{(L\hbar\beta)^3}\sum\limits_{k,k^\prime,k^{\prime\prime}}\sum\limits_{n^\prime,n^{\prime\prime},n^{\prime\prime\prime}}\frac{\hbar}{i\hbar\omega_{n^\prime}+\epsilon_k}\frac{\hbar}{i\hbar\omega_{n^{\prime\prime}}-\epsilon_{k^\prime}}\frac{i\hbar^2(\omega_{n^\prime}+\omega_{n^{\prime\prime}}+\omega_{n^{\prime\prime\prime}}-\Omega_n)}{\hbar^2(\omega_{n^\prime}+\omega_{n^{\prime\prime}}+\omega_{n^{\prime\prime\prime}}-\Omega_n)^2+\epsilon_{k^{\prime\prime}}^2}D_{bb}(i\omega_{n^{\prime\prime\prime}})\nonumber\\
    &=\frac{2(\pi v_Fg_\bot)^2}{(L\beta)^3}\sum\limits_{k,k^\prime,k^{\prime\prime}}\sum\limits_{n^\prime,n^{\prime\prime},n^{\prime\prime\prime}}\frac{1}{i\hbar\omega_{n^\prime}-\epsilon_k}\frac{1}{i\hbar\omega_{n^{\prime\prime}}-\epsilon_{k^\prime}}\frac{1}{i\hbar(\omega_{n^\prime}+\omega_{n^{\prime\prime}}+\omega_{n^{\prime\prime\prime}}-\Omega_n)-\epsilon_{k^{\prime\prime}}}D_{bb}(i\omega_{n^{\prime\prime\prime}})\;.
    \end{align}

    Having found an explicit formula for the three-loop diagram $C_{11}(i\Omega_n)$, we continue by evaluating two of the Matsubara sums. Using the Matsubara representation of the Fermi-Dirac distribution from Eq.~(\ref{eq:matsrep}), a simple partial fraction decomposition leads to the following identity:
    \begin{equation}
    \frac{1}{\beta}\sum\limits_{\mathclap{n=-\infty}}^\infty\frac{1}{i\hbar\omega_n-\epsilon}\frac{1}{i\hbar\omega_n-\epsilon^\prime}=\frac{n_F(\epsilon)-n_F(\epsilon^\prime)}{\epsilon-\epsilon^\prime}\;.\label{eq:matsum}
    \end{equation}
    Furthermore, it is straightforward to show that $n_F(\epsilon-i\hbar\Omega_n)=n_F(\epsilon)$ and $n_F(\epsilon-i\hbar\omega_n)=-n_B(\epsilon)$ for bosonic and fermionic Matsubara frequencies, respectively ($n_B(\epsilon)$ is the Bose-Einstein distribution). Applying Eq.~(\ref{eq:matsum}) twice and taking the continuum limit for all momentum sums, we obtain
    \begin{equation}
    C_{11}(i\Omega_n)=\frac{\Gamma}{8\pi\hbar^2\beta}\int\limits_{-\infty}^\infty\mathrm{d}\epsilon_k\int\limits_{-\infty}^\infty\mathrm{d}\epsilon_{k^\prime}\int\limits_{-\infty}^\infty\mathrm{d}\epsilon_{k^{\prime\prime}}\sum\limits_{\mathclap{n^\prime=-\infty}}^\infty\frac{\big(n_F(\epsilon_{k^\prime})-n_F(\epsilon_{k^{\prime\prime}})\big)\big(n_F(\epsilon_k)+n_B(\epsilon_{k^{\prime\prime}}-\epsilon_{k^\prime})\big)}{i\hbar\omega_{n^\prime-n}-(\epsilon_{k^{\prime\prime}}-\epsilon_k-\epsilon_{k^\prime})}D_{bb}(i\omega_{n^\prime})\;.
    \end{equation}
    Also switching to new variables $\epsilon\equiv(\epsilon_k+\epsilon_{k^\prime}-\epsilon_{k^{\prime\prime}})/2$, $\epsilon^\prime\equiv(\epsilon_k-\epsilon_{k^\prime}-\epsilon_{k^{\prime\prime}})/2$, $\epsilon^{\prime\prime}\equiv\epsilon_k+\epsilon_{k^\prime}+\epsilon_{k^{\prime\prime}}$:
    \begin{align}
    C_{11}(i\Omega_n)&=\frac{\Gamma}{8\pi\hbar^2\beta}\int\limits_{-\infty}^\infty\mathrm{d}\epsilon\int\limits_{-\infty}^\infty\mathrm{d}\epsilon^\prime\int\limits_{-\infty}^\infty\mathrm{d}\epsilon^{\prime\prime}\sum\limits_{\mathclap{n^\prime=-\infty}}^\infty\nonumber\\
    &\quad\,\frac{\big(n_F(\epsilon-\epsilon^\prime)-n_F(-\epsilon+\epsilon^{\prime\prime}/2)\big)\big(n_F(\epsilon^\prime+\epsilon^{\prime\prime}/2)+n_B(-2\epsilon+\epsilon^\prime+\epsilon^{\prime\prime}/2)\big)}{i\hbar\omega_{n^\prime-n}+2\epsilon}D_{bb}(i\omega_{n^\prime})\nonumber\\
    &=\frac{\Gamma}{4\pi\hbar^2\beta}\int\limits_{-\infty}^\infty\mathrm{d}\epsilon\int\limits_{-\infty}^\infty\mathrm{d}\epsilon^\prime\sum\limits_{\mathclap{n^\prime=-\infty}}^\infty\frac{(\epsilon+\epsilon^\prime)\cosh(\beta\epsilon)}{\sinh(\beta\epsilon)+\sinh(\beta\epsilon^\prime)}\frac{1}{i\hbar\omega_{n^\prime-n}+2\epsilon}D_{bb}(i\omega_{n^\prime})\nonumber\\
    &=\frac{\Gamma}{4\pi\hbar^2\beta}\int\limits_{-\infty}^\infty\mathrm{d}\epsilon\sum\limits_{\mathclap{n^\prime=-\infty}}^\infty\left(\frac{\pi^2}{2\beta^2}+2\epsilon^2\right)\frac{1}{i\hbar\omega_{n^\prime-n}+2\epsilon}D_{bb}(i\omega_{n^\prime})\nonumber\\
    &\rightarrow-\frac{\Gamma}{4\pi\hbar\beta}\int\limits_{-\Lambda^\prime}^{\Lambda^\prime}\mathrm{d}\epsilon\sum\limits_{\mathclap{n^\prime=-\infty}}^\infty\left(\frac{\pi^2}{2\beta^2}+2\epsilon^2\right)\frac{\hbar\omega_{n^\prime-n}}{(\hbar\omega_{n^\prime-n})^2+(2\epsilon)^2}\frac{1}{\hbar\omega_{n^\prime}+\text{sgn}(\omega_{n^\prime})\Gamma}\;,\label{eq:C11ex}
    \end{align}
    where $\Lambda^\prime=3\Lambda/2$ is the cut-off of the redefined variable $\epsilon$, and we used Eq.~(\ref{eq:eint}) with $B=0$ for the dot Green function. Moreover, we write out the Matsubara frequencies explicitly, perform the final integral, and take the limit $\Lambda^\prime\rightarrow\infty$ (see again footnote~\ref{fn:limits}) to find
    \begin{equation}
    C_{11}(i\Omega_n)=-\frac{\Gamma}{16\pi\hbar\beta^2}\sum\limits_{\mathclap{n^\prime=-\infty}}^\infty\frac{\pi^2\text{sgn}\!\left(n^\prime-n+\frac{1}{2}\right)\left(\frac{1}{2}-2\left(n^\prime-n+\frac{1}{2}\right)^2\right)+4\beta\Lambda^\prime\left(n^\prime-n+\frac{1}{2}\right)}{n^\prime+\frac{1}{2}+\text{sgn}\!\left(n^\prime+\frac{1}{2}\right)\frac{\beta\Gamma}{2\pi}} \;.
    \end{equation}

    We would like to calculate the linear susceptibility by expanding this current-current correlation function in $n$ and extracting the linear part. However, the above expression is not analytic due to the sign functions, and so we split the sum into different parts, each of which is analytic. Again restricting ourselves to $n>0$, the three different parts are: (i) $n^\prime<0$, with both sign functions equal to $-1$; (ii) $0\leq n^\prime<n$, where one of the sign functions is $-1$ while the other is $+1$; (iii) $n^\prime\geq n$, with both sign functions equal to $+1$. Writing $n^\prime\rightarrow-n^\prime-1$ in the first part, using $\sum_{n^\prime=n}^\infty=\sum_{n^\prime=0}^\infty-\sum_{n^\prime=0}^{n-1}$ in the third part, and subsequently combining the parts that sum over $n^\prime\in\{0,\ldots,n-1\}$, we obtain the following analytic form:
    \begin{align}
    C_{11}(i\Omega_{n>0})&=-\frac{\Gamma}{16\pi\hbar\beta^2}\Bigg(-2\pi^2\sum\limits_{n^\prime=0}^{n-1}\frac{\frac{1}{2}-2\left(n^\prime-n+\frac{1}{2}\right)^2}{n^\prime+\frac{1}{2}+\frac{\beta\Gamma}{2\pi}}\nonumber\\
    &\quad\,+\sum\limits_{n^\prime=0}^\infty\frac{\pi^2\left(\frac{1}{2}-2\left(n^\prime+n+\frac{1}{2}\right)^2\right)+4\beta\Lambda^\prime\left(n^\prime+n+\frac{1}{2}\right)}{n^\prime+\frac{1}{2}+\frac{\beta\Gamma}{2\pi}}\nonumber\\
    &\quad\,+\sum\limits_{n^\prime=0}^\infty\frac{\pi^2\left(\frac{1}{2}-2\left(n^\prime-n+\frac{1}{2}\right)^2\right)+4\beta\Lambda^\prime\left(n^\prime-n+\frac{1}{2}\right)}{n^\prime+\frac{1}{2}+\frac{\beta\Gamma}{2\pi}}\Bigg) \;.
    \end{align}

    Similar to the charge transport case, the second and third lines of the above expression each diverge, being proportional to $\Lambda^2$. However, combining the terms gives a result that is either constant or quadratic in $n$. For the purpose of finding the linear susceptibility, the above autocorrelator therefore simplifies to
    \begin{equation}
     C_{11}(i\Omega_{n>0})=\text{const.}+\frac{\pi\Gamma}{8\hbar\beta^2}\sum\limits_{n^\prime=0}^{n-1}\frac{\frac{1}{2}-2\left(n^\prime-n+\frac{1}{2}\right)^2}{n^\prime+\frac{1}{2}+\frac{\beta\Gamma}{2\pi}}+\mathcal{O}(\Omega_n^2) \;.
    \end{equation}
    Finally evaluating the remaining sum, expanding the result to linear order in $n$ (see Appendix~\ref{ap:heatcor}), and performing analytic continuation to real frequencies, we find
    \begin{equation}
    C^\text{R}_{11}(\omega)=\text{const.}-\frac{i\Gamma}{16\hbar\beta}\left[\frac{\beta\Gamma}{\pi}+\left(\frac{1}{2}-\frac{\beta^2\Gamma^2}{2\pi^2}\right)\psi^{(1)}\left(\frac{1}{2}+\frac{\beta\Gamma}{2\pi}\right)\right]\hbar\omega+\mathcal{O}(\omega^2) \;.
    \end{equation}
    Furthermore, we may identify $\beta\Gamma$ as $T_K/T\rightarrow\infty$ (at the NFL fixed point) and utilize the expansion of the trigamma function,
    \begin{equation}
    \frac{1}{x}\psi^{(1)}\left(\frac{1}{2}+\frac{1}{x}\right)=1-\frac{x^2}{12}+\mathcal{O}\left(x^4\right) \;.\label{eq:trigammaexp}
    \end{equation}
    This gives our final result for the heat current autocorrelator at the NFL fixed point,
    \begin{equation}
    C^\text{R}_{11}(\omega)=\text{const.}-\frac{i\pi\omega}{12\beta^2}+\mathcal{O}(\omega^2) \;.\label{eq:C11Tsquare}
    \end{equation}

To summarize, only the component $C_{11}$ has a linear term in frequency at the NFL fixed point,
such that the full NFL heat current autocorrelator can be written as
\begin{equation}
C^\text{R}_h(\omega)=\text{const.}+C^\text{R}_{11}(\omega)+\mathcal{O}(\omega^2) \;.
\end{equation}
As such, from Eq.~(\ref{eq:kubo}) we finally obtain the following exact result for the NFL heat susceptibility,
\begin{equation}
\chi_{h,\text{dc}}=\frac{\pi^2k_B^2T^2}{6h} \;.
\end{equation}

Returning to the discussion from the final paragraph of Sec.~\ref{sec:linres}, we briefly consider the off-diagonal terms from Eq.~(\ref{eq:offdiag}), which involve propagators of the form $\big\langle\hat{I}_c(\tau)\hat{I}_E(0)\big\rangle_0$ and $\big\langle\hat{I}_E(\tau)\hat{I}_c(0)\big\rangle_0$. Referring back to Eqs.~(\ref{eq:Ic}) and (\ref{eq:I1})-(\ref{eq:I5}), we immediately see that any terms involving $\hat{I}_1$, $\hat{I}_2$ or $\hat{I}_4$ are proportional to vanishing bubble diagrams (Appendix~\ref{ap:bubble}). Moreover, the charge current operator does not contain the $a$ Majorana fermion, such that the products of $\hat{I}_c$ with either $\hat{I}_3$ or $\hat{I}_5$ contain exactly one $a$ operator. At the NFL fixed point, the $a$ Majorana fermion is completely decoupled from all other modes, and all terms involving $\hat{I}_3$ and $\hat{I}_5$ are therefore equal to zero as well. We thus conclude that the off-diagonal terms from Eq.~(\ref{eq:offdiag}) are equal to zero at the NFL fixed point, and as such the temperature gradient does not induce thermopower. Consequently, the two choices $V=0$ and $I_c=0$ coincide, such that the heat conductance $\kappa$ is unambiguously given by
\begin{equation}
\kappa=\frac{\chi_{h,\text{dc}}}{T}=\frac{\pi^2k_B^2T}{6h}\label{eq:heatc}
\end{equation}
at the NFL fixed point of the C2CK model. This novel result is the main result of this work.

Finally, we comment on the heat conductance at the FL fixed point due to a symmetry-breaking perturbation (either channel asymmetry or magnetic field). In this case, for $T\ll T^*$ one of the two leads flows under RG to strong coupling, while the other asymptotically decouples. One can then argue that the conductances between the leads must vanish at the FL fixed point. We saw this by explicit calculation in the case of the charge conductance in Sec.~\ref{sec:charge}.
The full FL crossover in the heat conductance, even within linear response, is a more challenging calculation which we do not attempt here.


\section{Wiedemann-Franz law and CFT central charge}\label{sec:wf}
In this section, we unpack some of the implications of our results for the charge conductance in Sec.~\ref{sec:charge} and the heat conductance in Sec.~\ref{sec:heat}.

\begin{figure}[t]
  \centerline{\begin{tabular}{cc}
    \includegraphics[height=5 cm]{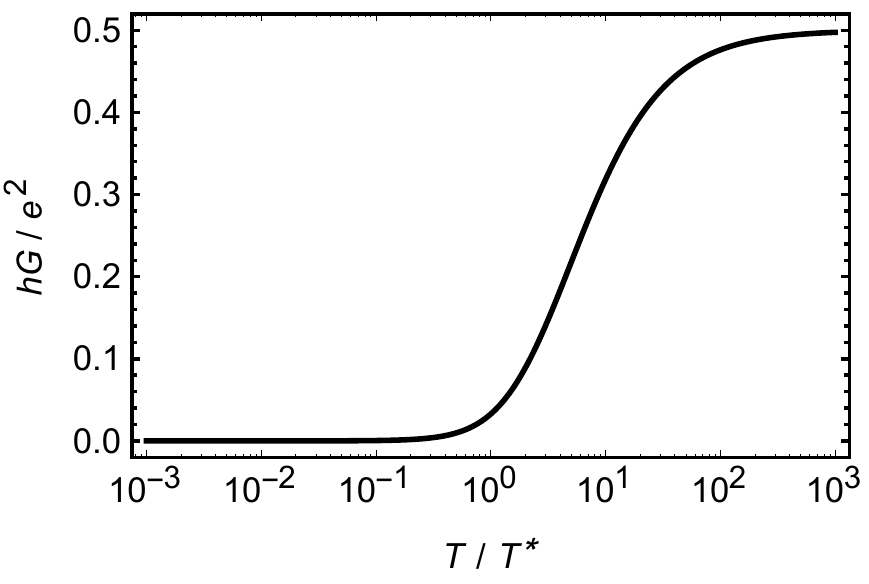} & \includegraphics[height=5 cm]{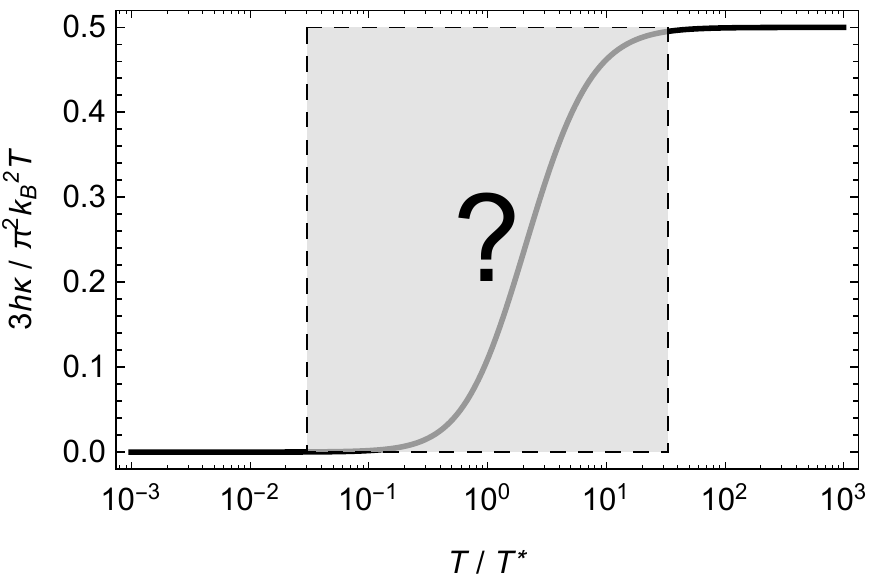}\end{tabular}}
\caption{\label{fig:cond}Linear response charge conductance $G$ and heat conductance $\kappa$ as a function of dimensionless temperature $T/T^*$, both in the limit $T_K\rightarrow\infty$. The full FL crossover region is not known exactly in the case of the heat conductance, although fixed point properties were obtained exactly.}
\end{figure}

In Fig.~\ref{fig:cond} we compare the behavior of the linear response charge and heat conductances along the FL crossover (assuming good scale separation $T^* \ll T_K$). The full FL crossover for the charge conductance (left panel) is given exactly by Eq.~(\ref{eq:chargec}). For the heat conductance (right panel), the NFL fixed point behavior is given exactly by Eq.~(\ref{eq:heatc}), while $\kappa/T=0$ at the FL fixed point due to the asymptotic decoupling of the leads for $T\ll T^*$. The intermediate crossover behavior of the heat conductance is presently unknown.

Note that the Kondo crossover from the local moment fixed point to the NFL fixed point on the scale of $T_K$ in the physical C2CK system cannot be described within this framework, since calculations are performed at the EK point (see Sec.~\ref{sec:lowtoulouse}). However, in Sec.~\ref{sec:pert} we access the incipient behavior near the NFL fixed point using perturbation theory around the EK solution, which gives corrections to our results in powers of $T/T_K$. These formally vanish at the NFL fixed point itself, and are negligible along the FL crossover given good scale separation $T^*\ll T_K$.

We now focus on the NFL fixed point linear susceptibilities, Eqs.~(\ref{eq:chargec}) and (\ref{eq:heatc}). In particular, we note that our results imply a non-trivial result for the dot central charge within the underlying conformal field theory (CFT) at the NFL fixed point. This follows from the fact that the heat current through a junction in a one-dimensional (1D) system is given by~\cite{centralcharge}
\begin{equation}
I_h=\frac{c\pi^2}{6h}k_B^2\left(T_l^2-T_r^2\right) \;,
\end{equation}
where $c$ is the CFT central charge of the degrees of freedom involved in the transport processes. Using the results from the previous sections, we can calculate this heat current for our system. In the FL regime there is no transport at all, such that $c$ trivially goes to zero. However, in the NFL region we find the following heat current for small $\Delta T/T$,
\begin{equation}
I_h=\kappa\Delta T=\frac{\pi^2}{6h}k_B^2T\Delta T=\frac{\pi^2}{12h}k_B^2\Delta(T^2)=\frac{\pi^2}{12h}k_B^2\left(T_l^2-T_r^2\right) \;.
\end{equation}
Our results thus imply a central charge of $c=1/2$, characteristic of the 1D Majorana fermions appearing in the tunneling term of the Hamiltonian, Eq.~(\ref{eq:Ham}). As the heat current is an observable quantity, this provides a way to experimentally verify the Majorana character of the dot in the NFL region.

Finally, let us consider the Lorenz ratio, defined as the ratio of the heat conductance to the charge conductance,\footnote{The Lorenz ratio specifically involves the heat conductance in absence of an electric current \cite{WiedemannFranz}; in terms of the conventions from Sec.~\ref{sec:linres}, the heat conductance should thus be read as $T\kappa=\chi_{22}-\chi_{12}\chi_{21}/\chi_{11}$. Although inconsequential at the NFL fixed point, in the general case it is therefore necessary to account for thermopower.} $L\equiv\kappa/TG$. The Wiedemann-Franz law states that this ratio reduces to a constant value $L_0=\pi^2k_B^2/3e^2$ in the case of ``normal" metals (mostly referring to Fermi liquids). For our setup, it immediately follows from Eqs.~(\ref{eq:chargec}) and (\ref{eq:heatc}) that the Wiedemann-Franz law is actually satisfied at the NFL fixed point, {\it i.e.}, $L_\text{NFL}=L_0$. It should be noted that both the charge and the heat conductance actually have an additional factor $1/2$ compared to most simple quantum dot setups, but these unconventional factors cancel in  $L_\text{NFL}$, leaving the ratio unchanged. In particular, note that this is the exact result at the NFL fixed point of the physical C2CK model. This is because both $\kappa$ and $G$ are finite at the NFL fixed point, to which the C2CK model flows for $T^* \ll T \ll T_K$. Corrections to the EK solution presented here may be obtained perturbatively around the EK point (this is done explicitly in Sec.~\ref{sec:pert}), with the additional terms controlled in powers of $T/T_K$. Therefore these corrections formally vanish at the NFL fixed point being considered here, and do not need to be considered in the calculation of $L_\text{NFL}$. Our result $L_\text{NFL}=L_0$ is therefore exact.\footnote{Certain other ratios of interest, such as the Wilson ratio which involves the ratio of the magnetic susceptibility to the specific heat, depend on quantities that are known to vanish at the NFL point. In such cases, one must compute the corrections around the EK point already to obtain the fixed point properties. By contrast, we emphasize again that this is \textit{not} required for the calculation of the NFL Lorenz ratio, since both charge and heat conductances remain finite at the NFL fixed point.}

Counter-intuitively, the Wiedemann-Franz law is expected to be violated in the FL regime of the C2CK model -- exactly opposite to naive expectation. This can be understood quantitatively by first expanding the conductances in terms of $T/T^*$. Doing so, the Lorenz ratio acquires the form $L=(\kappa_0+\kappa_2(T/T^*)^2+\ldots)/T(G_0+G_2(T/T^*)^2+\ldots)$, where $\kappa_{0,2}$ and $G_{0,2}$ are the first non-zero Taylor coefficients of the conductances.\footnote{The linear coefficient $G_1$ can explicitly be shown to vanish by expanding Eq.~(\ref{eq:chargec}), while $\kappa_1$ is expected to vanish in accordance with Fermi liquid theory.}
Since $\kappa_0=G_0=0$ at the FL fixed point, the Lorenz ratio has a well-controlled limit as $T\to 0$ of $L\to \kappa_2/T G_2$. As such, the Wiedemann-Franz law cannot be expected to hold despite the FL nature of the fixed point. Moreover, from a physical point of view the Lorenz ratio is expected to be enhanced, which can be understood by realizing that the FL ground state corresponds to a transmission node of the system~\cite{215}. Generally, the heat current (and by extension the entropy current) is less sensitive to transmission nodes than the charge current. This is due to the fact that the entropy current is inherently incoherent, while the charge current is not. Coherent currents are more easily blocked, and so the ratio of entropy (or heat) conductance to charge conductance is expected to be enhanced when approaching such a transmission node.

In conclusion, we have shown that the C2CK device at the NFL critical fixed point for $T^* \ll T \ll T_K$ is characterized by a CFT central charge $c=1/2$,  corresponding to a single Majorana fermion. But despite the unconventional Majorana degree of freedom on the dot that mediates quantum transport, the Wiedemann-Franz law is satisfied. Surprisingly, the Wiedemann-Franz law is expected to be violated instead at the FL fixed point for $T\ll T^*$, due to the transmission node in that limit. This elaborates on the results we presented in Ref.~\cite{dalum2020wf}.


\section{Perturbations away from the EK point}\label{sec:pert}
As discussed in Sec.~\ref{sec:lowtoulouse}, perturbation theory can be used to find the corrections to the NFL results in terms of a finite $T/T_K$. In this section, we will explicitly calculate the corrections to the linear response charge conductance away from the EK point to lowest order in $\lambda\equiv 2\pi\hbar v_F-J_z$ and $T/T_K$. Our starting point is the interaction term from Eq.~(\ref{eq:ham3}),
\begin{align}
\hat{H}_I&=\lambda:\psi_s^\dagger(0)\psi_s(0):\left(d^\dagger d-\frac{1}{2}\right)\nonumber\\
&=\frac{i\lambda}{L}ba\sum_{k,k^\prime}:\psi_{s,k}^\dagger\psi_{s,k^\prime}: \;,\label{eq:Hint}
\end{align}
which we treat as a perturbation to the non-interacting Hamiltonian from Eq.~(\ref{eq:Ham})~\cite{sh3}. To calculate the change in the linear susceptibility due to this interaction term, we must find the corrections to Eq.~(\ref{eq:Cchargeloop}), which should be understood as
\begin{align}
C_c^\tau(i\Omega_{n>0})=-\frac{e^2\Gamma}{8\pi\hbar^3\beta}\sum\limits_{n^\prime=-\infty}^\infty\int\limits_{-\infty}^\infty\mathrm{d}\epsilon_k\text{Tr}\!\left[\mathbf{L}_{0,k}(i\omega_{n^\prime})\right]D_{bb}(-i\omega_{n^\prime-n}) \;.\label{eq:corrD}
\end{align}
Since the interaction term does not involve spin-flavor modes, the bare propagators corresponding to those modes remain unchanged. Our first objective is thus to find the corrections to the $bb$ component of the dot Green function in presence of a non-zero $\lambda$. In doing so, we will set the magnetic field $B$ to zero ({\it i.e.}, setting $T^*=0$), restricting ourselves purely to the NFL regime and removing any effects from the FL regime in the process.

We approximate the full $bb$ component of the dot Green function $D_{bb}^\text{full}(i\omega_n)$ in presence of interactions by employing standard Feynman diagrammatic techniques. Utilizing the fact that the interaction Hamiltonian from Eq.~(\ref{eq:Hint}) provides a four-point vertex involving two $\psi_{s,k}$ legs, an $a$ leg and a $b$ leg, the Feynman rules lead to the following diagramatic expression for $D_{bb}^\text{full}(i\omega_n)$:
\begin{equation}
\parbox{20mm}{\begin{fmfgraph*}(50,60) \fmfleft{i} \fmfright{o} \fmf{double,label=$\omega_n$}{i,o}\end{fmfgraph*}}=\;\;\;\parbox{20mm}{\begin{fmfgraph*}(50,60) \fmfleft{i} \fmfright{o} \fmf{plain,label=$\omega_n$}{i,o} \end{fmfgraph*}}+\;\;\;\parbox{20mm}{\begin{fmfgraph*}(150,60) \fmfleft{i} \fmfright{o} \fmf{plain,label=$\omega_n$}{i,v1} \fmf{photon,label=$\omega_{n-l+m}$}{v1,v2} \fmf{dashes_arrow,left,label=$\omega_l$,tension=0}{v1,v2} \fmf{dashes_arrow,left,label=$\omega_m$,tension=0}{v2,v1}  \fmf{plain,label=$\omega_n$}{v2,o} \end{fmfgraph*}}\hspace{3.5cm}+\ldots\;.\label{eq:diagrams}
\end{equation}
Here, each vertex comes with a prefactor $i\lambda/\hbar^2\beta$ and a sum over Matsubara frequencies; the definitions of the other components can be found in Table~\ref{tab:feynman}. Explicitly, we find that the lowest order of the self-energy is given by,
\begin{equation}
\Sigma(i\omega_n)=-\frac{\lambda^2}{L^2\hbar^2}\frac{1}{(\hbar\beta)^2}\sum_{n^\prime,n^{\prime\prime}}\sum_{k,k^\prime}D_{aa}\big(-i(\omega_{n^\prime}-\omega_{n^{\prime\prime}}-\omega_n)\big)G_{s,k}(i\omega_{n^\prime})G_{s,k^\prime}(i\omega_{n^{\prime\prime}}) \;,\label{eq:selfenergyfull}
\end{equation}
where $G_{s,k}(i\omega_n)$ is shorthand notation for $G_{ss,kk,11}(i\omega_n)$. A more detailed derivation of this expression can be found in Appendix~\ref{ap:selfenergy}.
\begin{table}[t]
\centering
\begin{tabular}{|c|c||c|}
     \hline
     {Expression} & {\hspace{0.7cm}Diagram\hspace{0.7cm}} & Vertex \\
     \hline \hline
     {$D_{bb}^\text{full}(i\omega_n)$} & {\begin{fmfgraph*}(50,10) \fmfleft{i} \fmfright{o} \fmf{double,label=$\omega_n$}{i,o} \end{fmfgraph*}} & \\[0.2cm]
     \cline{1-2}
     {$D_{bb}(i\omega_n)$} & {\begin{fmfgraph*}(50,10) \fmfleft{i} \fmfright{o} \fmf{plain,label=$\omega_n$}{i,o} \end{fmfgraph*}} & \multirow{3}{*}{\begin{fmfgraph*}(50,40) \fmfleft{i1,i2} \fmfright{o1,o2} \fmf{dashes_arrow}{i1,v,o2} \fmf{plain}{i2,v} \fmf{photon}{v,o1} \end{fmfgraph*}}\\[0.2cm]
     \cline{1-2}
     {$D_{aa}(i\omega_n)$} & {\begin{fmfgraph*}(50,10) \fmfleft{i} \fmfright{o} \fmf{photon,label=$\omega_n$}{i,o} \end{fmfgraph*}} & \\[0.2cm]
     \cline{1-2}
     {$\frac{1}{L}\sum\limits_k G_{s,k}(i\omega_n)$} & {\begin{fmfgraph*}(50,10) \fmfleft{i} \fmfright{o} \fmf{dashes_arrow,label=$\omega_n$}{i,o} \end{fmfgraph*}} & \\[0.2cm]
     \hline
\end{tabular}
\caption{Definitions of the different components of the Feynman diagrams. The arrow in the fourth diagram indicates the propagation direction of $\psi_{s,k}$.}
\label{tab:feynman}
\end{table}

Using the fact that the $a$ and $\psi_{s,k}$ modes are completely isolated from the rest of the system if $B=0$ and $\lambda=0$, and taking the continuum limit of the $k,k^\prime$ sums, we have
\begin{equation}
\Sigma(i\omega_n)=-\frac{\lambda^2}{\hbar v_F^2}\frac{1}{(\hbar\beta)^2}\int\limits_{-\infty}^\infty\frac{\mathrm{d}\epsilon_k}{2\pi}\int\limits_{-\infty}^\infty\frac{\mathrm{d}\epsilon_{k^\prime}}{2\pi}\sum_{n^\prime,n^{\prime\prime}}\frac{1}{i\hbar(\omega_n-\omega_{n^\prime}+\omega_{n^{\prime\prime}})}\frac{1}{i\hbar\omega_{n^\prime}-\epsilon_k}\frac{1}{i\hbar\omega_{n^{\prime\prime}}-\epsilon_{k^\prime}} \;.
\end{equation}
Furthermore applying Eq.~(\ref{eq:matsum}) twice, together with the substitutions $\epsilon\equiv(\epsilon_k+\epsilon_{k^\prime})/2,\epsilon^\prime\equiv\epsilon_k-\epsilon_{k^\prime}$, the self-energy becomes
\begin{align}
\Sigma(i\omega_n)&=\frac{\lambda^2}{\hbar^3v_F^2}\int\limits_{-\infty}^\infty\frac{\mathrm{d}\epsilon_k}{2\pi}\int\limits_{-\infty}^\infty\frac{\mathrm{d}\epsilon_{k^\prime}}{2\pi}\frac{\big(n_F(0)-n_F(\epsilon_{k^\prime})\big)\big(n_F(\epsilon_k)+n_B(\epsilon_{k^\prime})\big)}{i\hbar\omega_n-(\epsilon_k-\epsilon_{k^\prime})}\nonumber\\
&\!\!\!\!\!\!\!\!\stackrel{\epsilon_{k^\prime}\rightarrow-\epsilon_{k^\prime}}{=}\frac{\lambda^2}{4\hbar^3v_F^2}\int\limits_{-\infty}^\infty\frac{\mathrm{d}\epsilon_k}{2\pi}\int\limits_{-\infty}^\infty\frac{\mathrm{d}\epsilon_{k^\prime}}{2\pi}\frac{\cosh\left(\beta(\epsilon_k+\epsilon_{k^\prime})/2\right)}{\cosh\left(\beta\epsilon_k/2\right)\cosh\left(\beta\epsilon_{k^\prime}/2\right)}\frac{1}{i\hbar\omega_n-(\epsilon_k+\epsilon_{k^\prime})}\nonumber\\
&=\frac{\lambda^2}{2\hbar^3v_F^2}\int\limits_{-\infty}^\infty\frac{\mathrm{d}\epsilon}{2\pi}\int\limits_{-\infty}^\infty\frac{\mathrm{d}\epsilon^\prime}{2\pi}\frac{\cosh\left(\beta\epsilon\right)}{\cosh\left(\beta\epsilon\right)+\cosh\left(\beta\epsilon^\prime/2\right)}\frac{1}{i\hbar\omega_n-2\epsilon}\nonumber\\
&=\frac{\lambda^2}{\pi\hbar^3v_F^2}\int\limits_{-\infty}^\infty\frac{\mathrm{d}\epsilon}{2\pi}\frac{\epsilon}{\tanh\left(\beta\epsilon\right)\left(i\hbar\omega_n-2\epsilon\right)} \;.
\end{align}
In order to deal with the remaining UV divergence, we again introduce the energy cut-off $\Lambda$. Noting that the real part of the integrand is odd in $\epsilon$, we obtain
\begin{equation}
\Sigma(i\omega_n)=-\frac{i\omega_n\lambda^2}{2\pi^2\hbar^2v_F^2}\int\limits_{-\Lambda}^\Lambda\mathrm{d}\epsilon\,\frac{\epsilon}{\tanh\left(\beta\epsilon\right)}\frac{1}{(\hbar\omega_n)^2+(2\epsilon)^2} \;,\label{eq:selfenergy}
\end{equation}
which diverges logarithmically as $\Lambda\rightarrow\infty$.

We now return to the correlation function from Eq.~(\ref{eq:corrD}), replacing $D_{bb}(i\omega_n)$ with $D_{bb}^\text{full}(i\omega_n)$ and evaluating the momentum integral. The methods of dealing with the momentum integrals are the same as in Sec.~\ref{sec:charge}, and the current-current correlator becomes
\begin{align}
C_c^\tau(i\Omega_{n>0})&=\frac{ie^2\Gamma}{4\hbar^2\beta}\sum\limits_{n^\prime=0}^\infty\left(D_{bb}^\text{full}(-i\omega_{n^\prime-n})-D_{bb}^\text{full}(i\omega_{n^\prime+n})\right)\nonumber\\
&=C_c^\tau(i\Omega_{n>0})\Big|_{\lambda=0}-\frac{ie^2\Gamma}{4\hbar^2\beta}\sum\limits_{n^\prime=0}^\infty\left(\left(D_{bb}(i\omega_{n^\prime-n})\right)^2\Sigma(i\omega_{n^\prime-n})+\left(D_{bb}(i\omega_{n^\prime+n})\right)^2\Sigma(i\omega_{n^\prime+n})\right)+\mathcal{O}(\lambda^4)\;,
\end{align}
where we used the fact that both $D_{bb}(i\omega_n)$ and $\Sigma(i\omega_n)$ are odd functions of $\omega_n$. Reading off $D_{bb}(i\omega_n)$ from Eq.~(\ref{eq:eint}) (taking the $B\rightarrow 0$ limit) and splitting the sum in the same way as in Eq.~(\ref{eq:Ccsplit}), the lowest order correction to the current-current correlator can be written as
\begin{align}
\Delta C_c^\tau(i\Omega_{n>0})&=\frac{ie^2\Gamma}{2\beta}\left(\sum\limits_{n^\prime=0}^\infty\frac{\Sigma(i\omega_{n^\prime})}{\left(\hbar\omega_{n^\prime}+\Gamma\right)^2}-\sum\limits_{n^\prime=0}^{n-1}\frac{\Sigma(i\omega_{n^\prime})}{\left(\hbar\omega_{n^\prime}+\Gamma\right)^2}\right)\nonumber\\
&=\text{const.}-\frac{e^2\Gamma\lambda^2}{4\pi^2\hbar^3v_F^2\beta}\int\limits_{-\Lambda}^\Lambda\mathrm{d}\epsilon\sum\limits_{n^\prime=0}^{n-1}\frac{\epsilon}{\tanh(\beta\epsilon)}\frac{1}{\left(\hbar\omega_{n^\prime}+\Gamma\right)^2}\frac{\hbar\omega_{n^\prime}}{(\hbar\omega_{n^\prime})^2+(2\epsilon)^2}\;.
\end{align}
The remaining sum can be evaluated by performing a partial fraction decomposition and applying the digamma identities from Appendix~\ref{ap:digamma}. Subsequently expanding the result to linear order in $\Omega_n$ and to lowest order in $1/\beta\Gamma$, we find
\begin{align}
\Delta C_c^\tau(i\Omega_{n>0})&=\text{const.}-\frac{e^2\beta\Gamma\lambda^2}{32\pi^4\hbar^3v_F^2}\int\limits_{-\beta\Lambda}^{\beta\Lambda}\mathrm{d}(\beta\epsilon)\frac{\beta\epsilon}{\tanh(\beta\epsilon)}\nonumber\\
&\quad\,\times\left[\left(\psi^{(1)}\left(\frac{1}{2}-\frac{i\beta\epsilon}{\pi}\right)+\psi^{(1)}\left(\frac{1}{2}+\frac{i\beta\epsilon}{\pi}\right)\right)\frac{\hbar\Omega_n}{(\beta\Gamma)^2}+\mathcal{O}\big(\Omega_n^2,(1/\beta\Gamma)^3\big)\right] \;.
\end{align}
Finally evaluating the remaining integral in the wide-band limit $\Lambda\rightarrow\infty$ and performing analytic continuation to real frequencies, we recover the lowest order correction to the linear dc charge susceptibility:
\begin{equation}
\Delta\chi_{c,\text{dc}}=-\frac{\pi^3e^2\lambda^2}{16h^3v_F^2}\frac{1}{\beta\Gamma}+\mathcal{O}\big((1/\beta\Gamma)^2\big) \;.
\end{equation}
Identifying $2\pi/\beta\Gamma$ as $T/T_K$ as before, the charge conductance in the vicinity of the NFL fixed point follows as
\begin{equation}
G=\frac{e^2}{2h}\left[1-\left(\frac{\pi\lambda}{4h v_F}\right)^2\frac{T}{T_K}+\ldots\right]\;.\label{eq:Gpert}
\end{equation}
Here, $+\ldots$ represents all higher order terms in products of $\lambda$ and $T/T_K$. This result has previously appeared in Refs.~\cite{pert1,pert2,pert3}; the above serves as a concise derivation of these perturbations.

A few remarks are now in order. (i) The linear order correction (in $T/T_K$) is found to remain finite as $\Lambda\rightarrow\infty$. However, if the cut-off is taken all the way to infinity, the higher order terms take over, as they are proportional to $\ln(\Lambda)$. This means that the lowest order correction to the dc conductance is only linear in $T/T_K$ if the cut-off is finite. Fortunately, this is always the case in real systems. (ii) The leading order correction to the NFL conductance precisely at the EK point is \emph{quadratic} in $T/T_K$, as can be read off from Eq.~(\ref{eq:chictou}) by using Eq.~(\ref{eq:trigammaexp}). This stands in contrast to the temperature dependence away from the EK point, which from Eq.~(\ref{eq:Gpert}) is seen to be linear. We thus find that the lowest order correction to the NFL conductance is linear in $T/T_K$ at a general point in parameter space (including the C2CK model). This linear behavior of the C2CK NFL conductance agrees with the known experimental and numerical results~\cite{experiment,andrew}. (iii) We found that the corrections to the conductance vanish as $T/T_K$ goes to zero, independent of $\lambda$. This is a manifestation of the irrelevance of the anisotropy $\Delta J_z\equiv J_z-J_\bot$: no matter the starting point (which is dictated by the parameter $\lambda$), the RG flow ensures that $\Delta J_z$ effectively goes to zero with the energy scale (in this case $T/T_K$), such that the EK point results become exact regardless of $\lambda$. We emphasize that perturbing away from the EK point emphatically does \textit{not} affect the NFL fixed point conductance itself, only the approach to this point. (iv) From the calculations performed in this section for the linear dc charge conductance, it is clear that the corresponding calculation for heat conductance would be extremely challenging (involving as it does five-loop diagrams), and is beyond the scope of this work. However, again we stress that the NFL fixed point properties themselves are not affected by these corrections, which are RG irrelevant.


\section{Conclusion}\label{sec:conclusion}
In this paper we studied theoretically the quantum transport properties of a model describing recent charge two-channel Kondo quantum dot experiments \cite{experiment,iftikhar2018tunable}. We used the Keldysh non-equilibrium technique and linear response to find exact analytic results for both charge and heat transport. Specifically, we employed the Emery-Kivelson effective theory, which is only valid at a special point in parameter space, but which we show yields asymptotically the true NFL fixed point behavior of the physical C2CK system, as well as its low-temperature FL crossover. In Sec.~\ref{sec:keldysh} we focused on the exact charge current due to a generally time-dependent bias voltage, within a general Keldysh framework. This allowed us to generalize existing exact expressions in the literature for the dc and ac expressions for the electrical current and conductance in the spin 2CK model to the case of the charge-Kondo quantum dot setup. The results successfully capture the FL crossover region, although energies of the order of the Kondo temperature scale are excluded due to the usage of the EK point. The framework used for these calculations is very general, and also paves the way for other time-dependences in the bias voltage that might be experimentally accessible. We also demonstrate the use of the Kubo formula for a direct calculation of the linear response charge conductance.

In Sec.~\ref{sec:heat}, we utilized linear response methods to study heat transport due to a temperature gradient between leads. We point out that the heat current operator is considerably more complicated than the charge current operator. Despite dealing with an effective free field theory at the EK point, three-loop diagrams must be calculated. We therefore restrict attention to the behavior of the C2CK model at NFL and FL fixed points, and obtain exact analytic results. We show that a heat transport measurement would give direct access to the central charge of the critical theory, thereby revealing the Majorana character of the effective model. We can furthermore show that, surprisingly, the Wiedemann-Franz law is satisfied at the NFL critical point, while the Lorenz ratio is expected to be enhanced as the system crosses over to the FL region. All of these $T\ll T_K$ results are quantities that are accessible experimentally, providing a way to bring theory and experiments together in the highly non-trivial context of strongly correlated quantum many-body nanoelectronics devices.


\section*{Acknowledgements}
We thank A.-P. Jauho for a very useful discussion concerning resonant tunneling devices. G.D. is thankful to B.F. McKeever for providing valuable support during the early stages of this research. This work is part of the D-ITP consortium, a program of the Netherlands Organisation for Scientific Research (NWO) that is funded by the Dutch Ministry of Education, Culture and Science (OCW). A.K.M. acknowledges funding from the Irish Research Council Laureate Awards 2017/2018 through grant IRCLA/2017/169.

\newpage
\appendix


\section{Bosonization}\label{ap:bosonization}
In this appendix, we derive the bosonization formulas used in Sec.~\ref{sec:emery}, following Refs.~\cite{senechal2004introduction,gogolin}. The first realization necessary to derive these relations is that the dispersion relation has been linearized around the Fermi level, such that the one-dimensional fermionic fields are only allowed to move with constant velocity $v_F$ in either direction. Introducing coordinates $z=i(x+v_Ft)=v_F\tau+ix$ and $\bar{z}=-i(x-v_Ft)=v_F\tau-ix$, the fields can be divided into left-movers $\psi(x,t)=\psi(z)$ and right-movers $\bar{\psi}(x,t)=\bar{\psi}(\bar{z})$. Since all fermionic fields in the anisotropic 2CK model are left-movers, we will now only consider fields of the form $\psi(z)$, noting that $\partial_x=i(\partial_z-\partial_{\bar{z}})=i\partial_z$ when acting on these fields.

The bosonization ansatz is that one can introduce a bosonic left-moving field $\Phi(z)$ that is related to its fermionic counterpart through the relations
\begin{equation}
\psi(z)=A\eta\,e^{-i\lambda\Phi(z)}\;,\qquad\psi^\dagger(z)=A\eta\,e^{i\lambda\Phi(z)}\;,
\end{equation}
where $A$ and $\lambda$ are real positive constants, and $\eta$ is a Klein factor that is introduced to ensure the anticommutation of different types of fermions ({\it i.e.}, $\left\{\eta_i,\eta_j\right\}=2\delta_{i,j}$ for fermion species $i$ and $j$). The operator exponentials in the above expressions are understood as normal ordered. In order to work with normal ordered operator exponentials of this form, we take notice of a useful formula:
\begin{equation}
:e^{i\alpha\Phi(z)}::e^{i\beta\Phi(z^\prime)}:\,=\,:e^{i\alpha\Phi(z)+i\beta\Phi(z^\prime)}:e^{-\alpha\beta\langle\Phi(z)\Phi(z^\prime)\rangle}\;,
\end{equation}
which is a direct consequence of the Campbell-Baker-Hausdorff formula (see, {\it e.g.}, Ref.~\cite{senechal2004introduction}). Furthermore, the fermionic and bosonic two-point functions can be calculated by performing mode expansions of the fields,\footnote{The same mode expansions can also be used to derive the correct (anti)commutation relations for the fields, including the one from Eq.~(\ref{eq:commutation}).} leading to
\begin{equation}
\big\langle\psi^\dagger(z)\psi(z^\prime)\big\rangle=\frac{1}{2\pi}\frac{1}{z-z^\prime}\;,\qquad\big\langle\Phi(z)\Phi(z^\prime)\big\rangle=\ln\left(\frac{a_0}{z-z^\prime}\right)
\end{equation}
for $\tau>\tau^\prime$; $a_0$ is a small regularization parameter that must be included in the mode expansion of the bosonic fields, and is often taken to be of the order of the lattice spacing of the lattice on which the calculation was performed~\cite{vondelft1998refermionization}. Combining all of the above equations and writing the Klein factor as an exponential, the desired bosonization identity is found to be
\begin{equation}
\psi(z)=\frac{1}{\sqrt{2\pi a_0}}e^{i\phi}e^{-i\Phi(z)}\;,
\end{equation}
which is the identity used in the Emery-Kivelson mapping procedure.

In the remainder of this section, we will use the above to derive Eqs.~(\ref{eq:bosonization1}), (\ref{eq:bosonization2}) and (\ref{eq:bosonization3}) from the main text. To do so, we make use of Wick's theorem, stating that a normal ordered product is equal to the corresponding time ordered product minus all possible contractions. For equal-time products, the use of time ordering and two-point functions requires point-splitting, which we implement by evaluating all creation operators at an imaginary time $\epsilon/v_F$ later than the annihilation operators. First considering a density term:
\begin{align}
:\psi^\dagger(z+\epsilon)\psi(z):&=T_\tau\psi^\dagger(z+\epsilon)\psi(z)-\big\langle\psi^\dagger(z+\epsilon)\psi(z)\big\rangle\nonumber\\
&=\frac{1}{2\pi a_0}:e^{i\Phi(z+\epsilon)-i\Phi(z)}:e^{\langle\Phi(z+\epsilon)\Phi(z)\rangle}-\frac{1}{2\pi\epsilon}\nonumber\\
&=\frac{1}{2\pi\epsilon}\left(:e^{i\partial_z\Phi(z)\,\epsilon+\mathcal{O}(\epsilon^2)}:-\,1\right)\nonumber\\
&=\frac{i}{2\pi}\partial_z\Phi(z)+\mathcal{O}(\epsilon)\;,
\end{align}
Taking the limit $\epsilon\rightarrow 0$ and using $i\partial_z=\partial_x$, this immediately leads to Eq.~(\ref{eq:bosonization1}). Evaluation of kinetic terms is more subtle, and we find them by considering an overall derivative instead:
\begin{align}
:\psi^\dagger(z+\epsilon)\partial_z\psi(z):&=\partial_z:\psi^\dagger(z^\prime)\psi(z):\Big|_{z^\prime\rightarrow z+\epsilon}\nonumber\\
&=\partial_z\left[\frac{1}{2\pi}\frac{1}{z^\prime-z}\left(:e^{i\Phi(z^\prime)-i\Phi(z)}:-\,1\right)\right]\Bigg|_{z^\prime\rightarrow z+\epsilon}\nonumber\\
&=\partial_z\left[\frac{1}{2\pi}\frac{1}{z^\prime-z}\left(i\big(\Phi(z^\prime)-\Phi(z)\big)-\frac{1}{2}\big(\Phi(z^\prime)-\Phi(z)\big)^2+\ldots\right)\right]\Bigg|_{z^\prime\rightarrow z+\epsilon}\nonumber\\
&=\frac{i}{4\pi}\partial_z^2\Phi(z)+\frac{1}{4\pi}\big(\partial_z\Phi(z)\big)^2+\mathcal{O}(\epsilon)\;.
\end{align}
The final line of this equation is obtained by first evaluating the derivative, then replacing $z^\prime\rightarrow z+\epsilon$ and expanding the result in $\epsilon$. Integrating over $x$ and again taking the limit $\epsilon\rightarrow 0$, we find
\begin{align}
\int\limits_{-\infty}^{\infty}\mathrm{d}x:\psi^\dagger(x)\partial_x\psi(x):&=-\frac{i}{4\pi}\int\limits_{-\infty}^{\infty}\mathrm{d}x\big(\partial_x\Phi(x)\big)^2+\frac{1}{4\pi}\partial_x\Phi(x)\Big|^\infty_{-\infty}\nonumber\\
&=-\frac{i}{4\pi}\int\limits_{-\infty}^{\infty}\mathrm{d}x\big(\partial_x\Phi(x)\big)^2+\frac{1}{2}:\psi^\dagger(x)\psi(x):\Big|^\infty_{-\infty}\;.
\end{align}
The boundary term on the right-hand side is equal to the difference in density at both ends of the infinite one-dimensional system and therefore equal to zero, such that the above is the same as Eq.~(\ref{eq:bosonization2}). Finally, the bosonization of interaction terms is a straightforward extension of the bosonization of density terms. Distinguishing between two different fermion species $\alpha$ and $\beta$ such that $\psi_\alpha(z)$ and $\psi^\dagger_\beta(z^\prime)$ anticommute, and $\Phi_\alpha(z)$ and $\Phi_\beta(z^\prime)$ commute:
\begin{align}
:\psi_\alpha^\dagger&(z+\epsilon)\psi_\alpha(z)\psi_\beta^\dagger(z+\epsilon)\psi_\beta(z):\nonumber\\
&=T_\tau\psi_\alpha^\dagger(z+\epsilon)\psi_\alpha(z)\psi_\beta^\dagger(z+\epsilon)\psi_\beta(z)\,-:\psi_\alpha^\dagger(z+\epsilon)\psi_\alpha(z):\big\langle\psi_\beta^\dagger(z+\epsilon)\psi_\beta(z)\big\rangle\nonumber\\
&\quad\,-:\psi_\beta^\dagger(z+\epsilon)\psi_\beta(z):\big\langle\psi_\alpha^\dagger(z+\epsilon)\psi_\alpha(z)\big\rangle-\big\langle\psi_\alpha^\dagger(z+\epsilon)\psi_\alpha(z)\big\rangle\big\langle\psi_\beta^\dagger(z+\epsilon)\psi_\beta(z)\big\rangle\nonumber\\
&=\frac{1}{(2\pi\epsilon)^2}\left(:e^{i\Phi_\alpha(z+\epsilon)-i\Phi_\alpha(z)}::e^{i\Phi_\beta(z+\epsilon)-i\Phi_\beta(z)}:-:e^{i\Phi_\alpha(z+\epsilon)-i\Phi_\alpha(z)}:-:e^{i\Phi_\beta(z+\epsilon)-i\Phi_\beta(z)}:+\,1\right)\nonumber\\
&=\frac{1}{(2\pi\epsilon)^2}\left(:e^{i\Phi_\alpha(z+\epsilon)-i\Phi_\alpha(z)}:-\,1\right)\left(:e^{i\Phi_\beta(z+\epsilon)-i\Phi_\beta(z)}:-\,1\right)\nonumber\\
&=-\frac{1}{4\pi^2}\big(\partial_z\Phi_\alpha(z)\big)\big(\partial_z\Phi_\beta(z)\big)+\mathcal{O}(\epsilon)\;.
\end{align}
Similar to the previous bosonization formulas, this straightforwardly leads to Eq.~(\ref{eq:bosonization3}).


\section{Kubo formula}\label{ap:kubo}
The Kubo formula from Eq.~(\ref{eq:kubo}) can be obtained by using functional integral formalism in imaginary time $\tau$. The contribution of the potential drop to the Euclidean action is $S_\phi=\int_0^{\hbar\beta}\mathrm{d}\tau\,\hat{H}_\phi(\tau)$, from where it follows that
\begin{align}
\big\langle\hat{I}_Q\big\rangle(\tau)&=\frac{1}{Z}\int\mathcal{D}\psi\,\hat{I}_Q(\tau)\,e^{-\frac{S_0}{\hbar}-\frac{S_\phi}{\hbar}}\nonumber\\
&=\frac{\int\mathcal{D}\psi\,\hat{I}_Q(\tau)\Big(1-\frac{1}{\hbar}\int\limits_0^{\hbar\beta}\mathrm{d}\tau^\prime\hat{Q}(\tau^\prime)\Delta\phi(\tau^\prime)+\mathcal{O}\left(\Delta\phi^2\right)\Big)e^{-\frac{S_0}{\hbar}}}{\int\mathcal{D}\psi\Big(1-\frac{1}{\hbar}\int\limits_0^{\hbar\beta}\mathrm{d}\tau^\prime\hat{Q}(\tau^\prime)\Delta\phi(\tau^\prime)+\mathcal{O}\left(\Delta\phi^2\right)\Big)e^{-\frac{S_0}{\hbar}}}\nonumber\\
&=\frac{\int\mathcal{D}\psi\,\hat{I}_Q(\tau)\Big(1-\frac{1}{\hbar}\int\limits_0^{\hbar\beta}\mathrm{d}\tau^\prime\hat{Q}(\tau^\prime)\Delta\phi(\tau^\prime)+\mathcal{O}\left(\Delta\phi^2\right)\Big)e^{-\frac{S_0}{\hbar}}}{Z_0\Big(1-\frac{1}{\hbar}\int\limits_0^{\hbar\beta}\mathrm{d}\tau^\prime\big\langle\hat{Q}(\tau^\prime)\big\rangle_0\Delta\phi(\tau^\prime)+\mathcal{O}\left(\Delta\phi^2\right)\Big)}\nonumber\\
&=\big\langle\hat{I}_Q(\tau)\big\rangle_0-\frac{1}{\hbar}\int\limits_0^{\hbar\beta}\mathrm{d}\tau^\prime\left(\big\langle T_\tau\hat{I}_Q(\tau)\hat{Q}(\tau^\prime)\big\rangle_0-\big\langle\hat{I}_Q(\tau)\big\rangle_0\big\langle\hat{Q}(\tau^\prime)\big\rangle_0\right)\Delta\phi(\tau^\prime)+\mathcal{O}\left(\Delta\phi^2\right)\nonumber\\
&=\int\limits_0^{\hbar\beta}\mathrm{d}\tau^\prime\chi(\tau,\tau^\prime)\Delta\phi(\tau^\prime)+\mathcal{O}\left(\Delta\phi^2\right)\;,\qquad\chi(\tau,\tau^\prime)\equiv-\frac{1}{\hbar}\big\langle T_\tau\hat{I}_Q(\tau)\hat{Q}(\tau^\prime)\big\rangle_0\;,\label{eq:linres}
\end{align}
with $\mathcal{D}\psi$ referring to all Grassmann fields $\psi_{\alpha\sigma}$ and $\bar{\psi}_{\alpha\sigma}$, and those contained in the dot spin operator $\pmb{\tau}$. Here, $\langle\ldots\rangle$ denotes the full expectation value, while $\langle\ldots\rangle_0$ refers to the expectation value in absence of a potential gradient; the first term vanishes because there is no transport if the potential gradient is zero, and the bubble diagrams $\sim\big\langle\hat{I}_Q(\tau)\big\rangle_0\big\langle\hat{Q}(\tau^\prime)\big\rangle_0$ vanish for the same reason. Moreover, the time ordering operator $T_\tau$ originates from the slicing procedure used in the derivation of the functional integral formalism. It is now important to note that the susceptibility is an imaginary time Green function,
\begin{equation}
\chi(\tau,\tau^\prime)=\frac{1}{\hbar}C^\tau_{IQ}(\tau-\tau^\prime)\;,
\end{equation}
where we used the fact that the bare Hamiltonian ({\it i.e.}, in absence of $\Delta\phi$) is time-independent to justify the statement that any bare two-point function $\big\langle\hat{A}(\tau)\hat{B}(\tau^\prime)\big\rangle_0$ can be written as a function only depending on the time difference $\left(\tau-\tau^\prime\right)$. Fourier transforming Eq.~(\ref{eq:linres}) to Matsubara frequencies, analytically continuing to real frequencies and Fourier transforming back to time, Eq.~(\ref{eq:linres}) becomes
\begin{equation}
\big\langle\hat{I}_Q\big\rangle(t)=\int\limits_{-\infty}^{\infty}\mathrm{d}t^\prime\chi(t,t^\prime)\Delta\phi(t^\prime)+\mathcal{O}\left(\Delta\phi^2\right)\;,\label{eq:linresreal}
\end{equation}
with
\begin{equation}\label{eq:chiCIQ}
\chi(t,t^\prime)=\frac{1}{\hbar}C^\text{R}_{IQ}(t-t^\prime)=-\frac{i}{\hbar}\theta(t-t^\prime)\big\langle\big[\hat{I}_Q(t),\hat{Q}(t^\prime)\big]\big\rangle_0\;.
\end{equation}
This gives us an expression for the linear response susceptibility of the current of charge $Q$ due to a potential drop $\Delta\phi$, in terms of a ``bare'' expectation value.

The above susceptibility can also be written as a current-current correlation function. To do so, we introduce another potential $A$ such that $\Delta\phi(t)=-\partial_tA(t)$. Integrating Eq.~(\ref{eq:linresreal}) by parts and using the definition of the current from Eq.~(\ref{eq:currentdef}), we find
\begin{align}
\big\langle\hat{I}_Q\big\rangle(t)&=\int\limits_{-\infty}^{\infty}\mathrm{d}t^\prime\left(\frac{i}{\hbar}\theta(t-t^\prime)\big\langle\big[\hat{I}_Q(t),\hat{I}_Q(t^\prime)\big]\big\rangle_0+\frac{i}{\hbar}\delta(t-t^\prime)\big\langle\big[\hat{I}_Q(t),\hat{Q}(t)\big]\big\rangle_0\right)A(t^\prime)+\mathcal{O}\left(\Delta\phi^2\right)\nonumber\\
&=\int\limits_{-\infty}^{\infty}\mathrm{d}t^\prime\left(-\frac{1}{\hbar}C^\text{R}(t-t^\prime)+\frac{i}{\hbar}\delta(t-t^\prime)\big\langle\big[\hat{I}_Q(t),\hat{Q}(t)\big]\big\rangle_0\right)A(t^\prime)+\mathcal{O}\left(\Delta\phi^2\right)\;,
\end{align}
Note that the boundary terms from integrating by parts vanish due to the procedure of adiabatically switching the potential on and off in the distant past and future (also used in the Keldysh formalism from Sec.~\ref{sec:keldysh}), such that $A(t\rightarrow-\infty)=A(t\rightarrow\infty)=0$.
Furthermore realizing that the Fourier transform of the equation $\Delta\phi(t)=-\partial_tA(t)$ is simply $\Delta\phi(\omega)=i\omega A(\omega)$, we obtain
\begin{align}
\big\langle\hat{I}_Q\big\rangle(\omega)&=\left(-\frac{1}{\hbar}C^\text{R}(\omega)+\frac{i}{\hbar}\big\langle\big[\hat{I}_Q(t),\hat{Q}(t)\big]\big\rangle_0\right)A(\omega)+\mathcal{O}\left(\Delta\phi^2\right)\nonumber\\
&=\frac{i}{\hbar\omega}\left(C^\text{R}(\omega)-i\big\langle\big[\hat{I}_Q(t),\hat{Q}(t)\big]\big\rangle_0\right)\Delta\phi(\omega)+\mathcal{O}\left(\Delta\phi^2\right)\;,\label{eq:Iomega}
\end{align}
where it should be noted that the boundary term $i\langle[\hat{I}_Q(t),\hat{Q}(t)]\rangle_0$ is a (real) constant, again due to the fact that the bare system is time-independent. Starting from Eq.~(\ref{eq:CRdef}) and using the fact that the current operator is defined as minus the time derivative of the charge operator, we can relate this constant to the zero frequency current autocorrelator:
\begin{align}
C^\text{R}(\omega=0)&=-i\int\limits_{-\infty}^{0}\mathrm{d}\Delta t\,\big\langle\big[\hat{I}_Q(t),\hat{I}_Q(t+\Delta t)\big]\big\rangle_0\nonumber\\
&=i\big\langle\big[\hat{I}_Q(t),\hat{Q}(t)\big]\big\rangle_0-i\big\langle\big[\hat{I}_Q(t),\hat{Q}(-\infty)\big]\big\rangle_0\;,
\end{align}
the final term being zero. With this, we finally arrive at
\begin{equation}
\chi(\omega)=\frac{i}{\hbar\omega}\left(C^\text{R}(\omega)-C^\text{R}(0)\right)\;,
\end{equation}
which is the expression used in the main text.

In the dc case, the potential drop is given by $\Delta\phi(\omega)=2\pi\Delta\phi\,\delta(\omega)$. As a result, the dc susceptibility is equal to the $\omega\rightarrow 0$ limit of $\chi(\omega)$, as can be seen by inverse Fourier transforming Eq.~(\ref{eq:Iomega}). In this limit, the above result can be further simplified by noting that $\text{Re}[C^\text{R}(\omega)]=\text{Re}[C^\text{R}(-\omega)]$ and $\text{Im}[C^\text{R}(\omega)]=-\text{Im}[C^\text{R}(-\omega)]$~\cite{kubo2}. The real part of $(C^\text{R}(\omega)-C^\text{R}(0))$ is therefore at least quadratic in $\omega$, such that the dc susceptibility becomes
\begin{equation}
\chi_\text{dc}=\lim\limits_{\omega\rightarrow 0}\frac{1}{\hbar\omega}\left(-\text{Im}\!\left[C^\text{R}(\omega)\right]\right)\;.
\end{equation}
This is the form of the Kubo formula often found in literature.


\section{Majorana Green functions on the dot}\label{ap:dotgreen}
In the following Appendix, we calculate the necessary components of the dot Green function $\mathbf{D}(t,t^\prime)$. Referring back to the self-energy from Eq.~(\ref{eq:sigmad}) and the Hamiltonian from Eq.~(\ref{eq:HamKel}), and using structures from Eqs.~(\ref{eq:gbotm}) and (\ref{eq:greensform}), we find
\begin{align}
\mathbf{D}&=\left(\mathbf{D}_0^{-1}-\frac{g_\bot^2}{\hbar^2}\,\mathbf{g}^\dagger\cdot\mathbf{L}_0^\prime\cdot\mathbf{g}\right)^{-1}\nonumber\\
&=\left(\mathbf{D}_0^{-1}\Big|_{B=0}-\frac{B}{\hbar}\,\delta(t-t^\prime)\begin{pmatrix}1 & 0 \\ 0 & -1\end{pmatrix}-\frac{g_\bot^2}{\hbar^2}\left(L_{0,1}^\prime+L_{0,2}^\prime\right)\begin{pmatrix}1 & -1 \\ -1 & 1\end{pmatrix}\right)^{-1}\;.\label{eq:Dfull}
\end{align}
Let us now look at the retarded and advanced components of this Green function, using Eq.~(\ref{eq:L0pRA}):
\begin{align}
\mathbf{D}^\text{R/A}(t,t^\prime)&=\left(\delta(t-t^\prime)\begin{pmatrix}i\partial_{t^\prime}-\frac{B}{\hbar}\pm\frac{ig_\bot^2}{\hbar^2v_F} & \mp\frac{ig_\bot^2}{\hbar^2v_F}\\ \mp\frac{ig_\bot^2}{\hbar^2v_F} & i\partial_{t^\prime}+\frac{B}{\hbar}\pm\frac{ig_\bot^2}{\hbar^2v_F}\end{pmatrix}\right)^{-1}\nonumber\\
&=\left(\delta(t-t^\prime)\begin{pmatrix}i\partial_{t^\prime}\pm\frac{1}{\hbar}\left(\mp B+\frac{i}{2}\Gamma\right) & \mp\frac{i\Gamma}{2\hbar}\\ \mp\frac{i\Gamma}{2\hbar} & i\partial_{t^\prime}\pm\frac{1}{\hbar}\left(\pm B+\frac{i}{2}\Gamma\right)\end{pmatrix}\right)^{-1}\;,
\end{align}
where the convergence factor $i0^+$ has been omitted due to the presence of a non-zero imaginary part. As seen in the main text, it is convenient to consider the Fourier transform of the dot Green function as well. In particular, we work out the following object, which also appears in the expressions for the currents:
\begin{align}
\mathbf{D}^\text{R/A}(\epsilon)&=\int\limits_{-\infty}^{\infty}\mathrm{d}\Delta t\,e^{\frac{i\epsilon}{\hbar}\Delta t}\mathbf{D}^\text{R/A}(\Delta t)\nonumber\\
&=\int\limits_{-\infty}^{\infty}\mathrm{d}t^\prime\,e^{-\frac{i\epsilon}{\hbar}(t-t^\prime)}\mathbf{D}^\text{R/A}(t^\prime,t)\;,
\end{align}
using that $\mathbf{D}^\text{R/A}(t,t^\prime)=\mathbf{D}^\text{R/A}(t-t^\prime)$, and where in the second line we wrote $\Delta t\equiv t^\prime-t$, with $t$ a constant. This is allowed due to the fact that the time-dependence of the Hamiltonian has no influence on any of the necessary components (see Eq.~(\ref{eq:L0pRA})) and can therefore be treated as fully time-independent. First calculating the Fourier transform of the inverse of $\mathbf{D}^\text{R/A}(t,t^\prime)$:
\begin{align}
\big(\mathbf{D}^\text{R/A}\big)^{-1}(\epsilon)&=\int\limits_{-\infty}^{\infty}\mathrm{d}t^\prime\,\delta(t^\prime-t)\begin{pmatrix}i\partial_t\pm\frac{1}{\hbar}\left(\mp B+\frac{i}{2}\Gamma\right) & \mp\frac{i\Gamma}{2\hbar}\\ \mp\frac{i\Gamma}{2\hbar} & i\partial_t\pm\frac{1}{\hbar}\left(\pm B+\frac{i}{2}\Gamma\right)\end{pmatrix}e^{-\frac{i\epsilon}{\hbar}(t-t^\prime)}\nonumber\\
&=\frac{1}{\hbar}\begin{pmatrix}\epsilon-B\pm\frac{i}{2}\Gamma & \mp\frac{i}{2}\Gamma\\ \mp\frac{i}{2}\Gamma & \epsilon+B\pm\frac{i}{2}\Gamma\end{pmatrix}\;.
\end{align}
Inverting this matrix, we obtain
\begin{equation}
\mathbf{D}^\text{R/A}(\epsilon)=\frac{\hbar}{\epsilon(\epsilon\pm i\Gamma)-B^2}\begin{pmatrix}\epsilon+B\pm\frac{i}{2}\Gamma & \pm\frac{i}{2}\Gamma\\ \pm\frac{i}{2}\Gamma & \epsilon-B\pm\frac{i}{2}\Gamma\end{pmatrix}\;.
\end{equation}
Referring back to Eqs.~(\ref{eq:majoranagreens1})-(\ref{eq:majoranagreens2}), it also immediately follows that the Fourier transformed Majorana Green functions are given by
\begin{align}
D^\text{R/A}_{aa}(\epsilon)&=\frac{\hbar(\epsilon\pm i\Gamma)}{\epsilon(\epsilon\pm i\Gamma)-B^2}\;,\label{eq:Daae}\\
D^\text{R/A}_{bb}(\epsilon)&=\frac{\hbar\epsilon}{\epsilon(\epsilon\pm i\Gamma)-B^2}\;,\\
D^\text{R/A}_{ab}(\epsilon)&=\frac{-i\hbar B}{\epsilon(\epsilon\pm i\Gamma)-B^2}=-D^\text{R/A}_{ba}(\epsilon)\;.\label{eq:Dabe}
\end{align}

Although for the examples in the main text it suffices to know the Fourier transform of the dot Green function, it is often necessary to have an expression for $\mathbf{D}^\text{R/A}(t,t^\prime)$ itself. This expression is found by evaluating the following integral:
\begin{equation}
\mathbf{D}^\text{R/A}(t,t^\prime)=\int\limits_{-\infty}^{\infty}\frac{\mathrm{d}\epsilon}{2\pi\hbar}e^{-\frac{i\epsilon}{\hbar}(t-t^\prime)}\frac{\hbar}{\epsilon(\epsilon\pm i\Gamma)-B^2}\begin{pmatrix}\epsilon+B\pm\frac{i}{2}\Gamma & \pm\frac{i}{2}\Gamma\\ \pm\frac{i}{2}\Gamma & \epsilon-B\pm\frac{i}{2}\Gamma\end{pmatrix}\;.
\end{equation}
Let us now turn to contour integration to evaluate this integral. The poles of the integrand are located at $\epsilon_{\pm,n}=\mp i\Gamma/2+(-1)^n\sqrt{B^2-\Gamma^2/4}=\mp i\big(\Gamma/2\mp(-1)^n\sqrt{\Gamma^2/4-B^2}\big)$ (where $n=0,1$), and so in the case of the retarded (advanced) components, they are both located in the negative (positive) imaginary plane for any real $B$. Properly closing the contour, we thus find that $\mathbf{D}^{\text{R}(\text{A})}(t,t^\prime)$ is zero if $t<t^\prime$ ($t>t^\prime$), and non-zero otherwise. This confirms that it is indeed a retarded (advanced) function. Meanwhile, the denominator in the integrand can be written as $(-1)^ni\sqrt{\Gamma^2-4B^2}(\epsilon-\epsilon_{\pm,n})$ to linear order around the poles. Applying the residue theorem:
\begin{align}
\mathbf{D}^\text{R/A}(t,t^\prime)&=\mp i\theta\left(\pm(t-t^\prime)\right)\sum_n\text{Res}\!\left(\frac{e^{-\frac{i\epsilon}{\hbar}(t-t^\prime)}}{\epsilon(\epsilon\pm i\Gamma)-B^2}\begin{pmatrix}\epsilon+B\pm\frac{i}{2}\Gamma & \pm\frac{i}{2}\Gamma\\ \pm\frac{i}{2}\Gamma & \epsilon-B\pm\frac{i}{2}\Gamma\end{pmatrix},\epsilon_{\pm,n}\right)\nonumber\\
&=\mp i\theta\left(\pm(t-t^\prime)\right)\frac{-i}{\sqrt{\Gamma^2-4B^2}}\bigg[e^{-\frac{i\epsilon_{\pm,0}}{\hbar}(t-t^\prime)}\begin{pmatrix}\epsilon_{\pm,0}+B\pm\frac{i}{2}\Gamma & \pm\frac{i}{2}\Gamma\\ \pm\frac{i}{2}\Gamma & \epsilon_{\pm,0}-B\pm\frac{i}{2}\Gamma\end{pmatrix}\nonumber\\
&\quad\,-e^{-\frac{i\epsilon_{\pm,1}}{\hbar}(t-t^\prime)}\begin{pmatrix}\epsilon_{\pm,1}+B\pm\frac{i}{2}\Gamma & \pm\frac{i}{2}\Gamma\\ \pm\frac{i}{2}\Gamma & \epsilon_{\pm,1}-B\pm\frac{i}{2}\Gamma\end{pmatrix}\bigg]\;,
\end{align}
where the overall sign in front emerges from the integration direction. Writing out the components of the above matrix, we find
\begin{align}
D^\text{R/A}_{11}(t,t^\prime)&=\mp i\theta\left(\pm(t-t^\prime)\right)e^{\mp\frac{\Gamma}{2\hbar}(t-t^\prime)}\left[\cosh\left(\frac{1}{2\hbar}\sqrt{\Gamma^2-4B^2}(t-t^\prime)\right)\right.\nonumber\\
&\quad\,-\left.\frac{2iB}{\sqrt{\Gamma^2-4B^2}}\sinh\left(\frac{1}{2\hbar}\sqrt{\Gamma^2-4B^2}(t-t^\prime)\right)\right]\;,\\
D_{12}^\text{R/A}(t,t^\prime)&=-i\theta\left(\pm(t-t^\prime)\right)\frac{\Gamma}{\sqrt{\Gamma^2-4B^2}}e^{\mp\frac{\Gamma}{2\hbar}(t-t^\prime)}\sinh\left(\frac{1}{2\hbar}\sqrt{\Gamma^2-4B^2}(t-t^\prime)\right)\;,
\end{align}
and
\begin{equation}
D_{21}^\text{R/A}(t,t^\prime)=D_{12}^\text{R/A}(t,t^\prime)\;,\qquad D_{22}^\text{R/A}(t,t^\prime)=D_{11}^\text{R/A}(t,t^\prime)\Big|_{B\rightarrow -B}\;.
\end{equation}
Meanwhile, the Majorana Green functions are given by
\begin{align}
D^\text{R/A}_{aa}(t,t^\prime)&=\mp i\theta\left(\pm(t-t^\prime)\right)e^{\mp\frac{\Gamma}{2\hbar}(t-t^\prime)}\left[\cosh\left(\frac{1}{2\hbar}\sqrt{\Gamma^2-4B^2}(t-t^\prime)\right)\right.\nonumber\\
&\quad\,\pm\left.\frac{\Gamma}{\sqrt{\Gamma^2-4B^2}}\sinh\left(\frac{1}{2\hbar}\sqrt{\Gamma^2-4B^2}(t-t^\prime)\right)\right]\;,\label{eq:DaaRF}\\
D^\text{R/A}_{bb}(t,t^\prime)&=\mp i\theta\left(\pm(t-t^\prime)\right)e^{\mp\frac{\Gamma}{2\hbar}(t-t^\prime)}\left[\cosh\left(\frac{1}{2\hbar}\sqrt{\Gamma^2-4B^2}(t-t^\prime)\right)\right.\nonumber\\
&\quad\,\mp\left.\frac{\Gamma}{\sqrt{\Gamma^2-4B^2}}\sinh\left(\frac{1}{2\hbar}\sqrt{\Gamma^2-4B^2}(t-t^\prime)\right)\right]\;,\\
D^\text{R/A}_{ab}(t,t^\prime)&=\pm i\theta\left(\pm(t-t^\prime)\right)e^{\mp\frac{\Gamma}{2\hbar}(t-t^\prime)}\frac{2B}{\sqrt{\Gamma^2-4B^2}}\sinh\left(\frac{1}{2\hbar}\sqrt{\Gamma^2-4B^2}(t-t^\prime)\right)\;,\label{eq:DabRF}
\end{align}
and $D^\text{R/A}_{ba}(t,t^\prime)=-D^\text{R/A}_{ab}(t,t^\prime)$. For future reference, note in particular that $D^\text{R/A}_{ba}(t,t^\prime)$ is proportional to the magnetic field $B$. This remains true upon analytically continuing to imaginary time $\tau$, such that the time-ordered expectation value $\big\langle T_\tau b(\tau)a(\tau^\prime)\big\rangle$ goes to zero with the magnetic field.


\section{The Keldysh structure}\label{ap:kel}
In the case of fermions, the Keldysh formalism for non-equilibrium problems~\cite{kamenev} starts with the action as a function of Grassmann fields $\psi$ and $\bar{\psi}$, involving integration over the closed time contour $\mathcal{C}$ (which consists of a forward branch from $-\infty$ to $+\infty$, and a backward branch from $+\infty$ to $-\infty$). The fields are then doubled: one field for the forward branch, and one for the backward branch. Following the Keldysh prescription of doubling and rotating the fields (such that we have two new fields $\psi_1$ and $\psi_2$ for each field $\psi$), the action becomes
\begin{equation}
S=\hbar\int\limits_{-\infty}^{\infty}\mathrm{d}t\,\bar{\psi}\cdot G^{-1}\otimes\gamma^\text{cl}\cdot\psi\;,
\end{equation}
with
\begin{equation}
\psi\equiv\left(\psi_1\;\psi_2\right)^T\;,\qquad\gamma^\text{cl}\equiv\begin{pmatrix}1 & 0\\ 0 & 1\end{pmatrix}\;.
\end{equation}
In this Keldysh rotated (1,2)-basis, the full Green function $\mathbf{G}$ assumes a triangular structure:
\begin{equation}
\mathbf{G}=\begin{pmatrix}G^\text{R} & G^\text{K}\\ 0 & G^\text{A}\end{pmatrix}\;,
\end{equation}
where $G^\text{R/A}$ are the retarded and advanced Green functions, and $G^\text{K}$ is the Keldysh component of the Green functions. In general, expectation values relate to these components in the following way:
\begin{equation}
\big\langle \psi_\alpha \psi^\dagger_\beta\big\rangle=\frac{1}{2}\big\langle\psi_{\alpha,1}\bar{\psi}_{\beta,2}\big\rangle=\frac{i}{2}G^\text{K}_{\alpha\beta}\;,
\end{equation}
where the labels $\alpha,\beta$ refer to the different fields that might be present in the system.

As discussed in the main text, the Keldysh Green function of a system in thermal equilibrium can be found by using the FDT. However, the dot region is not in equilibrium, so the Keldysh component on the dot has to be found in a different way. In order to find this component, we apply block inversion to the triangular Keldysh structure:
\begin{equation}
\left(\mathbf{G}^{-1}\right)^\text{R/A}=\big(\mathbf{G}^\text{R/A}\big)^{-1}\;,\qquad \left(\mathbf{G}^{-1}\right)^\text{K}=-\left(\mathbf{G}^\text{R}\right)^{-1}\cdot \mathbf{G}^\text{K}\cdot\left(\mathbf{G}^\text{A}\right)^{-1}\;,\label{eq:compinv}
\end{equation}
such that
\begin{equation}
\mathbf{D}^\text{K}=-\mathbf{D}^\text{R}\cdot\left(\left(\mathbf{D}^{-1}_0\right)^\text{K}-\pmb{\Sigma}_d^\text{K}\right)\cdot \mathbf{D}^\text{A}\;.
\end{equation}
Combining Eqs.~(\ref{eq:keldyshcomp}), (\ref{eq:FDT}) and (\ref{eq:compinv}) with the general properties of the (bare) retarded and advanced Green functions, we see that the term
\begin{equation}
\left(\mathbf{D}^{-1}_0\right)^\text{K}=2i0^+\mathbf{F}
\end{equation}
is just a regulator. Therefore,
\begin{equation}
\mathbf{D}^\text{K}=\mathbf{D}^\text{R}\cdot\pmb{\Sigma}_d^\text{K}\cdot \mathbf{D}^\text{A}\;,\label{eq:dk}
\end{equation}
eliminating the need to find the out-of-equilibrium matrix $\mathbf{F}$ on the dot.

Another thing to note is that the FDT from Eq.~(\ref{eq:FDT}) should in general be applied to Wigner transformed Green functions. To see why the FDT manifests itself in the way that it does in Sec.~\ref{sec:keldysh}, consider a simple time-independent and homogeneous model. In this case, the Wigner transform reduces simply to the usual Fourier transform. Applying the FDT:
\begin{align}
G_{0,k}^\text{K}(\omega)&=f(\omega)\left(G_{0,k}^\text{R}(\omega)-G_{0,k}^\text{A}(\omega)\right)\nonumber\\
&=f(\omega)\left(\frac{\hbar}{\omega-\epsilon_k+i0^+}-\frac{\hbar}{\omega-\epsilon_k-i0^+}\right)\nonumber\\
&=-2i\hbar f(\omega)\frac{0^+}{(\omega-\epsilon_k)^2+(0^+)^2}\nonumber\\
&=-2\pi i\hbar f(\omega)\delta(\omega-\epsilon_k)\;.
\end{align}
Fourier transforming the frequency back to time:
\begin{align}
G_{0,k}^\text{K}(t,t^\prime)&=\int\limits_{-\infty}^{\infty}\frac{\mathrm{d}\omega}{2\pi\hbar}G_{0,k}^\text{K}(\omega)e^{-\frac{i\omega}{\hbar}(t-t^\prime)}\nonumber\\
&=-if(\epsilon_k)e^{-\frac{i\epsilon_k}{\hbar}(t-t^\prime)}\;.
\end{align}
To obtain the additional time-dependence due to the bias voltage as seen in Eq.~(\ref{eq:L0K}), we subsequently return to the assumption that the bias voltage only acts on the junction without influencing the flavor and spin-flavor modes themselves ({\it i.e.}, keeping them in thermal equilibrium), such that it can be incorporated by simply replacing $\epsilon_k$ by $\epsilon_k-eV(t)/2$.


\section{Properties of the Digamma function}\label{ap:digamma}
The digamma function is defined as
\begin{equation}
\Psi(z)\equiv\frac{\mathrm{d}\ln\Gamma(z)}{\mathrm{d}z}=\frac{1}{\Gamma(z)}\frac{\mathrm{d}\Gamma(z)}{\mathrm{d}z}\;,
\end{equation}
where $\Gamma(z)$ is the gamma function. Differentiating the relation $\Gamma(z+1)=z\Gamma(z)$ (where $\text{Re}(z)>0$) with respect to $z$, it immediately follows that the digamma function satisfies a somewhat similar relation, namely $\Psi(z+1)=\Psi(z)+1/z$. Consider now the following sum, with $\text{Re}(a)>0$ and $\text{Re}(b)>0$:
\begin{equation}
(a-b)\sum\limits_{\mathclap{n=0}}^\infty\frac{1}{(n+a)(n+b)}=\sum\limits_{\mathclap{n=0}}^\infty\left(\frac{1}{n+b}-\frac{1}{n+a}\right)=\sum\limits_{\mathclap{n=0}}^\infty\int\limits_0^1\mathrm{d}x\left(x^{n+b-1}-x^{n+a-1}\right)=\int\limits_0^1\mathrm{d}x\,\frac{x^{b-1}-x^{a-1}}{1-x}\;.
\end{equation}
In order to evaluate this integral, we introduce a more general version of this integral,
\begin{equation}
I(\epsilon)\equiv\int\limits_0^1\mathrm{d}x\left(x^{b-1}-x^{a-1}\right)(1-x)^{\epsilon-1}=B(b,\epsilon)-B(a,\epsilon)\;,
\end{equation}
with $\epsilon>0$, and $B(x,y)$ being the beta function, satisfying the well-known relation $\Gamma(x)\Gamma(y)=B(x,y)\Gamma(x+y)$. The above sum can now be evaluated:
\begin{align}
(a-b)\sum\limits_{\mathclap{n=0}}^\infty\frac{1}{(n+a)(n+b)}&=\lim_{\mathclap{\epsilon\rightarrow 0^+}}I(\epsilon)=\lim_{\mathclap{\epsilon\rightarrow 0^+}}\left(B(b,\epsilon)-B(a,\epsilon)\right)\nonumber\\
&=\lim_{\mathclap{\epsilon\rightarrow 0^+}}\Gamma(\epsilon)\left(\frac{\Gamma(b)}{\Gamma(b+\epsilon)}-\frac{\Gamma(a)}{\Gamma(a+\epsilon)}\right)\nonumber\\
&=\lim_{\mathclap{\epsilon\rightarrow 0^+}}\Gamma(\epsilon+1)\left(\frac{\Gamma(b)-\Gamma(b+\epsilon)}{\epsilon}\frac{1}{\Gamma(b+\epsilon)}-\frac{\Gamma(a)-\Gamma(a+\epsilon)}{\epsilon}\frac{1}{\Gamma(a+\epsilon)}\right)\nonumber\\
&=\Psi(a)-\Psi(b)\;.\label{eq:digammasum}
\end{align}
In addition, we can recursively apply the relation $\Psi(a+1)=\Psi(a)+1/a$ to find another sum:
\begin{align}
\Psi(a+n)&=\Psi(a+n-1)+\frac{1}{a+n-1}\nonumber\\
&=\Psi(a+n-2)+\frac{1}{a+n-2}+\frac{1}{a+n-1}\nonumber\\
&\hspace{2cm}\vdots\nonumber\\
&=\Psi(a)+\sum\limits_{n^\prime=0}^{n-1}\frac{1}{n^\prime+a}\;,\label{eq:sumn}
\end{align}
where $n\in\mathbb{N}_{>0}$ and $\text{Re}(a)>0$. Using this, we can evaluate the following sum as well:
\begin{align}
\sum\limits_{n^\prime=0}^{n-1}\frac{1}{n^\prime+a}\frac{1}{n^\prime-b-n+1}&=-\frac{1}{a+b+n-1}\left(\,\sum\limits_{n^\prime=0}^{n-1}\frac{1}{n^\prime+a}-\sum\limits_{n^\prime=0}^{n-1}\frac{1}{n^\prime-b-n+1}\right)\nonumber\\
&=-\frac{1}{a+b+n-1}\left(\,\sum\limits_{n^\prime=0}^{n-1}\frac{1}{n^\prime+a}+\sum\limits_{n^\prime=0}^{n-1}\frac{1}{n^\prime+b}\right)\nonumber\\
&=\frac{1}{a+b+n-1}\big(\Psi\left(a\right)-\Psi\left(a+n\right)+\Psi\left(b\right)-\Psi\left(b+n\right)\big)\;,\label{eq:digammasum2}
\end{align}
where the second term of the second line involves a redefinition according to $n^\prime\rightarrow-n^\prime+n-1$. Finally, we use Eq.~(\ref{eq:sumn}) to evaluate one final sum:
\begin{align}
\sum\limits_{n^\prime=0}^{n-1}\frac{1}{(n^\prime+a)^2}&=\lim\limits_{\delta\rightarrow 0}\sum\limits_{n^\prime=0}^{n-1}\frac{1}{(n^\prime+a)(n^\prime+a+\delta)}\nonumber\\
&=\lim\limits_{\delta\rightarrow 0}-\frac{1}{\delta}\sum\limits_{n^\prime=0}^{n-1}\left(\frac{1}{n^\prime+a+\delta}-\frac{1}{n^\prime+a}\right)\nonumber\\
&=\lim\limits_{\delta\rightarrow 0}-\frac{1}{\delta}\big(\Psi(a+n+\delta)-\Psi(a+\delta)-\Psi(a+n)+\Psi(a)\big)\nonumber\\
&=\psi^{(1)}(a)-\psi^{(1)}(a+n)\;,
\end{align}
where $\psi^{(1)}(z)$ is the trigamma function ({\it i.e.}, the derivative of the digamma function).

Let us also briefly consider two specific expansions of the trigamma function. Using the known values of the gamma function and its derivatives at $z=1/2$, the first one is simple:
\begin{equation}
x\,\psi^{(1)}\left(\frac{1}{2}+x\right)=x\,\psi^{(1)}\left(\frac{1}{2}\right)+\mathcal{O}\left(x^2\right)=\frac{\pi^2x}{2}+\mathcal{O}\left(x^2\right)\;.\label{eq:approx1}
\end{equation}
The second one is more complicated, and requires the asymptotic series
\begin{equation}
\Psi(x)\sim\ln(x)-\frac{1}{2x}-\frac{1}{12x^2}+\mathcal{O}\left((1/x)^4\right)\;,
\end{equation}
valid for real variable $x\gg 1$, which can be derived from the Stirling series. Taking the derivative of this asymptotic series and plugging in the argument we are interested in:
\begin{equation}
\frac{1}{x}\psi^{(1)}\left(\frac{1}{2}+\frac{1}{x}\right)=\frac{1}{x}\left(\frac{x}{1+x/2}+\frac{x^2}{2(1+x/2)^2}+\frac{x^3}{6(1+x/2)^3}\right)+\mathcal{O}\left(x^4\right)\;.\label{eq:asseries}
\end{equation}
Also applying the binomial series,
\begin{equation}
(1+x)^n=\sum\limits_{\mathclap{k=0}}^\infty\frac{n!}{(n-k)!\,k!}x^k\;,
\end{equation}
it is straightforward to show that Eq.~(\ref{eq:asseries}) reduces to
\begin{equation}
\frac{1}{x}\psi^{(1)}\left(\frac{1}{2}+\frac{1}{x}\right)=1-\frac{x^2}{12}+\mathcal{O}\left(x^4\right)\;.\label{eq:approx2}
\end{equation}


\section{Properties of Bessel functions of the first kind}\label{ap:bessel}
For integers $n$, the Bessel functions of the first kind are defined as
\begin{equation}
J_n(\alpha)\equiv\sum\limits_{m=0}^\infty\frac{(-1)^m}{m!(m+n)!}\left(\frac{\alpha}{2}\right)^{2m+n}\;.
\end{equation}
Using this expression, the following sum can be evaluated:
\begin{align}
\sum\limits_{\mathclap{n=-\infty}}^\infty e^{-in\omega t}J_n(\alpha)&=\sum\limits_{\mathclap{n=-\infty}}^\infty e^{-in\omega t}\sum\limits_{m=0}^\infty\frac{(-1)^m}{m!(m+n)!}\left(\frac{\alpha}{2}\right)^{2m+n}\nonumber\\
&=\sum\limits_{m=0}^\infty\frac{1}{m!}\left(-\frac{\alpha}{2}\right)^m\sum\limits_{\mathclap{n=-\infty}}^\infty\frac{1}{(m+n)!}\left(\frac{\alpha}{2}\right)^{m+n}e^{-in\omega t}\nonumber\\
&=\sum\limits_{m=0}^\infty\frac{1}{m!}\left(-\frac{\alpha}{2}e^{i\omega t}\right)^m\sum\limits_{\mathclap{n=-\infty}}^\infty\frac{1}{(m+n)!}\left(\frac{\alpha}{2}e^{-i\omega t}\right)^{m+n}\nonumber\\
&=\sum\limits_{m=0}^\infty\frac{1}{m!}\left(-\frac{\alpha}{2}e^{i\omega t}\right)^m\sum\limits_{n^\prime=0}^\infty\frac{1}{n^\prime!}\left(\frac{\alpha}{2}e^{-i\omega t}\right)^{n^\prime}\nonumber\\
&=e^{-\frac{\alpha}{2}\left(e^{i\omega t}-e^{-i\omega t}\right)}\nonumber\\
&=e^{-i\alpha\sin(\omega t)}\;,
\end{align}
where in the fourth line we used the fact that the terms corresponding to $m+n<0$ vanish (due to the fact that $1/n!=0$ if $n$ is a negative integer), and defining a new dummy index $n^\prime$ by using the observation that the second infinite sum must be independent of $m$. From the definition of the Bessel functions, it also follows that they can be expanded as
\begin{equation}
J_0(\alpha)=1+\mathcal{O}(\alpha^2)\;,\qquad J_{n\neq 0}(\alpha)=\frac{\left(\text{sgn}(n)\right)^{|n|}}{|n|!}\left(\frac{\alpha}{2}\right)^{|n|}+\mathcal{O}(\alpha^{|n|+2})\;.
\end{equation}
As such, the expansion of a general $J_n(\alpha)$ is given by
\begin{equation}
J_n(\alpha)=\delta_{n,0}+\frac{\alpha}{2}(\delta_{n,1}-\delta_{n,-1})+\mathcal{O}(\alpha^2)\;.
\end{equation}


\section{Ac charge current for $B=0$}\label{ap:accurrent}
In this appendix, we will evaluate the remaining integral from Eq.~(\ref{eq:Inint}), for $B=0$. We have two integrals to evaluate:
\begin{equation}
\Gamma\int\limits_{-\infty}^\infty\mathrm{d}\epsilon\frac{n_F(\epsilon)}{\epsilon^2+\Gamma^2}\;,\qquad\int\limits_{-\infty}^\infty\mathrm{d}\epsilon\left(\frac{n_F(\epsilon-\epsilon_0)}{\epsilon+i\Gamma}-\frac{n_F(\epsilon-\epsilon_0-n\hbar\omega_0)}{\epsilon-i\Gamma}\right)\;,
\end{equation}
where $\epsilon_0\equiv eV_0/2+n^\prime\hbar\omega_0$. To perform these integrals, we again use the Matsubara representation of the Fermi-Dirac distribution, Eq.~(\ref{eq:matsrep}). Let us look at the first integral, employing the same techniques as for the dc case:
\begin{align}
\Gamma\int\limits_{-\infty}^\infty\mathrm{d}\epsilon\frac{n_F(\epsilon)}{\epsilon^2+\Gamma^2}&=\frac{2}{\beta}\sum\limits_{j=0}^\infty\text{Re}\!\left[\int\limits_{-\infty}^\infty\mathrm{d}\epsilon\frac{\Gamma}{\epsilon^2+\Gamma^2}\frac{e^{i\omega_j0^+}}{i\hbar\omega_j-\epsilon}\right]\nonumber\\
&=\frac{2\pi}{\beta}\sum\limits_{j=0}^\infty\text{Re}\!\left[\frac{e^{i\omega_j0^+}}{i\hbar\omega_j+i\Gamma}\right]\;.\label{eq:firstint}
\end{align}
The second integral can be evaluated in a similar fashion:
\begin{align}
\int\limits_{-\infty}^\infty\mathrm{d}\epsilon&\left(\frac{n_F(\epsilon)}{\epsilon+\epsilon_0+i\Gamma}-\frac{n_F(\epsilon)}{\epsilon+\epsilon_0+n\hbar\omega_0-i\Gamma}\right)\nonumber\\
&=\frac{n\hbar\omega_0-2i\Gamma}{\beta}\int\limits_{-\infty}^\infty\mathrm{d}\epsilon\sum\limits_{j=0}^\infty\frac{1}{\left(\epsilon+\epsilon_0+i\Gamma\right)\left(\epsilon+\epsilon_0+n\hbar\omega_0-i\Gamma\right)}\left(\frac{e^{i\omega_j0^+}}{i\hbar\omega_j-\epsilon}+\frac{e^{-i\omega_j0^+}}{-i\hbar\omega_j-\epsilon}\right)\nonumber\\
&=-\frac{2\pi}{\beta}\sum\limits_{j=0}^\infty\left(\frac{1}{\hbar\omega_j+\Gamma-i\epsilon_0}-\frac{1}{\hbar\omega_j+\Gamma+i\left(\epsilon_0+n\hbar\omega_0\right)}\right)\nonumber\\
&\quad\,-\frac{2\pi}{\beta}\sum\limits_{j=0}^\infty\left(\frac{e^{i\omega_j0^+}-1}{\hbar\omega_j+\Gamma-i\epsilon_0}-\frac{e^{-i\omega_j0^+}-1}{\hbar\omega_j+\Gamma+i\left(\epsilon_0+n\hbar\omega_0\right)}\right)\;,\label{eq:secondint}
\end{align}
where we wrote $e^{\pm i\omega_j0^+}\rightarrow 1+(e^{\pm i\omega_j0^+}-1)$ for reasons that will become clear in a moment. For the final line, we can use the fact that it vanishes unless $j\rightarrow\infty$ to replace the summand with its $j\rightarrow\infty$ value. Doing so, we can recognize this second term to be equal to $-2i$ times the integral from Eq.~(\ref{eq:firstint}). Looking at Eq.~(\ref{eq:Inint}), we see that the two cancel each other, and so we only need to evaluate the first term of Eq.~(\ref{eq:secondint}):
\begin{align}
-\frac{2\pi}{\beta}\sum\limits_{j=0}^\infty&\left(\frac{1}{\hbar\omega_j+\Gamma-i\epsilon_0}-\frac{1}{\hbar\omega_j+\Gamma+i\left(\epsilon_0+n\hbar\omega_0\right)}\right)\nonumber\\
&=-\frac{i\beta}{2\pi}\sum\limits_{j=0}^\infty\frac{2\epsilon_0+n\hbar\omega_0}{\left(j+\frac{1}{2}+\frac{\beta\left(\Gamma-i\epsilon_0\right)}{2\pi}\right)\left(j+\frac{1}{2}+\frac{\beta\left(\Gamma+i\left(\epsilon_0+n\hbar\omega_0\right)\right)}{2\pi}\right)}\nonumber\\
&=\Psi\left(\frac{1}{2}+\frac{\Gamma-i\epsilon_0}{2\pi k_BT}\right)-\Psi\left(\frac{1}{2}+\frac{\Gamma+i\left(\epsilon_0+n\hbar\omega_0\right)}{2\pi k_BT}\right)\;.
\end{align}
Plugging this back into the expression for $I_n$, we obtain
\begin{align}
I_n&=\frac{ie\Gamma}{4\pi\hbar}\sum_{\mathclap{n^\prime=-\infty}}^\infty J_{n^\prime}\left(\frac{e\Delta V}{2\hbar\omega_0}\right)J_{n^\prime+n}\left(\frac{e\Delta V}{2\hbar\omega_0}\right)\nonumber\\
&\quad\,\times\left(\Psi\left(\frac{1}{2}+\frac{\Gamma-i\left(eV_0/2+n^\prime\hbar\omega_0\right)}{2\pi k_BT}\right)-\Psi\left(\frac{1}{2}+\frac{\Gamma+i\left(eV_0/2+\left(n^\prime+n\right)\hbar\omega_0\right)}{2\pi k_BT}\right)\right)\;.\label{eq:Inacfinal}
\end{align}
Note that in the limit $\Delta V\rightarrow 0$, the Bessel functions reduce to $J_n(0)=\delta_{n,0}$. As a result, only $I_0$ remains non-zero, and $\omega_0$ completely vanishes from the expressions, such that we indeed recover the correct dc limit.


\section{Vanishing bubble diagrams}\label{ap:bubble}
In the first part of Sec.~\ref{sec:heat}, we utilized the fact that several bubble diagrams appearing in the calculation of the heat current autocorrelator vanish. In the following Appendix, we demonstrate this important property explicitly. First, let us consider the bubble diagram $\sum_{k,k^\prime}\big\langle\psi^\dagger_k(\tau)\psi_{k^\prime}(\tau)\big\rangle_0$, where $\psi_k$ is a fermionic field operator that only appears in the kinetic term of the Hamiltonian. In terms of Matsubara frequencies and disregarding the now redundant Nambu basis, the Green function corresponding to this field is given by
\begin{equation}
G_{kk^\prime}(i\omega_n)=\delta_{k,k^\prime}\frac{\hbar}{i\hbar\omega_n-\epsilon_k}\;.
\end{equation}
In order to make sense of the bubble diagram, it is necessary to implement point-splitting. The only physical way ({\it i.e.}, without breaking causality) to achieve this is to first annihilate a fundamental particle, then create one in the new state. For a fundamental field $c$, the point-splitting procedure we adopt is therefore $c^\dagger(\tau)c(\tau)\rightarrow c^\dagger(\tau^+)c(\tau)$, with $\tau^+\equiv\tau+0^+$. At this point, it is crucial to remember that the field $\psi_k$ is not fundamental, but instead describes excitations about the ground state.\footnote{In Appendix~\ref{ap:bosonization} it was not necessary to take this subtlety into account. The reason for this is that we only used the point-splitting as a mathematical tool in the previous appendix, whereas we presently use it for the calculation of observable quantities.} For fermions, this is the state in which all states up to the Fermi energy are filled, and all states above the Fermi energy are empty. Given that the momentum $k$ is measured with respect to the Fermi momentum $k_F$, the excitation field $\psi_k$ is thus defined as\footnote{This particular example is about right-movers. Left-movers can be considered using the same methods, and leads to the same conclusions.}
\begin{equation}
\psi_k(\tau)\equiv\begin{cases}c_k(\tau)\;, & k>0\;,\\c^\dagger_k(\tau)\;, & k\leq 0\;,\end{cases}
\end{equation}
where $c_k$ is the field describing the fundamental particles. Using this, we find that the causal point-splitting procedure of the excitation field is given by
\begin{align}
\psi^\dagger_k(\tau)\psi_k(\tau)&\rightarrow\theta(k)c^\dagger_k(\tau^+)c_k(\tau)+\theta(-k)c_k(\tau)c^\dagger_k(\tau^+)\nonumber\\
&=\theta(k)\psi^\dagger_k(\tau^+)\psi_k(\tau)-\theta(-k)\psi_k(\tau^+)\psi^\dagger_k(\tau)\;.
\end{align}
We now apply this point-splitting procedure to the required expectation value:
\begin{align}
\big\langle\psi^\dagger_k(\tau)\psi_{k^\prime}(\tau)\big\rangle_0&\rightarrow\theta(k)\big\langle\psi^\dagger_k(\tau^+)\psi_k(\tau)\big\rangle_0-\theta(-k)\big\langle\psi_k(\tau^+)\psi^\dagger_k(\tau)\big\rangle_0\nonumber\\
&=\theta(k)G_{kk}(-0^+)+\theta(-k)G_{kk}(0^+)\nonumber\\
&=\frac{\theta(k)}{\hbar\beta}\sum_{\mathclap{n=-\infty}}^\infty\frac{\hbar}{i\hbar\omega_n-\epsilon_k}e^{i\omega_n 0^+}-\frac{\theta(-k)}{\hbar\beta}\sum_{\mathclap{n=-\infty}}^\infty\frac{\hbar}{i\hbar\omega_n+\epsilon_k}e^{i\omega_n 0^+}\nonumber\\
&=\theta(k)n_F(\epsilon_k)-\theta(-k)n_F(-\epsilon_k)\;,
\end{align}
where we used Eq.~(\ref{eq:matsrep}) to introduce the Fermi-Dirac distribution. Employing the continuum limit, the desired bubble diagram can now be written as
\begin{equation}
\sum_{k,k^\prime}\big\langle\psi^\dagger_k(\tau)\psi_{k^\prime}(\tau)\big\rangle_0=\frac{L}{2\pi}\int\limits_0^\infty\mathrm{d}k\,\big(n_F(\epsilon_k)-n_F(-\epsilon_{-k})\big)\;.
\end{equation}
In the present case, there is particle-hole symmetry ({\it i.e.}, $\epsilon_k=-\epsilon_{-k}$), such that this bubble diagram trivially vanishes.

In addition, we prove that the combination $(C_{35}+C_{53})$ is indeed equal to zero in absence of a magnetic field. Using that the $a$ Majorana fermion is decoupled from the rest of the system, this combination can be expressed as
\begin{align}
C_{35}(\tau)+C_{53}(\tau)&=-\frac{i\pi v_F\Lambda g_\bot}{2^{5/2}\hbar L^{3/2}}\sum\limits_{\mathclap{k,k^\prime,k^{\prime\prime}}}(\epsilon_{k^\prime}-\epsilon_k)\big\langle a(\tau)a(0)\big\rangle_0\bigg(\left\langle\big(\psi^\dagger_{sf,k^{\prime\prime}}(\tau)+\psi_{sf,k^{\prime\prime}}(\tau)\big)\psi^\dagger_{sf,k}(0)\psi_{sf,k^\prime}(0)b(0)\right\rangle_0\nonumber\\
&\quad\,+\left\langle\psi^\dagger_{sf,k}(\tau)\psi_{sf,k^\prime}(\tau)b(\tau)\big(\psi^\dagger_{sf,k^{\prime\prime}}(0)+\psi_{sf,k^{\prime\prime}}(0)\big)\right\rangle_0\bigg)\;.
\end{align}
The expectation values inside the brackets can be evaluated using Wick's theorem. Before doing so, we first note that the diagonal components of the propagator $\mathbf{L}_{kk^\prime}$ remain the same upon interchanging $k$ and $k^\prime$, which can straightforwardly be shown by working out the matrix multiplications in the Green functions from Sec.~\ref{sec:propagators}. Multiplied with $(\epsilon_{k^\prime}-\epsilon_k)$ and summed over $k$ and $k^\prime$, this cancels all terms proportional to $\big\langle\psi^\dagger_{sf,k}(\tau)\psi_{sf,k^\prime}(\tau)\big\rangle_0$. Secondly, the propagator $\mathbf{G}_{ld,k}(i\omega_n)$ is proportional to $\omega_n^{-2}$ as $n\rightarrow\pm\infty$. As a result, this Green function is independent of the point-splitting procedure, such that we can simply write $\mathbf{G}_{ld,k}(0)=(\hbar\beta)^{-1}\sum_n\mathbf{G}_{ld,k}(i\omega_n)$. Again working out the matrix multiplications, and also using that $D_{bb}(i\omega_n)$ is odd in $\omega_n$ to remove the odd part of the summand, we find
\begin{equation}
\big\langle\psi_{sf,k}(\tau)b(\tau)\big\rangle_0=\big\langle\psi^\dagger_{sf,k}(\tau)b(\tau)\big\rangle_0=-\frac{\sqrt{2}g_\bot}{\sqrt{L}}\frac{1}{\hbar\beta}\sum\limits_{n=-\infty}^\infty\frac{\hbar\omega_n}{(\hbar\omega_n)^2+\epsilon_k^2}D_{bb}(i\omega_n)\;.\label{eq:bubblenz}
\end{equation}
Next, we apply Wick's theorem. After a series of simplifications, this gives
\begin{equation}
C_{35}(\tau)+C_{53}(\tau)=\frac{i\pi v_F\Lambda g_\bot^2}{2\hbar L^2}\frac{1}{\hbar\beta}\sum\limits_{\mathclap{k,k^\prime,k^{\prime\prime}}}(\epsilon_{k^\prime}-\epsilon_k)D_{aa}(\tau)\sum\limits_{n=-\infty}^\infty\frac{\hbar\omega_n}{(\hbar\omega_n)^2+\epsilon_{k^\prime}^2}D_{bb}(i\omega_n)\sum_{\mu\nu}G_{ll,kk^{\prime\prime},\mu\nu}(\tau)\;.
\end{equation}
It is straightforward to check that the object following the sum over $n$ is even in all momenta, such that all terms of the overall momentum sum are odd in either $k$ or $k^\prime$. This immediately leads to the conclusion that the combination $(C_{35}+C_{53})$ is equal to zero.


\section{Details of the heat current autocorrelator calculation}\label{ap:heatcor}

In this Appendix, we elaborate on some of the steps from Sec.~\ref{sec:heat} and show explicitly that the remaining terms do not contribute to the heat conductance. We take $n>0$ throughout.

\begin{itemize}[leftmargin=*,listparindent=\parindent,
  parsep=0pt]

    \item \textbf{Diagonal component $\pmb{C_{11}}$: the sum $\sum_{n^\prime=0}^{n-1}\frac{\frac{1}{2}-2\left(n^\prime-n+\frac{1}{2}\right)^2}{n^\prime+\frac{1}{2}+\frac{\beta\Gamma}{2\pi}}$}\\
    In order to evaluate this object, we first refer back to Eq.~(\ref{eq:sumn}) and use it to calculate two related sums:
    \begin{align}
    \sum\limits_{n^\prime=0}^{n-1}\frac{n^\prime}{n^\prime+a}&=\sum\limits_{\mathclap{n^\prime=1}}^{n-1}\left(1-\frac{a}{n^\prime+a}\right)=n-1-a\big(\Psi(a+n)-\Psi(a+1)\big)\;,\\
    \sum\limits_{n^\prime=0}^{n-1}\frac{n^{\prime 2}}{n^\prime+a}&=\sum\limits_{\mathclap{n^\prime=1}}^{n-1}\left(n^\prime-a+\frac{a^2}{n^\prime+a}\right)=\frac{(n-2a)(n-1)}{2}+a^2\big(\Psi(a+n)-\Psi(a+1)\big)\;,
    \end{align}
    where $n>0$ and $\text{Re}(a)>0$. Using the above, a straightforward calculation gives
    \begin{align}
    \sum\limits_{n^\prime=0}^{n-1}\frac{\frac{1}{2}-2\left(n^\prime-n+\frac{1}{2}\right)^2}{n^\prime+\frac{1}{2}+\frac{\beta\Gamma}{2\pi}}&=-2n(n-1)\sum\limits_{n^\prime=0}^{n-1}\frac{1}{n^\prime+\frac{1}{2}+\frac{\beta\Gamma}{2\pi}}+2(2n-1)\sum\limits_{n^\prime=0}^{n-1}\frac{n^\prime}{n^\prime+\frac{1}{2}+\frac{\beta\Gamma}{2\pi}}-2\sum\limits_{n^\prime=0}^{n-1}\frac{n^{\prime 2}}{n^\prime+\frac{1}{2}+\frac{\beta\Gamma}{2\pi}}\nonumber\\
    &=3n^2+\frac{\beta\Gamma n}{\pi}+\left(\frac{1}{2}-2\left(n+\frac{\beta\Gamma}{2\pi}\right)^2\right)\left(\Psi\left(\frac{1}{2}+\frac{\beta\Gamma}{2\pi}+n\right)-\Psi\left(\frac{1}{2}+\frac{\beta\Gamma}{2\pi}\right)\right)\;.
    \end{align}
    Expanded to linear order in $n$, this can be written as
    \begin{equation}
    \sum\limits_{n^\prime=0}^{n-1}\frac{\frac{1}{2}-2\left(n^\prime-n+\frac{1}{2}\right)^2}{n^\prime+\frac{1}{2}+\frac{\beta\Gamma}{2\pi}}=\left[\frac{\beta\Gamma}{\pi}+\left(\frac{1}{2}-\frac{\beta^2\Gamma^2}{2\pi^2}\right)\psi^{(1)}\left(\frac{1}{2}+\frac{\beta\Gamma}{2\pi}\right)\right]n+\mathcal{O}(n^2)\;.
    \end{equation}
    $~$


    \item \textbf{The flavor terms: $\pmb{C_{22}+C_{44}+C_{24}+C_{42}}$}\\
    In order to evaluate these terms, four more Green functions are required. Referring back to Sec.~\ref{sec:propagators} and working out the matrix multiplications, they are given by
    \begin{align}
    \sum_{\mu\nu} G_{dd,\mu\nu}&(i\omega_n)=2D_{aa}(i\omega_n)=\frac{2}{i\omega_n}\;,\qquad
    \sum_{\mu\nu} G_{ld,k,\mu\nu}(i\omega_n)=0\;,\nonumber\\
    \sum_{\mu\nu} G_{ll,kk^\prime,\mu\nu}&(i\omega_n)=-2i\hbar\,\delta_{k,k^\prime}\frac{\hbar\omega_n}{(\hbar\omega_n)^2+\epsilon_k^2}-\frac{8g_\bot^2}{L}\frac{\hbar\omega_n}{(\hbar\omega_n)^2+\epsilon_k^2}\frac{\hbar\omega_n}{(\hbar\omega_n)^2+\epsilon_{k^\prime}^2}D_{bb}(i\omega_n)\;,\nonumber\\
    G_{ld,k,11}(i\omega_n)&-G_{ld,k,22}(i\omega_n)-G_{ld,k,12}(i\omega_n)+G_{ld,k,21}(i\omega_n)=\frac{4ig_\bot}{\sqrt{L}}\frac{\hbar\omega_n}{(\hbar\omega_n)^2+\epsilon_k^2}D_{bb}(i\omega_n)\;.\label{eq:unsignedsums}
    \end{align}
    Here, we used that the $a$ Majorana fermion is completely free in absence of a magnetic field, and has a zero energy. Going through the same procedure as for $C_{11}$ and using that the sum over all components of $\mathbf{G}_{ld,k}(i\omega_n)$ is equal to zero, we find
    \begin{align}
    C_{22}(i\Omega_n)&=-\frac{(\pi v_Fg_\bot)^2}{4L^3}\frac{1}{(\hbar\beta)^3}\sum\limits_{\substack{k,k^\prime,k^{\prime\prime}\\q,q^\prime,q^{\prime\prime}}}\sum_{\mu\nu}\sum_{\rho\sigma}\sum\limits_{n^\prime,n^{\prime\prime},n^{\prime\prime\prime}}G_{ff,k^\prime q^{\prime\prime},22}(i\omega_{n^\prime})G_{ff,k^{\prime\prime}q^\prime,11}(i\omega_{n^{\prime\prime}})\nonumber\\
    &\quad\,\times G_{ll,kq,\mu\nu}(i\omega_{n^{\prime\prime\prime}})G_{dd,\rho\sigma}\big(-i(\omega_{n^\prime}+\omega_{n^{\prime\prime}}+\omega_{n^{\prime\prime\prime}}-\Omega_n)\big)\nonumber\\
    &=\frac{\hbar(\pi v_Fg_\bot)^2}{(L\beta)^3}\sum\limits_{k,k^\prime,k^{\prime\prime}}\sum\limits_{n^\prime,n^{\prime\prime},n^{\prime\prime\prime}}\frac{1}{i\hbar\omega_{n^\prime}-\epsilon_k}\frac{1}{i\hbar\omega_{n^{\prime\prime}}-\epsilon_{k^\prime}}\frac{1}{i\hbar\omega_{n^{\prime\prime\prime}}-\epsilon_{k^{\prime\prime}}}\frac{1}{i\hbar(\omega_{n^\prime}+\omega_{n^{\prime\prime}}+\omega_{n^{\prime\prime\prime}}-\Omega_n)}\nonumber\\
    &\quad\,\times\left(1+\frac{4g_\bot^2}{\hbar L}\sum\limits_{k^{\prime\prime\prime}}\frac{1}{i\hbar\omega_{n^{\prime\prime\prime}}-\epsilon_{k^{\prime\prime\prime}}}D_{bb}(i\omega_{n^{\prime\prime\prime}})\right)\;.
    \end{align}
    Also evaluating the sums in the same way as for $C_{11}$ ({\it i.e.}, performing two frequency sums using Eq.~(\ref{eq:matsum}), going to the continuum limit for the momentum sums, introducing the coordinates $\epsilon\equiv(\epsilon_k+\epsilon_{k^\prime})/2$, $\epsilon^\prime\equiv\epsilon_k-\epsilon_{k^\prime}$, and evaluating the integrals over $\epsilon_{k^{\prime\prime}}$, $\epsilon_{k^{\prime\prime\prime}}$ and $\epsilon^\prime$):
    \begin{equation}
    C_{22}(i\Omega_n)=-\frac{\Gamma}{8\hbar\beta}\int\limits_{-\Lambda}^{\Lambda}\mathrm{d}\epsilon\sum\limits_{\mathclap{n^\prime=-\infty}}^\infty\frac{\epsilon}{\tanh(\beta\epsilon)}\frac{\hbar\omega_{n^\prime-n}}{(\hbar\omega_{n^\prime-n})^2+(2\epsilon)^2}\frac{\hbar\omega_{n^\prime}}{|\hbar\omega_{n^\prime}|+\Gamma}\;.\label{eq:C22}
    \end{equation}
    Before going any further, we also calculate the component
    \begin{equation}
    C_{44}(\tau)=\frac{(\pi v_F)^2}{4L^2}\sum\limits_{\mathclap{\substack{k,k^\prime\\q,q^\prime}}}(\epsilon_{k^\prime}-\epsilon_k)(\epsilon_{q^\prime}-\epsilon_q)\big\langle a(\tau)a(0)\big\rangle_0\big\langle b(\tau)b(0)\big\rangle_0\left\langle\psi_{f,k}^\dagger(\tau)\psi_{f,k^\prime}(\tau)\psi_{f,q}^\dagger(0)\psi_{f,q^\prime}(0)\right\rangle_0\;.
    \end{equation}
    Once again following the same procedure as for the previous components, this becomes
    \begin{align}
    C_{44}(i\Omega_n)&=\frac{(\pi v_F)^2}{4L^2\beta^3}\sum\limits_{\mathclap{k,k^\prime}}\sum\limits_{n^\prime,n^{\prime\prime},n^{\prime\prime\prime}}(\epsilon_{k^\prime}-\epsilon_k)^2\frac{1}{i\hbar\omega_{n^\prime}+\epsilon_k}\frac{1}{i\hbar\omega_{n^{\prime\prime}}-\epsilon_{k^\prime}}\frac{1}{i\hbar(\omega_{n^\prime}+\omega_{n^{\prime\prime}}+\omega_{n^{\prime\prime\prime}}-\Omega_n)}D_{bb}(i\omega_{n^{\prime\prime\prime}})\nonumber\\
    &=-\frac{1}{2\hbar\beta}\int\limits_{-\Lambda}^{\Lambda}\mathrm{d}\epsilon\sum\limits_{\mathclap{n^\prime=-\infty}}^\infty\frac{\epsilon^3}{\tanh(\beta\epsilon)}\frac{\hbar\omega_{n^\prime-n}}{(\hbar\omega_{n^\prime-n})^2+(2\epsilon)^2}\frac{1}{\hbar\omega_{n^\prime}+\text{sgn}(\omega_{n^\prime})\Gamma}\;.\label{eq:C44int}
    \end{align}
    Finally, without explicitly going through the calculation, the combination $(C_{24}+C_{42})$ can analogously be derived to be equal to
    \begin{equation}
    C_{24}(i\Omega_n)+C_{42}(i\Omega_n)=-\frac{\Gamma}{\hbar\beta}\int\limits_{-\Lambda}^{\Lambda}\mathrm{d}\epsilon\sum\limits_{\mathclap{n^\prime=-\infty}}^\infty\frac{\epsilon^3}{\tanh(\beta\epsilon)}\frac{1}{(\hbar\omega_{n^\prime-n})^2+(2\epsilon)^2}\frac{1}{|\hbar\omega_{n^\prime}|+\Gamma}\;.\label{eq:C2442}
    \end{equation}

    In order to extract the contribution of the above components to the linear susceptibility, we combine the above four components and discuss them together, starting with Eqs.~(\ref{eq:C22}) and (\ref{eq:C2442}). Combined, these terms can be written as
    \begin{align}
    C&_{22}(i\Omega_n)+C_{24}(i\Omega_n)+C_{42}(i\Omega_n)=-\frac{\Gamma}{2\hbar\beta}\int\limits_{-\Lambda}^{\Lambda}\mathrm{d}\epsilon\sum\limits_{\mathclap{n^\prime=-\infty}}^\infty\frac{\epsilon^3}{\tanh(\beta\epsilon)}\frac{1}{(\hbar\omega_{n^\prime-n})^2+(2\epsilon)^2}\frac{1}{|\hbar\omega_{n^\prime}|+\Gamma}\nonumber\\
    &\!\!-\frac{\Gamma}{8\hbar\beta}\int\limits_{-\Lambda}^{\Lambda}\mathrm{d}\epsilon\sum\limits_{\mathclap{n^\prime=-\infty}}^\infty\frac{\epsilon}{\tanh(\beta\epsilon)}\frac{1}{|\hbar\omega_{n^\prime}|+\Gamma}-\frac{\Gamma\Omega_n}{8\beta}\int\limits_{-\Lambda}^{\Lambda}\mathrm{d}\epsilon\sum\limits_{\mathclap{n^\prime=-\infty}}^\infty\frac{\epsilon}{\tanh(\beta\epsilon)}\frac{\hbar\omega_{n^\prime-n}}{(\hbar\omega_{n^\prime-n})^2+(2\epsilon)^2}\frac{1}{|\hbar\omega_{n^\prime}|+\Gamma}\;.\label{eq:C222442}
    \end{align}
    The second line of this expression clearly does not contribute to the linear susceptibility: the first term does not depend on $n$ at all, while the second term is at least quadratic on $\Omega_n$ (to see this, simply note that the summand is odd in $\omega_{n^\prime}$ if $n=0$, while the sum goes over all $\omega_{n^\prime}$). With that in mind, we unite the four components. Splitting the remaining sums over $n^\prime$ into an $n^\prime<0$ part and an $n^\prime\geq 0$ part, and writing $n^\prime\rightarrow-n^\prime-1$ in the former, we find
    \begin{align}
    C&_{22}(i\Omega_n)+C_{44}(i\Omega_n)+C_{24}(i\Omega_n)+C_{42}(i\Omega_n)=\nonumber\\
    &\text{const.}-\frac{1}{2\hbar\beta}\int\limits_{-\Lambda}^{\Lambda}\mathrm{d}\epsilon\sum\limits_{n^\prime=0}^\infty\frac{\epsilon^3}{\tanh(\beta\epsilon)}\left(\frac{\hbar\omega_{n^\prime+n}+\Gamma}{(\hbar\omega_{n^\prime+n})^2+(2\epsilon)^2}+\frac{\hbar\omega_{n^\prime-n}+\Gamma}{(\hbar\omega_{n^\prime-n})^2+(2\epsilon)^2}\right)\frac{1}{\hbar\omega_{n^\prime}+\Gamma}+\mathcal{O}(\Omega_n^2)\;.\label{eq:Ccombi}
    \end{align}
    Contrary to the previously calculated autocorrelators, the remaining integral cannot be evaluated exactly. As such, we are required to expand in $n$ before having evaluated all of the sums and integrals. Formally, this is the incorrect order of operations, therefore leading to incorrect results if not done carefully. For example, although Eq.~(\ref{eq:Ccombi}) suggests that the remaining sum only contributes to even powers in $n$, this is not necessarily true. The reason for this is hidden in the fact that $\omega_{n^\prime-n}<0$ for some of the terms, such that the identities from Appendix~\ref{ap:digamma} cannot be applied directly. As a result, the sum over the terms containing $\omega_{n^\prime-n}$ evaluates to a different function than the function that emerges from the sum over the terms with $\omega_{n^\prime+n}$. Taking this into account, we have to explicitly evaluate the sum before expanding it in $n$. This is done below. We find that the resulting power series does indeed contain odd powers in $n$, but the linear term is missing. As such, this combination of components does not contribute to the linear susceptibility.\\


 \noindent\textit{\textbf{Expanding terms in $\pmb{n}$}:}\\
    As was discussed above, the combination $C_{22}+C_{44}+C_{24}+C_{42}$ cannot be calculated exactly, such that we have to expand Eq.~(\ref{eq:Ccombi}) in $n$ before evaluating the integral over $\epsilon$. First, the sum over the terms involving $\omega_{n^\prime-n}$ has to be split into two parts. This is necessary due to the conditions $\text{Re}(a)>0$, $\text{Re}(b)>0$ of Eqs.~(\ref{eq:digammasum}) and (\ref{eq:digammasum2}) not being satisfied whenever $n^\prime<n$. Doing so, we find
    \begin{align}
    \sum\limits_{n^\prime=0}^\infty&\left(\frac{\hbar\omega_{n^\prime+n}+\Gamma}{(\hbar\omega_{n^\prime+n})^2+(2\epsilon)^2}+\frac{\hbar\omega_{n^\prime-n}+\Gamma}{(\hbar\omega_{n^\prime-n})^2+(2\epsilon)^2}\right)\frac{1}{\hbar\omega_{n^\prime}+\Gamma}=\nonumber\\
    &\quad\left(\frac{\beta}{2\pi}\right)^2\Bigg(\sum\limits_{n^\prime=0}^\infty\frac{n^\prime+n+\frac{1}{2}+\frac{\beta\Gamma}{2\pi}}{\left(n^\prime+n+\frac{1}{2}\right)^2+\big(\frac{\beta\epsilon}{\pi}\big)^2}\frac{1}{n^\prime+\frac{1}{2}+\frac{\beta\Gamma}{2\pi}}\nonumber\\
    &\quad+\sum\limits_{n^\prime=0}^\infty\frac{n^\prime+\frac{1}{2}+\frac{\beta\Gamma}{2\pi}}{\left(n^\prime+\frac{1}{2}\right)^2+\big(\frac{\beta\epsilon}{\pi}\big)^2}\frac{1}{n^\prime+n+\frac{1}{2}+\frac{\beta\Gamma}{2\pi}}+\sum\limits_{n^\prime=0}^{n-1}\frac{n^\prime+\frac{1}{2}-\frac{\beta\Gamma}{2\pi}}{\left(n^\prime+\frac{1}{2}\right)^2+\big(\frac{\beta\epsilon}{\pi}\big)^2}\frac{1}{n^\prime-n+\frac{1}{2}-\frac{\beta\Gamma}{2\pi}}\Bigg)\;,
    \end{align}
    where we have written $n^\prime\rightarrow n^\prime+n$ in the second sum on the right-hand side, and $n^\prime\rightarrow-n^\prime+n-1$ in the third sum. Furthermore rewriting all of the sums by using
    \begin{equation}
    \frac{x}{x^2+y^2}=\frac{1}{2}\left(\frac{1}{x-iy}+\frac{1}{x+iy}\right)\;,\qquad\frac{1}{x^2+y^2}=\frac{1}{2iy}\left(\frac{1}{x-iy}-\frac{1}{x+iy}\right)\;,
    \end{equation}
    every sum is now of the form of one of the sums from Appendix~\ref{ap:digamma}:
    \begin{align}
    \sum\limits_{n^\prime=0}^\infty&\left(\frac{\hbar\omega_{n^\prime+n}+\Gamma}{(\hbar\omega_{n^\prime+n})^2+(2\epsilon)^2}+\frac{\hbar\omega_{n^\prime-n}+\Gamma}{(\hbar\omega_{n^\prime-n})^2+(2\epsilon)^2}\right)\frac{1}{\hbar\omega_{n^\prime}+\Gamma}=\nonumber\\
    &\quad\frac{\beta^2}{8\pi^2}\left(1-\frac{i\Gamma}{2\epsilon}\right)\Bigg(\sum\limits_{n^\prime=0}^\infty\frac{1}{n^\prime+n+\frac{1}{2}-\frac{i\beta\epsilon}{\pi}}\frac{1}{n^\prime+\frac{1}{2}+\frac{\beta\Gamma}{2\pi}}\nonumber\\
    &\quad+\sum\limits_{n^\prime=0}^\infty\frac{1}{n^\prime+\frac{1}{2}-\frac{i\beta\epsilon}{\pi}}\frac{1}{n^\prime+n+\frac{1}{2}+\frac{\beta\Gamma}{2\pi}}+\sum\limits_{n^\prime=0}^{n-1}\frac{1}{n^\prime+\frac{1}{2}+\frac{i\beta\epsilon}{\pi}}\frac{1}{n^\prime-n+\frac{1}{2}-\frac{\beta\Gamma}{2\pi}}\Bigg)+\text{c.c.}\;.\label{eq:sum24}
    \end{align}
    Evaluating the sums and expanding the result in $n$, one finds that there is no linear term, despite the fact that every other power does appear in the expansion.\\


    \item \textbf{Diagonal component $\pmb{C_{33}}$}\\
    The component $C_{33}$ is very similar to $C_{22}$, so we can straightforwardly modify the previous steps to find
    \begin{align}
    C_{33}(i\Omega_n)&=-\frac{(\Lambda g_\bot)^2}{16\hbar^2L}\frac{1}{\hbar\beta}\sum\limits_{\mathclap{k,k^\prime}}\sum_{\mu\nu}\sum_{\rho\sigma}\sum\limits_{n^\prime=-\infty}^\infty G_{ll,kk^\prime,\mu\nu}(i\omega_{n^\prime})G_{dd,\rho\sigma}(-i\omega_{n^\prime-n})\nonumber\\
    &=-\frac{\Gamma\Lambda^2}{16\hbar\beta}\sum\limits_{\mathclap{n^\prime=-\infty}}^\infty\frac{1}{\hbar\omega_{n^\prime-n}}\frac{\hbar\omega_{n^\prime}}{|\hbar\omega_{n^\prime}|+\Gamma}\;.
    \end{align}
    As is shown below, the latter sum does not contain a linear term in $n$ after evaluation. Consequently, this component does also not contribute to the linear susceptibility.\\


\noindent\textit{\textbf{The sum $\sum_{n^\prime=-\infty}^\infty\frac{1}{n^\prime-n+\frac{1}{2}}\frac{n^\prime+\frac{1}{2}}{|n^\prime+\frac{1}{2}|+\frac{\beta\Gamma}{2\pi}}$:}}\\
    As usual, the first step in the evaluation of this sum is to split it into three parts:
    \begin{align}
    \sum\limits_{\mathclap{n^\prime=-\infty}}^\infty&\frac{1}{n^\prime-n+\frac{1}{2}}\frac{n^\prime+\frac{1}{2}}{|n^\prime+\frac{1}{2}|+\frac{\beta\Gamma}{2\pi}}=\nonumber\\
    &\!\!\sum\limits_{n^\prime=0}^\infty\frac{1}{n^\prime+n+\frac{1}{2}}\frac{n^\prime+\frac{1}{2}}{n^\prime+\frac{1}{2}+\frac{\beta\Gamma}{2\pi}}+\sum\limits_{n^\prime=0}^\infty\frac{1}{n^\prime+\frac{1}{2}}\frac{n^\prime+n+\frac{1}{2}}{n^\prime+n+\frac{1}{2}+\frac{\beta\Gamma}{2\pi}}-\sum\limits_{n^\prime=0}^{n-1}\frac{1}{n^\prime+\frac{1}{2}}\frac{n^\prime-n+\frac{1}{2}}{n^\prime-n+\frac{1}{2}-\frac{\beta\Gamma}{2\pi}}\;.
    \end{align}
    Here, we have again written $n^\prime\rightarrow-n^\prime-1$ in the first part, $n^\prime\rightarrow n^\prime+n$ in the second part, and $n^\prime\rightarrow-n^\prime+n-1$ in the third part. Each of these sums is subsequently split into two more sums, {\it e.g.},
    \begin{equation}
    \sum\limits_{n^\prime=0}^\infty\frac{1}{n^\prime+n+\frac{1}{2}}\frac{n^\prime+\frac{1}{2}}{n^\prime+\frac{1}{2}+\frac{\beta\Gamma}{2\pi}}=\sum\limits_{n^\prime=0}^\infty\frac{1}{n^\prime+n+\frac{1}{2}}-\frac{\beta\Gamma}{2\pi}\sum\limits_{n^\prime=0}^\infty\frac{1}{n^\prime+n+\frac{1}{2}}\frac{1}{n^\prime+\frac{1}{2}+\frac{\beta\Gamma}{2\pi}}\;.
    \end{equation}
    Two of the six resulting sums diverge, and should be understood to have a finite cut-off $N\sim\beta\Lambda$ (see footnote~\ref{fn:limits}). Implementing this cut-off, all of the six sums are now of the form of one of the those discussed in Appendix~\ref{ap:digamma}, such that we find
    \begin{align}
    &\sum\limits_{\mathclap{n^\prime=-N}}^{N-1}\frac{1}{n^\prime-n+\frac{1}{2}}\frac{n^\prime+\frac{1}{2}}{|n^\prime+\frac{1}{2}|+\frac{\beta\Gamma}{2\pi}}=\nonumber\\
    &\,\,\Psi\left(\frac{1}{2}-n+N\right)+\Psi\left(\frac{1}{2}+n+N\right)-\frac{2}{n^2-\big(\frac{\beta\Gamma}{2\pi}\big)^2}\left(n^2\Psi\left(\frac{1}{2}+n\right)-\left(\frac{\beta\Gamma}{2\pi}\right)^2\Psi\left(\frac{1}{2}+\frac{\beta\Gamma}{2\pi}\right)\right)\;,
    \end{align}
    where we have taken $N\rightarrow\infty$ whenever possible. Again, although this expression is not even in $n$, a Taylor expansion in $n$ reveals that it does not contain a linear term.


    \item \textbf{Diagonal component $\pmb{C_{55}}$}\\
    This component is by far the most complicated due to the fact that the spin-flavor modes are coupled to the $b$ Majorana mode, combined with the fact that the propagators corresponding to these modes contain non-zero off-diagonal components. Keeping that in mind, Wick's theorem gives gives us 15 terms to consider. As is discussed further below, five of these terms are vanishing bubble diagrams, while the remaining four bubble diagrams do not have a linear term. For the purpose of finding the linear susceptibility, we therefore only have to consider six terms. Without explicitly performing the lengthy calculation, we note that these combined terms can be written as
    \begin{align}
    C_{55}(i\Omega_n)&=\text{const.}+C_{44}(i\Omega_n)-\frac{(\pi\hbar v_Fg_\bot)^2}{(L\hbar\beta)^3}\sum\limits_{k,k^\prime,k^{\prime\prime}}\sum\limits_{n,n^\prime,n^{\prime\prime}}(\epsilon_k-\epsilon_{k^\prime})(\epsilon_k-\epsilon_{k^{\prime\prime}})\frac{1}{i\hbar(\omega_{n^\prime}+\omega_{n^{\prime\prime}}+\omega_{n^{\prime\prime\prime}}-\Omega_n)}\nonumber\\
    &\quad\,\times\frac{1}{i\hbar\omega_{n^{\prime\prime\prime}}-\epsilon_k}\left(\frac{1}{i\hbar\omega_{n^\prime}+\epsilon_{k^\prime}}-\frac{1}{i\hbar\omega_{n^{\prime\prime}}+\epsilon_{k^\prime}}\right)\frac{1}{i\hbar\omega_{n^{\prime\prime}}+\epsilon_{k^{\prime\prime}}}D_{bb}(i\omega_{n^\prime})D_{bb}(i\omega_{n^{\prime\prime}})+\mathcal{O}(\Omega_n^2)\;.
    \end{align}
    As we show below, the isolated component $C_{44}$ does in fact contain a linear term in $\Omega_n$, however, this term goes to zero at the NFL fixed point. As such, $C_{44}$ does not contribute to the linear susceptibility at this point, and we can instead focus on the other term.
    \begin{figure}[t]
    \centerline{\begin{tabular}{ccc}
    \includegraphics[width=7cm]{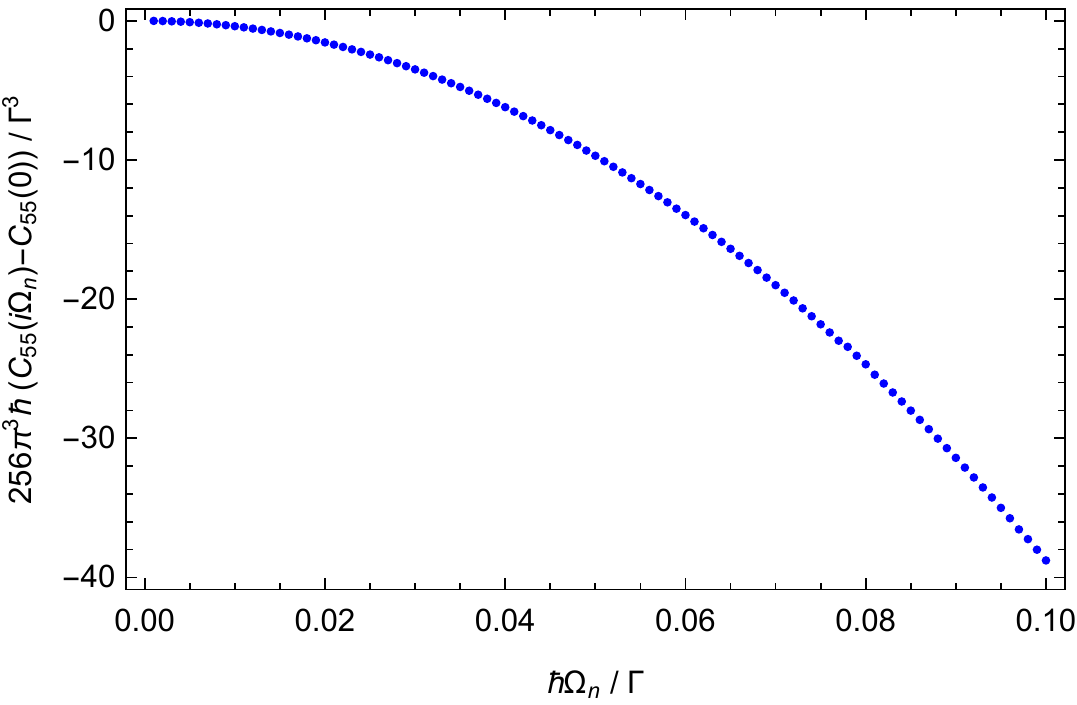} & & \includegraphics[width=7cm]{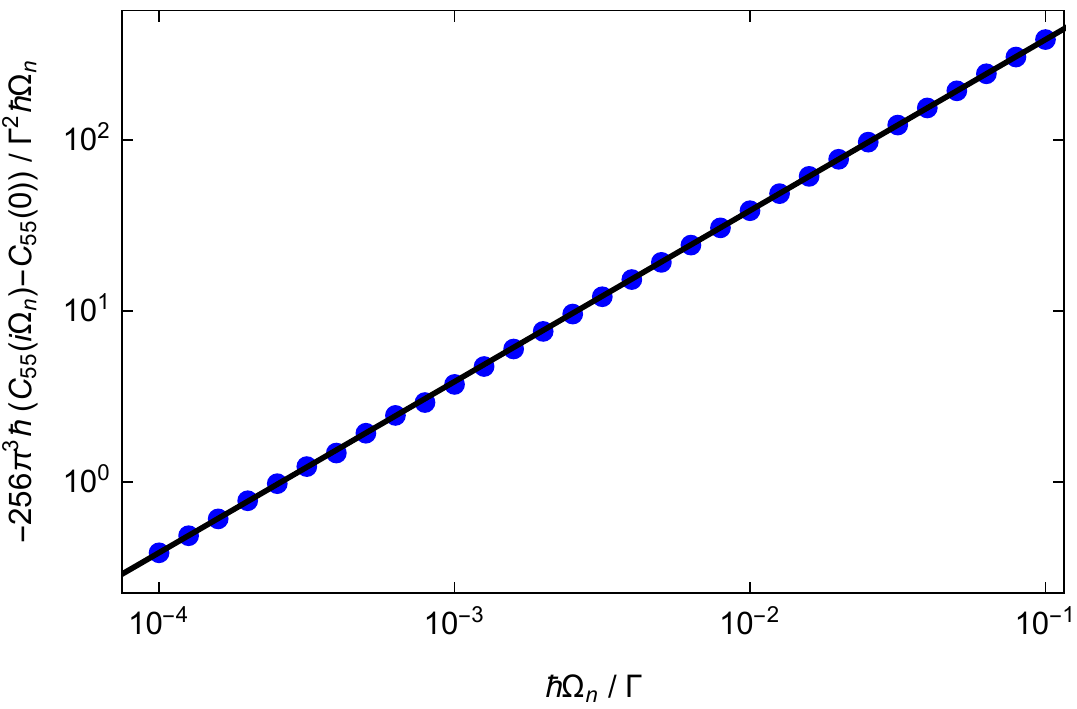}
    \end{tabular}}
    \caption{\label{fig:C55}The component $C_{55}(i\Omega_n)$ at the NFL fixed point, numerically calculated as a function of dimensionless Matsubara frequency $\hbar\Omega_n/\Gamma$ with $\Lambda/\Gamma=10^2$. Left: $C_{55}(i\Omega_n)$ minus its zeroth order term, rescaled with a constant prefactor to make it dimensionless. Right: log-log plot of minus the same object, divided by the dimensionless frequency. The solid line is a function of the form $y=ax$ (its slope in the log-log plot therefore being equal to one), confirming that the susceptibility is perfectly linear in the frequency over this domain. Note that these curves are independent of temperature in the regime $T\ll T_K$.}
    \end{figure}

    Contrary to all of the previously calculated terms, the remaining term cannot be calculated exactly, nor can it be successfully expanded in $\Omega_n$ before evaluation. The reason for this is the presence of an additional $D_{bb}$ propagator that is interwoven in the sums. Instead of using analytical methods, we therefore calculate the sums numerically as a function of $\Omega_n$, and show that the corresponding contribution to the linear susceptibility goes to zero at the NFL fixed point, see below for further details. The results for $\beta\Gamma\rightarrow\infty$ ({\it i.e.}, at the NFL fixed point) are shown in Fig.~\ref{fig:C55}, where we have set the only remaining parameter $\Lambda/\Gamma$ to $10^2$ as an example. As can be seen in the left panel, the lowest non-trivial order term of the component $C_{55}(i\Omega_n)$ is quadratic in $\Omega_n$, similar to what we have seen for most of the other components. Moreover, the right panel shows a log-log plot of the corresponding linear susceptibility $\chi_{55}(i\Omega_n)$ up to a constant prefactor. Analytically continuing to real frequencies, the plot confirms that this contribution to the susceptibility is perfectly linear in $\omega$ over the entire small-$\omega$ region, such that it indeed goes to zero in the dc limit $\omega\rightarrow 0$.\\


\noindent\textit{\textbf{Analytic expression:}}\\
    Five of the terms appearing in $C_{55}$ are bubble diagrams that are proportional to
    \begin{equation}
    \sum\limits_{\mathclap{k,k^\prime}}(\epsilon_{k^\prime}-\epsilon_k)\big\langle\psi_{sf,k}^\dagger(\tau)\psi_{sf,k^\prime}(\tau)\big\rangle_0\;.
    \end{equation}
    As was discussed in Appendix~\ref{ap:bubble}, the expectation value appearing in this expression is invariant under $k\leftrightarrow k^\prime$, such that this entire bubble diagram vanishes after summing over the momenta. In addition to these vanishing bubble diagrams, there are four terms involving the bubble diagrams from Eq.~(\ref{eq:bubblenz}). Carefully combining these terms, relabelling the momenta wherever necessary and discarding the terms that are odd in any of the momenta, they are given by
    \begin{equation}
    \frac{(\pi v_Fg_\bot)^2}{2L^3}\frac{1}{(\hbar\beta)^2}\sum\limits_{\mathclap{k,k^\prime,k^{\prime\prime}}}\epsilon_k^2 D_{aa}(\tau)\sum_{\mu\nu}L_{0,k,\mu\nu}(\tau)\sum\limits_{\mathclap{n,n^\prime}}\frac{\hbar\omega_n}{(\hbar\omega_n)^2+\epsilon_{k^\prime}^2}D_{bb}(i\omega_n)\frac{\hbar\omega_{n^\prime}}{(\hbar\omega_{n^\prime})^2+\epsilon_{k^{\prime\prime}}^2}D_{bb}(i\omega_{n^\prime})\;,
    \end{equation}
    where $L_{0,k,\mu\nu}$ is the $\mu\nu$ component of the spin-flavor propagator $\mathbf{L}_{0,k}$ in absence of tunneling. In terms of Matsubara frequencies, the object we have to calculate is thus given by
    \begin{equation}
    \sum_k\epsilon_k^2\sum\limits_{\mathclap{n^\prime=-\infty}}^\infty D_{aa}(i\omega_{n^\prime})\sum_{\mu\nu}L_{0,k,\mu\nu}(-i\omega_{n^\prime-n})\propto\sum_k\epsilon_k^2\sum\limits_{\mathclap{n^\prime=-\infty}}^\infty\frac{1}{n^\prime+\frac{1}{2}}\frac{n^\prime-n+\frac{1}{2}}{\left(n^\prime-n+\frac{1}{2}\right)^2+\big(\frac{\beta\epsilon_k}{2\pi}\big)^2}\;,
    \end{equation}
    everything else simply being a constant prefactor. Splitting this sum into an $n^\prime<0$ part (sending $n^\prime\rightarrow-n^\prime-1$) and an $n^\prime\geq 0$ part, it is essentially identical to Eq.~(\ref{eq:sum24}) with $\Gamma\rightarrow 0$. Consequently, the above terms do not contain a linear term. The linear contribution of the component $C_{55}$ can thus be calculated from the remaining six terms:
    \begin{align}
    C_{55}(\tau)&\cong\frac{(\pi v_F)^2}{4L^2}\sum\limits_{\mathclap{k,k^\prime,k^{\prime\prime},k^{\prime\prime\prime}}}(\epsilon_{k^{\prime\prime\prime}}-\epsilon_{k^{\prime\prime}})(\epsilon_{k^\prime}-\epsilon_k)\big\langle a(\tau)a(0)\big\rangle_0\,\nonumber\\
    &\quad\,\times\Bigg(\big\langle\psi_{sf,k}^\dagger(\tau)\psi_{sf,k^{\prime\prime\prime}}(0)\big\rangle_0\,\big\langle\psi_{sf,k^\prime}(\tau)\psi_{sf,k^{\prime\prime}}^\dagger(0)\big\rangle_0\big\langle b(\tau)b(0)\big\rangle_0\nonumber\\
    &\quad\,-\big\langle\psi_{sf,k}^\dagger(\tau)\psi_{sf,k^{\prime\prime}}^\dagger(0)\big\rangle_0\,\big\langle\psi_{sf,k^\prime}(\tau)\psi_{sf,k^{\prime\prime\prime}}(0)\big\rangle_0\,\big\langle b(\tau)b(0)\big\rangle_0\nonumber\\
    &\quad\,+\big\langle\psi_{sf,k}^\dagger(\tau)\psi_{sf,k^{\prime\prime\prime}}(0)\big\rangle_0\,\big\langle\psi_{sf,k^\prime}(\tau)b(0)\big\rangle_0\,\big\langle\psi_{sf,k^{\prime\prime}}^\dagger(0)b(\tau)\big\rangle_0\nonumber\\
    &\quad\,-\big\langle\psi_{sf,k}^\dagger(\tau)\psi_{sf,k^{\prime\prime}}^\dagger(0)\big\rangle_0\,\big\langle\psi_{sf,k^\prime}(\tau)b(0)\big\rangle_0\,\big\langle\psi_{sf,k^{\prime\prime\prime}}(0)b(\tau)\big\rangle_0\nonumber\\
    &\quad\,+\big\langle\psi_{sf,k}^\dagger(\tau)b(0)\big\rangle_0\,\big\langle\psi_{sf,k^\prime}(\tau)\psi_{sf,k^{\prime\prime}}^\dagger(0)\big\rangle_0\,\big\langle\psi_{sf,k^{\prime\prime\prime}}(0)b(\tau)\big\rangle_0\nonumber\\
    &\quad\,-\big\langle\psi_{sf,k}^\dagger(\tau)b(0)\big\rangle_0\,\big\langle\psi_{sf,k^\prime}(\tau)\psi_{sf,k^{\prime\prime\prime}}(0)\big\rangle_0\,\big\langle\psi_{sf,k^{\prime\prime}}^\dagger(0)b(\tau)\big\rangle_0\Bigg)\;.
    \end{align}

    At this point, it is a matter of plugging in the propagators from Sec.~\ref{sec:propagators} and simplifying the result, mostly by relabelling indices and using that any term that is odd in any of the momenta vanishes after summation. A straightforward but very lengthy calculation leads to the following result:
    \begin{align}
    &C_{55}(i\Omega_n)\cong\nonumber\\
    &\,-\frac{i}{16\hbar\beta^3}\int\limits_{-\Lambda}^\Lambda\mathrm{d}\epsilon_k\int\limits_{-\Lambda}^\Lambda\mathrm{d}\epsilon_{k^\prime}\sum\limits_{\mathclap{n^\prime,n^{\prime\prime},n^{\prime\prime\prime}}}(\epsilon_k+\epsilon_{k^\prime})^2\frac{1}{i\hbar\omega_{n^{\prime\prime\prime}}-\epsilon_k}\frac{1}{i\hbar\omega_{n^{\prime\prime}}-\epsilon_{k^\prime}}\frac{1}{i\hbar(\omega_{n^\prime}+\omega_{n^{\prime\prime}}+\omega_{n^{\prime\prime\prime}}-\Omega_n)}\frac{1}{\hbar\omega_{n^\prime}+\text{sgn}(\omega_{n^\prime})\Gamma}\nonumber\\
    &\,+\frac{\Gamma}{16\pi\hbar\beta^3}\int\limits_{-\Lambda}^\Lambda\mathrm{d}\epsilon_k\int\limits_{-\Lambda}^\Lambda\mathrm{d}\epsilon_{k^\prime}\int\limits_{-\Lambda}^\Lambda\mathrm{d}\epsilon_{k^{\prime\prime}}\sum\limits_{\mathclap{n^\prime,n^{\prime\prime},n^{\prime\prime\prime}}}(\epsilon_k+\epsilon_{k^\prime})(\epsilon_k+\epsilon_{k^{\prime\prime}})\left(\frac{1}{i\hbar\omega_{n^\prime}-\epsilon_{k^{\prime\prime}}}-\frac{1}{i\hbar\omega_{n^{\prime\prime}}-\epsilon_{k^{\prime\prime}}}\right)\frac{1}{i\hbar\omega_{n^{\prime\prime\prime}}-\epsilon_k}\nonumber\\
    &\,\times\frac{1}{i\hbar\omega_{n^{\prime\prime}}-\epsilon_{k^\prime}}\frac{1}{i\hbar(\omega_{n^\prime}+\omega_{n^{\prime\prime}}+\omega_{n^{\prime\prime\prime}}-\Omega_n)}\frac{1}{\hbar\omega_{n^\prime}+\text{sgn}(\omega_{n^\prime})\Gamma}\frac{1}{\hbar\omega_{n^{\prime\prime}}+\text{sgn}(\omega_{n^{\prime\prime}})\Gamma}\;,
    \end{align}
    where the first line can be identified as $C_{44}(i\Omega_n)$, see Eq.~(\ref{eq:C44int}). Splitting the sum from Eq.~(\ref{eq:C44int}) into two parts, we recognize that $C_{44}$ can be evaluated by using Eq.~(\ref{eq:sum24}), only with $\Gamma\rightarrow 0$ in the numerators on the left-hand side and $(1-i\Gamma/2\epsilon)\rightarrow 1$ on the right-hand side. Going through the familiar procedure to evaluate the sum and expanding the result in $n$, we obtain
    \begin{align}
    &C_{44}(i\Omega_n)=\nonumber\\
    &\,\text{const.}+\frac{\beta}{8\pi^2\hbar}\int\limits_{-\Lambda}^{\Lambda}\mathrm{d}\epsilon\frac{\epsilon^3}{\tanh(\beta\epsilon)}\frac{\frac{\beta\Gamma}{2\pi}}{\big(\frac{\beta\Gamma}{2\pi}\big)^2+\big(\frac{\beta\epsilon}{\pi}\big)^2}\left[\psi^{(1)}\left(\frac{1}{2}-\frac{i\beta\epsilon}{\pi}\right)+\psi^{(1)}\left(\frac{1}{2}+\frac{i\beta\epsilon}{\pi}\right)\right]n+\mathcal{O}(n^2)\;.
    \end{align}
    Since we are interested in the NFL fixed point, we furthermore expand to lowest order in $T/T_K\sim 1/\beta\Gamma$, allowing us to evaluate the remaining integral:
    \begin{equation}
    C_{44}(i\Omega_n)=\text{const.}+\frac{\pi^4\Omega_n}{256\beta^2}\frac{1}{\beta\Gamma}+\mathcal{O}\big(\Omega_n^2,1/(\beta\Gamma)^2\big)\;.
    \end{equation}
    Recognizing that $\beta\Gamma\rightarrow\infty$ at the NFL fixed point, we conclude that $C_{44}$ does not have a linear term at this point.

    For the remaining terms, we first evaluate the sum over $n^{\prime\prime}$ and the integrals over $\epsilon_{k^\prime}$ and $\epsilon_{k^{\prime\prime}}$ while keeping $\Lambda$ finite. Taking into account that $C_{44}$ can be discarded at the NFL fixed point, $C_{55}$ becomes
    \begin{align}
    &C_{55}(i\Omega_n)\cong-\frac{\Gamma}{64\pi\hbar\beta^2}\int\limits_{-\Lambda}^\Lambda\mathrm{d}\epsilon_k\sum\limits_{\mathclap{n^\prime,n^{\prime\prime}}}\tanh\left(\frac{\beta\epsilon_k}{2}\right)\frac{1}{i\hbar(\omega_{n^\prime}+\omega_{n^{\prime\prime}}-\Omega_n)+\epsilon_k}\frac{1}{\hbar\omega_{n^\prime}+\text{sgn}(\omega_{n^\prime})\Gamma}\frac{1}{\hbar\omega_{n^{\prime\prime}}+\text{sgn}(\omega_{n^{\prime\prime}})\Gamma}\nonumber\\
    &\,\times\left((\epsilon_k+i\hbar\omega_{n^\prime})\left(2\arctan\left(\frac{\hbar\omega_{n^\prime}}{\Lambda}\right)-\text{sgn}(\omega_{n^\prime})\pi\right)-(\epsilon_k+i\hbar\omega_{n^{\prime\prime}})\left(2\arctan\left(\frac{\hbar\omega_{n^{\prime\prime}}}{\Lambda}\right)-\text{sgn}(\omega_{n^{\prime\prime}})\pi\right)\right)^2\;.\label{eq:C55equiv}
    \end{align}
    This is as far as we will go without falling back to numerical methods, performed below.\\


\noindent \textit{\textbf{Numerical evaluation:}}\\
    In order to calculate Eq.~(\ref{eq:C55equiv}) numerically at the NFL fixed point, we first switch to dimensionless variables. In particular, we define $\epsilon\equiv\epsilon_k/\Gamma$, $\omega^\prime\equiv\hbar\omega_{n^\prime}/\Gamma$, $\omega^{\prime\prime}\equiv\hbar\omega_{n^{\prime\prime}}/\Gamma$. This choice allows us to take the continuum limit of the sums over the Matsubara frequencies: the step sizes are $\Delta\omega^\prime=\Delta\omega^{\prime\prime}=2\pi/\beta\Gamma\rightarrow 0$, such that we can write
    \begin{equation}
    \sum\limits_{n^\prime=-\infty}^\infty\rightarrow\frac{\beta\Gamma}{2\pi}\int\limits_{-\infty}^\infty\mathrm{d}\omega^\prime\;,
    \end{equation}
    and similarly for $\omega_{n^{\prime\prime}}$. In terms of these dimensionless variables, we have
    \begin{align}
    &C_{55}(i\Omega_n)\cong-\frac{\Gamma^3}{256\pi^3\hbar}\int\limits_{-\Lambda/\Gamma}^{\Lambda/\Gamma}\mathrm{d}\epsilon\int\limits_{-\infty}^\infty\mathrm{d}\omega^\prime\int\limits_{-\infty}^\infty\mathrm{d}\omega^{\prime\prime}\tanh\left(\frac{\beta\Gamma\epsilon}{2}\right)\frac{1}{i\left(\omega^\prime+\omega^{\prime\prime}-\frac{\hbar\Omega_n}{\Gamma}\right)+\epsilon}\frac{1}{\omega^\prime+\text{sgn}(\omega^\prime)}\frac{1}{\omega^{\prime\prime}+\text{sgn}(\omega^{\prime\prime})}\nonumber\\
    &\,\times\left((\epsilon+i\omega^\prime)\left(2\arctan\left(\frac{\omega^\prime}{\Lambda/\Gamma}\right)-\text{sgn}(\omega^\prime)\pi\right)-(\epsilon+i\omega^{\prime\prime})\left(2\arctan\left(\frac{\omega^{\prime\prime}}{\Lambda/\Gamma}\right)-\text{sgn}(\omega^{\prime\prime})\pi\right)\right)^2\;.\label{eq:C55num}
    \end{align}
    We can now treat this as a function of $\hbar\Omega_n/\Gamma$, and depending on two parameters $\Lambda/\Gamma$ and $\beta\Gamma$. The former parameter should be interpreted as large but finite, while the latter is sent to infinity at the NFL fixed point, such that we can write $\tanh(\beta\Gamma\epsilon/2)\rightarrow\text{sgn}(\epsilon)$. Fixing the parameter $\Lambda/\Gamma$ ({\it i.e.}, the ratio of the cut-off energy scale to the Kondo energy scale), we have numerically calculated Eq.~(\ref{eq:C55num}) for many small values of the dimensionless frequency $\hbar\Omega_n/\Gamma$, leading to the plots shown in Fig.~\ref{fig:C55}. Note that Eq.~(\ref{eq:C55num}) with $\tanh(\beta\Gamma\epsilon/2)\rightarrow\text{sgn}(\epsilon)$ ({\it i.e.}, the NFL fixed point) is valid for \emph{finite} temperatures, meaning that this component is independent of temperature for any finite temperatures $T\ll T_K$. This is different from the qualitative behavior of the contributing component $C_{11}$ from Eq.~(\ref{eq:C11Tsquare}), which is proportional to $T^2$ in this regime.
\end{itemize}


\section{Derivation of the self-energy away from the EK point}\label{ap:selfenergy}
In this Appendix, we derive the diagramatic expression from Eq.~(\ref{eq:diagrams}) explicitly. To do so, we will use the fact that $\big\langle T_\tau b(\tau)a(\tau^\prime)\big\rangle_0$ is proportional to the magnetic field $B$ and therefore vanishes in the limit $T\gg T^*$, significantly simplifying the process. Starting with the partition function and using the interaction Hamiltonian from Eq.~(\ref{eq:Hint}) together with Wick's theorem:
\begin{align}
Z^\text{full}&=Z\bigg(1-\frac{i\lambda}{L\hbar}\int\limits_0^{\hbar\beta}\mathrm{d}\tau\big\langle b(\tau)a(\tau)\sum_{k,k^\prime}:\psi^\dagger_{s,k}(\tau)\psi_{s,k^\prime}(\tau):\big\rangle_0\nonumber\\
&\quad\,-\frac{\lambda^2}{2L^2\hbar^2}\int\limits_0^{\hbar\beta}\mathrm{d}\tau\int\limits_0^{\hbar\beta}\mathrm{d}\tau^\prime\big\langle T_\tau b(\tau)a(\tau)\sum_{k,k^\prime}:\psi^\dagger_{s,k}(\tau)\psi_{s,k^\prime}(\tau):b(\tau^\prime)a(\tau^\prime)\sum_{k^{\prime\prime},k^{\prime\prime\prime}}:\psi^\dagger_{s,k^{\prime\prime}}(\tau^\prime)\psi_{s,k^{\prime\prime\prime}}(\tau^\prime):\big\rangle_0+\mathcal{O}(\lambda^3)\bigg)\nonumber\\
&=Z\bigg(1+\frac{\lambda^2}{2L^2\hbar^2}\int\limits_0^{\hbar\beta}\mathrm{d}\tau\int\limits_0^{\hbar\beta}\mathrm{d}\tau^\prime D_{aa}(\tau-\tau^\prime)D_{bb}(\tau-\tau^\prime)\sum_{k,k^\prime}\left(G_{s,k}(0)G_{s,k^\prime}(0)-G_{s,k}(\tau-\tau^\prime)G_{s,k^\prime}(\tau^\prime-\tau)\right)+\mathcal{O}(\lambda^3)\bigg)\nonumber\\
&\equiv ZZ^\prime\;.
\end{align}
Moreover, $\sum_kG_{s,k}(0)=0$, see Appendix~\ref{ap:bubble}.

With this information, we can calculate full propagator of interest as well. Using the same methods as before and utilizing $D_{aa/bb}(\tau)=-D_{aa/bb}(-\tau)$, we find
\begin{align}
D_{bb}^\text{full}(\tau-\tau^\prime)&=\frac{1}{Z^\prime}\bigg(D_{bb}(\tau-\tau^\prime)+\frac{\lambda^2}{2L^2\hbar^2}\int\limits_0^{\hbar\beta}\mathrm{d}\tau^{\prime\prime}\int\limits_0^{\hbar\beta}\mathrm{d}\tau^{\prime\prime\prime}\nonumber\\
&\quad\,\big\langle T_\tau b(\tau)b(\tau^\prime)b(\tau^{\prime\prime})a(\tau^{\prime\prime})\sum_{k,k^\prime}:\psi^\dagger_{s,k}(\tau^{\prime\prime})\psi_{s,k^\prime}(\tau^{\prime\prime}):b(\tau^{\prime\prime\prime})a(\tau^{\prime\prime\prime})\sum_{k^{\prime\prime},k^{\prime\prime\prime}}:\psi^\dagger_{s,k^{\prime\prime}}(\tau^{\prime\prime\prime})\psi_{s,k^{\prime\prime\prime}}(\tau^{\prime\prime\prime}):\big\rangle_0+\mathcal{O}(\lambda^3)\bigg)\nonumber\\
&=\frac{1}{Z^\prime}\bigg(D_{bb}(\tau-\tau^\prime)-\frac{\lambda^2}{2L^2\hbar^2}\int\limits_0^{\hbar\beta}\mathrm{d}\tau^{\prime\prime}\int\limits_0^{\hbar\beta}\mathrm{d}\tau^{\prime\prime\prime}D_{aa}(\tau^{\prime\prime}-\tau^{\prime\prime\prime})\sum_{k,k^\prime}G_{s,k}(\tau^{\prime\prime}-\tau^{\prime\prime\prime})G_{s,k^\prime}(\tau^{\prime\prime\prime}-\tau^{\prime\prime})\nonumber\\
&\quad\,\times\big(D_{bb}(\tau-\tau^\prime)D_{bb}(\tau^{\prime\prime}-\tau^{\prime\prime\prime})+D_{bb}(\tau-\tau^{\prime\prime})D_{bb}(\tau^{\prime\prime\prime}-\tau^\prime)-D_{bb}(\tau-\tau^{\prime\prime\prime})D_{bb}(\tau^{\prime\prime}-\tau^\prime)\big)+\mathcal{O}(\lambda^3)\bigg)\nonumber\\
&=D_{bb}(\tau-\tau^\prime)-\frac{\lambda^2}{L^2\hbar^2}\int\limits_0^{\hbar\beta}\mathrm{d}\tau^{\prime\prime}\int\limits_0^{\hbar\beta}\mathrm{d}\tau^{\prime\prime\prime}D_{aa}(\tau^{\prime\prime}-\tau^{\prime\prime\prime})\sum_{k,k^\prime}G_{s,k}(\tau^{\prime\prime}-\tau^{\prime\prime\prime})G_{s,k^\prime}(\tau^{\prime\prime\prime}-\tau^{\prime\prime})\nonumber\\
&\quad\,\times D_{bb}(\tau-\tau^{\prime\prime})D_{bb}(\tau^{\prime\prime\prime}-\tau^\prime)+\mathcal{O}(\lambda^3)\;.
\end{align}
Finally, we rewrite this in terms of Matsubara frequencies, similar to Eq.~(\ref{eq:Ctmat}). Using the $\tau$ integrals to obtain Kronecker deltas, we indeed find
\begin{equation}
D_{bb}^\text{full}(i\omega_n)=D_{bb}(i\omega_n)+D_{bb}(i\omega_n)\Sigma(i\omega_n)D_{bb}(i\omega_n)+\mathcal{O}(\lambda^3)\;,
\end{equation}
where the self-energy is given by Eq.~(\ref{eq:selfenergyfull}).

\clearpage

\nocite{apsrev41Control}
\bibliographystyle{apsrev4-1}

\bookmarksetup{startatroot}
\stepcounter{newsections}
\addcontentsline{toc}{section}{References}

\end{fmffile}

\end{document}